\documentclass[
  reprint,
  amsmath,
  amssymb,
  aps,
  floatfix,
]{revtex4-2}

\usepackage{graphicx}
\usepackage{dcolumn}
\usepackage{bm}
\usepackage{nccmath}
\usepackage{amsmath}
\usepackage{empheq}
\usepackage{subfigure}
\usepackage{makecell}
\usepackage{comment}
\usepackage{multirow}
\usepackage{tabularx}
\usepackage{booktabs}
\usepackage{float}

\begin{document}

\preprint{APS/123-QED}

\title{Black holes immersed in modified Chaplygin-like dark fluid and cloud of strings: geodesics, shadows, and images}%

\author{Xiang-Qian Li$^{a,b}$} \email[Corresponding author: ]{lixiangqian@tyut.edu.cn}
\author{Yoonbai Kim $^{b}$} \email[Corresponding author: ]{yoonbai@skku.edu}
\author{Bum-Hoon Lee $^{c,d}$} \email[Corresponding author: ]{bhl@sogang.ac.kr}
\author{Hao-Peng Yan$^{a}$} \email{yanhaopeng@tyut.edu.cn}
\author{Xiao-Jun Yue$^{a}$} \email{yuexiaojun@tyut.edu.cn}
\affiliation{$^{a}$ College of Physics, Taiyuan University of Technology, Taiyuan 030024, China}
\affiliation{$^{b}$ Department of Physics, Sungkyunkwan University, Suwon 16419, Korea}
\affiliation{$^{c}$ Center for Quantum Spacetime, Sogang University, Seoul, 121-742, Korea}
\affiliation{$^{d}$ Department of Physics, Sogang University, Seoul, 121-742, Korea}

\begin{abstract}
This study investigates a black hole surrounded by a cloud of strings and a cosmological dark fluid characterized by a modified Chaplygin-like equation of state (MCDF), $p=A\rho-B/\rho^{\beta}$. We analyze its geodesic structure, shadow, and optical appearance. Analysis of the effective potential and epicyclic frequencies reveals that the existence of innermost/outermost stable circular orbits (ISCOs/OSCOs) for timelike particles is controlled by the parameters of the MCDF and the cloud of strings. The behavior of orbital conserved quantities and the Keplerian frequency are also examined. By equating the influence of the MCDF on the spacetime metric at spatial infinity with that of a cosmological constant, we constrain the MCDF parameters using the observed shadow radii of Sgr A* and M87*. We investigate the effects of the cloud of strings and MCDF on the black hole's shadows and optical images, assuming various thin disk accretion profiles. Using the method developed by Wald and collaborators, light trajectories are classified by their impact parameters into direct emission, the lensing ring, and the photon ring. The presence of OSCOs can lead to the existence of outer edges in the direct emission and lensing ring images. Observed brightness primarily originates from direct emission, with a minor contribution from the lensing ring, while the photon ring's contribution is negligible due to extreme demagnification. The influence of the cloud of strings and MCDF parameters on all results is analyzed throughout the study.
\end{abstract}

\maketitle


\section{Introduction}
\label{sec:intro}
The Event Horizon Telescope (EHT) collaboration has significantly advanced our understanding of supermassive black holes by releasing groundbreaking Very Long Baseline Interferometry (VLBI) observations of the black hole at the center of the Messier 87 galaxy, known as M87$^{\ast}$~\cite{Narayan:2019imo,EventHorizonTelescope:2019dse,EventHorizonTelescope:2019uob,EventHorizonTelescope:2019jan,EventHorizonTelescope:2019ths,EventHorizonTelescope:2019pgp,EventHorizonTelescope:2019ggy}, as well as that at the center of the Milky Way, referred to as Sgr A$^{\ast}$~\cite{EventHorizonTelescope:2022wkp}. These observations, achieving angular resolutions comparable to the event horizon scale, have revealed a central dark region—commonly referred to as the black hole shadow—surrounded by a bright photon ring~\cite{Takahashi:2004xh}. The theoretical study of light deflection in strong gravitational fields traces back to Synge’s seminal work~\cite{Synge:1966okc}, and was later extended by Bardeen, who calculated the shadow radius of a static Schwarzschild black hole to be $r_{\rm sh}=3M$, and further demonstrated that rotation deforms the shadow from a perfect circle~\cite{Bardeen:1972fi}. It is now widely accepted that astrophysical black holes are not isolated but are instead embedded within luminous accretion flows, which strongly influence the observed image. The first theoretical image of a geometrically thin accretion disk around a Schwarzschild black hole was proposed by Luminet in 1979~\cite{Luminet:1979nyg}, followed by studies of spherical accretion confirming the robustness of the shadow structure~\cite{Falcke:1999pj}. More recently, Perlick, Tsupko, and collaborators analytically investigated the shadow of a Schwarzschild black hole in an expanding universe driven by a positive cosmological constant~\cite{Perlick:2018iye}. These developments have spurred a surge of interest in exploring black hole shadows in various gravitational and cosmological settings~\cite{Gibbons:2008rj,Werner:2012rc,Crisnejo:2018uyn,Kumar:2018ple,Guo:2019lur,Gralla:2019drh, Jusufi:2020cpn,Kumar:2020owy,Zeng:2020dco,Gan:2021pwu,Guo:2021bwr,Jusufi:2020zln,Saurabh:2020zqg, Gralla:2019xty,Peng:2020wun,Chakhchi:2022fls,Guo:2021bhr,He:2022yse,Li:2021riw,Zeng:2021dlj,Zeng:2021mok,He:2021htq, Guo:2022rql, Guerrero:2021ues,Yan:2021ygy,Guerrero:2022qkh,Atamurotov:2021hck,Kumar:2019ohr,Kumar:2017tdw,Hu:2022lek, Heydari-Fard:2023ent,Pulice:2023dqw,Yang:2022btw,Ma:2022jsy,Wang:2023rjl,Kumaran:2023brp,Huang:2023ilm,Zhang:2022osx,Hu:2023bzy,Zeng:2024ptv,Wang:2025ihg,Shu:2024tut,Zheng:2024ftk,Li:2024owp,Cao:2024kht,Meng:2024puu,Chen:2023qic,Chen:2023wzv,Hu:2023pyd,Zeng:2023tjb,Gao:2023mjb,Hu:2024ogh,Aslam:2024bmx,Hu:2024yhh,Zeng:2025xoe,He:2025rjq,He:2024amh,He:2025qmq,He:2025hbu}.

Astronomical observations have confirmed that the universe is currently undergoing an accelerated expansion, a phenomenon widely attributed to an unknown component termed dark energy, which is characterized by negative pressure and positive energy density~\cite{SupernovaCosmologyProject:1998vns,SupernovaSearchTeam:1998fmf,SupernovaSearchTeam:1998cav,Kochappan:2024jyf,Kim:2024gfn}. One plausible explanation for this negative pressure is quintessence dark energy, described by the equation of state $p = \omega\rho$, where the state parameter $\omega$ lies in the range $-1 < \omega < -1/3$~\cite{Wang:1999fa,Bahcall:1999xn,Liu:2025ulr}. A static, spherically symmetric black hole solution incorporating quintessence matter was first proposed by Kiselev~\cite{Kiselev:2002dx}, paving the way for further investigations into the influence of quintessence dark energy on black hole shadows~\cite{Lacroix:2012nz,Haroon:2018ryd,Khan:2020ngg,Zeng:2020vsj,He:2021aeo,Heydari-Fard:2022jdu}. In addition, unified models that aim to describe both dark matter and dark energy have been developed, with the Chaplygin gas and its generalizations emerging as prominent candidates. These models have garnered considerable interest for their ability to account for the observed accelerated expansion~\cite{Kamenshchik:2001cp,Bilic:2001cg,Bento:2002ps}, address the Hubble tension~\cite{Sengupta:2023yxh}, and describe the evolution of cosmological perturbations~\cite{Abdullah:2021tee}. Although commonly employed in cosmological modeling, the Chaplygin gas equation of state, $p = -\frac{B}{\rho}$, is not merely phenomenological; rather, it arises naturally within the framework of string theory~\cite{Ogawa:2000gj,Bordemann:1993ep,Jackiw:2000cc}. Recently, analytical solutions and corresponding thermodynamic quantities were derived for charged, static, spherically symmetric black holes surrounded by Chaplygin-like dark fluid (CDF) within Lovelock gravity~\cite{Li:2019lhr}. This framework has since been extended to include the modified Chaplygin gas (MCG), with an equation of state $p = A\rho - \frac{B}{\rho^\beta}$, to examine the stability of MCG-surrounded black holes in both Einstein-Gauss-Bonnet~\cite{Li:2019ndh} and Lovelock~\cite{Li:2022csn} gravity theories. Inspired by thermodynamic studies, Ref.~\cite{Li:2023zfl} investigated the phase transitions and critical phenomena of static, spherically symmetric AdS black holes surrounded by CDF in general relativity, highlighting the correspondence between black hole thermodynamics and optical features. Furthermore, the geodesic structure, shadow, and optical appearance of black holes immersed in CDF were systematically explored in Ref.~\cite{Li:2024abk}. Additional research on black holes in CDF or MCG backgrounds can be found in Refs.~\cite{Ali:2020omz,Ali:2024rrm,Arora:2023mve,Sekhmani:2023plr,Zhang:2024fxj,Javed:2024lyd,Javed:2024jad,Zahid:2024kyf,Rahmatov:2024foj,Zare:2024zmg,Sekhmani:2024udl,Fathi:2024kda,Becar:2024agj}.

Although the term modified Chaplygin gas (MCG) was used in our previous works~\cite{Li:2019ndh,Li:2022csn}, the effect of cosmic expansion was not considered in those studies. Therefore, to distinguish our model from the cosmological usage of MCG, we refer to it as modified Chaplygin-like dark fluid (MCDF) in the present work. In this paper, we investigate the geodesic structures, shadows, and optical appearances of black holes immersed in MCDF and a cloud of strings under different accretion conditions. In particular, we focus on both geometrically thin and optically thin disk accretion models. It is anticipated that the shadow and optical appearance of black holes in an MCDF background could offer observational constraints on the MCDF and cloud of strings models, particularly through EHT observations. In many accretion disk models, the innermost stable circular orbit plays a crucial role in determining observable features, motivating a detailed study of geodesic structures around such black holes. Assuming that the MCDF behaves similarly to a cosmological constant at spatial infinity and that a static observer is located at the Earth's position observing Sgr A$^{\ast}$ and M87$^{\ast}$, we utilize observational data on shadow radii to constrain the parameters of both the MCDF and the cloud of strings. When analyzing the optical appearances of black holes, we assume a static observer situated near the pseudo-cosmological horizon, enabling us to construct black hole images using the impact parameter as the coordinate scale.

The structure of this paper is organized as follows: In Section~\ref{secmetric}, we derive a static spherically symmetric black hole solution within Einstein gravity in the presence of both a MCDF and a cloud of strings. Section~\ref{secgeodesic} explores the properties of both timelike and null geodesics in the resulting spacetime. In Section~\ref{disk}, we investigate the optical appearances of black holes surrounded by geometrically and optically thin accretion disks, employing three distinct emission profiles proposed by Gralla–Lupsasca–Marrone (GLM). Finally, Section~\ref{conclusion} concludes the paper with a summary of the main results and a discussion of their implications.

\section{Static spherically-symmetric black holes immersed in MCDF and cloud of strings}
\label{secmetric}
We begin by considering a black hole spacetime in the background of MCDF and a cloud of strings. We assume the following ansatz for the metric
\begin{equation}
ds^2 = -f(r)\,dt^2 + \frac{1}{f(r)}\,dr^2 + r^2 d\Omega^2, \label{dsf}
\end{equation}
where $f(r)$ is a general function of the radial coordinate $r$, and $d\Omega^2=d\theta^2+{\rm sin}^2\theta d\phi^2$ denotes the standard line element on the two-sphere $S^2$.

Recent theoretical developments favor modeling the universe as a collection of extended objects rather than point-like particles, with one-dimensional strings being the most natural and widely accepted candidates. The general spherically symmetric solution for a string cloud was first obtained by Letelier~\cite{Letelier:1979ej}. The energy-momentum tensor of the cloud of strings in 4-dimension can be written as~\cite{Letelier:1979ej,Ghosh:2014pga}
\begin{equation}
T^{\nu}_{\mu} = \frac{a}{r^{2}} \, \mathrm{Diag}[1, 1, 0, 0],
\end{equation}
which leads to the following form of the lapse function
\begin{equation}
f(r) = 1 - a - \frac{2M}{r}, \label{metricofsc}
\end{equation}
where the constant \( a \) characterizes the influence of the cloud of strings on the spacetime geometry.

We next consider the interaction between the MCDF and the spacetime geometry. In our given context, the MCDF is characterized by a non-linear equation of state, expressed as $p=A{\rho}-\frac{B}{{\rho}^{\beta}}$, where $A$ and $B$ are positive parameters and $\beta$ stays in the interval $0\leq\beta\leq1$. As shown in~\cite{Li:2022csn}, the energy-momentum tensor for MCDF in four-dimensional spacetime takes the following form
\begin{equation}
T^t_t = T^r_r = -\rho(r), \label{eq:Ttt}
\end{equation}
\begin{equation}
T^\theta_\theta = T^\phi_\phi = \left(1 + 3A\right)\frac{\rho(r)}{2} - \frac{3B}{2\left[\rho(r)\right]^\beta}. \label{eq:Ttheta}
\end{equation}
By solving Einstein's field equations, the energy density distribution of the MCDF can be obtained as
\begin{equation}
\rho(r) = \left[ \frac{1}{1 + A} \left( B + \left[ \frac{Q}{r^3} \right]^{\omega} \right) \right]^{\frac{1}{1 + \beta}}, \label{eq:rho}
\end{equation}
with $\omega=(1+A)(1+\beta)$ and $Q>0$ a normalization factor representing the intensity of the MCDF. And the corresponding lapse function takes the form
\begin{equation}
f(r)=1-\frac{2M}{r}-\frac{r^2}{3}\Big(\frac{B}{1+A}\Big)^{\frac{1}{1+\beta}}\mathcal{F}(r), \label{metricofMCDF}
\end{equation}
with $\mathcal{F}(r)$ given by the Gauss hypergeometric function
\begin{equation}\label{hypergeometric}
\mathcal{F}(r)=_{2}F_{1}\left(\left[-\frac{1}{1+\beta},-\frac{1}{w}\right],1-\frac{1}{w},-\frac{1}{B}\left(\frac{Q}{r^3}\right)^{w}\right).
\end{equation}
In this study, we assume that the MCDF leads to the emergence of a de Sitter spacetime, and therefore the cosmological constant is not introduced in Eq.~(\ref{metricofMCDF}).

\begin{figure*}[htbp]
  \centering
  \begin{minipage}{0.32\textwidth}
    \includegraphics[width=\linewidth]{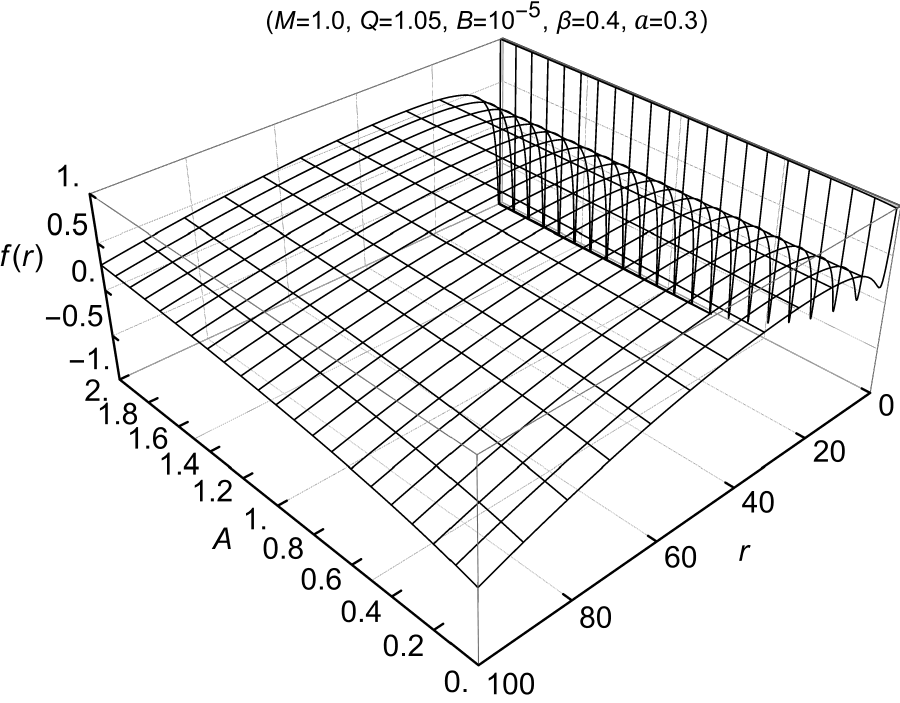}
  \end{minipage}
  \begin{minipage}{0.32\textwidth}
    \includegraphics[width=\linewidth]{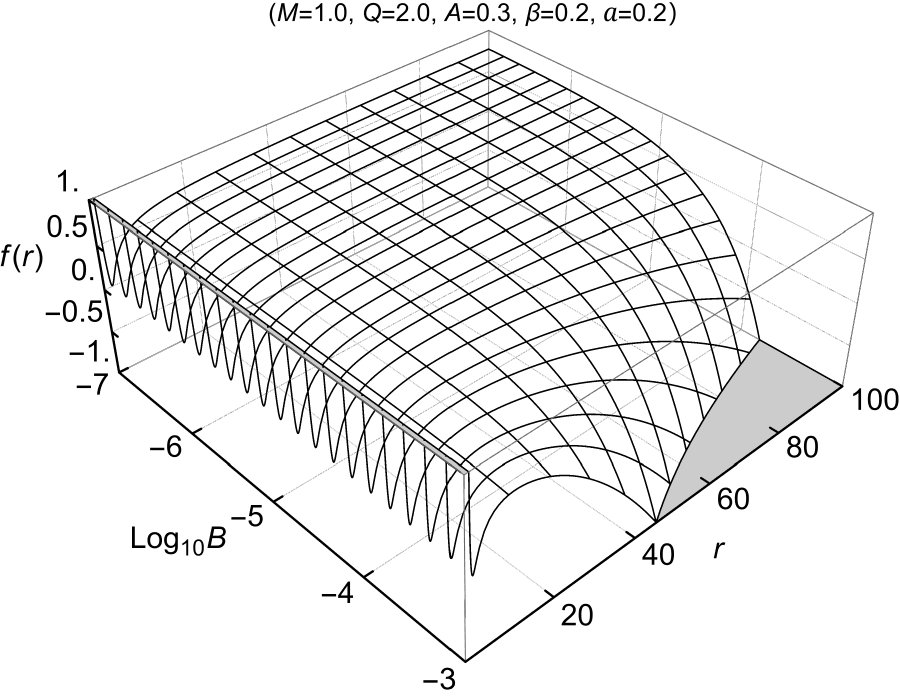}
  \end{minipage}
 \begin{minipage}{0.32\textwidth}
    \includegraphics[width=\linewidth]{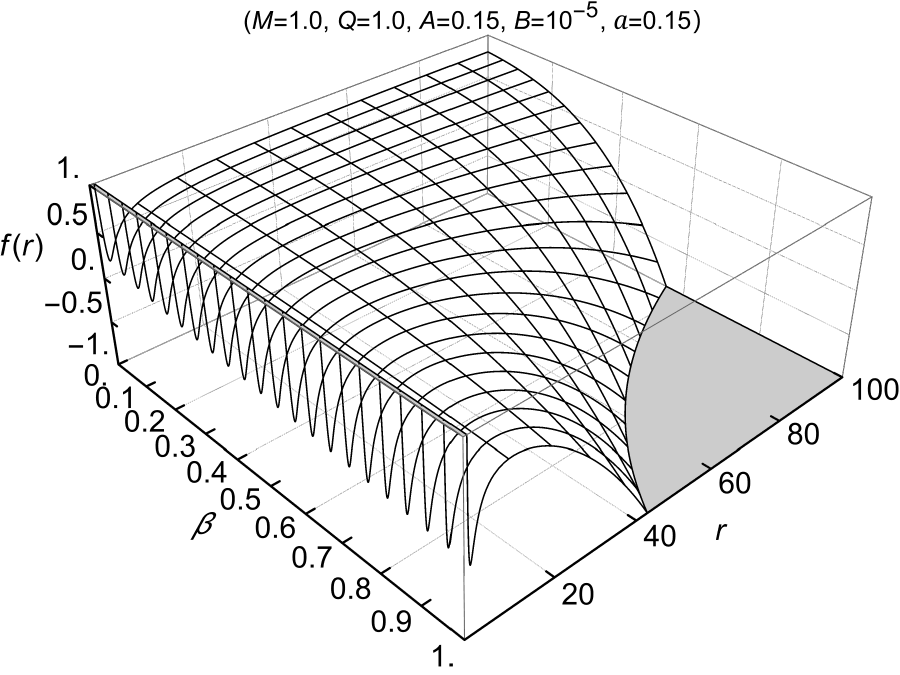}
  \end{minipage}
  \begin{minipage}{0.32\textwidth}
    \includegraphics[width=\linewidth]{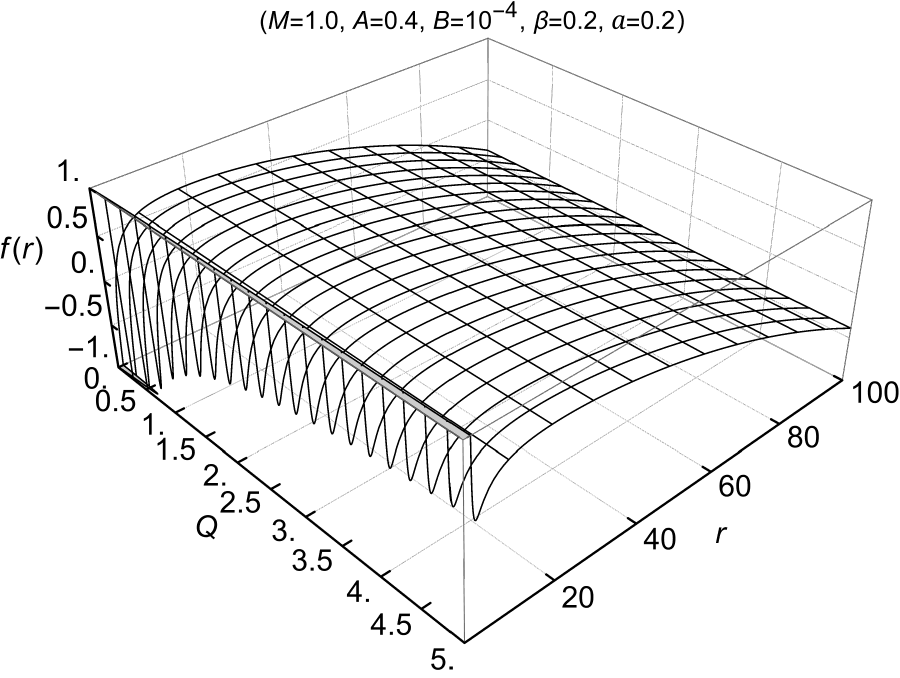}
  \end{minipage}
    \begin{minipage}{0.32\textwidth}
    \includegraphics[width=\linewidth]{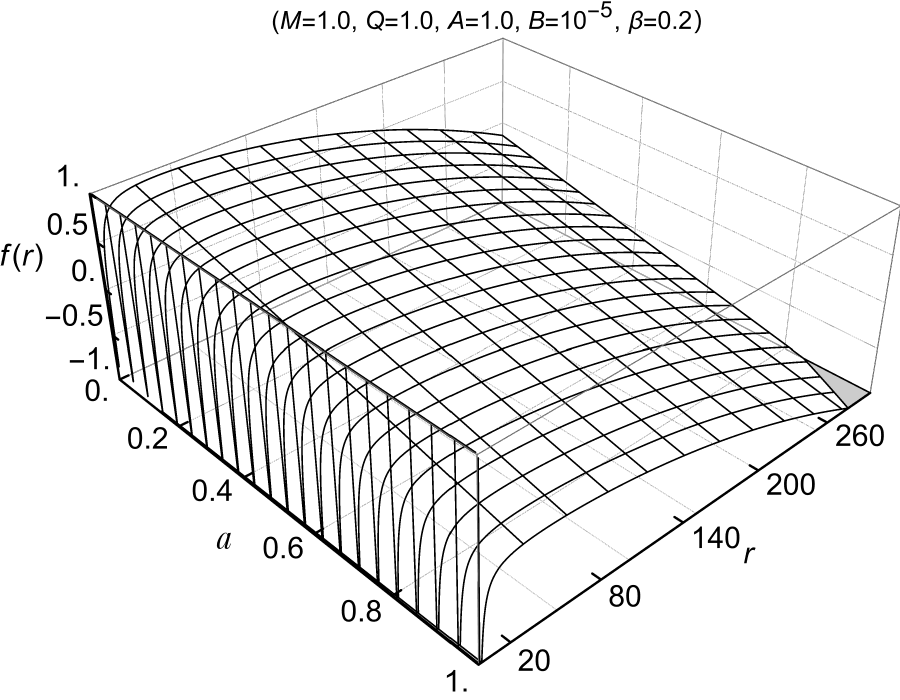}
  \end{minipage}
  \caption{\label{figoffr} The lapse function $f(r)$ with varying $A$, $B$, $\beta$, $Q$ and $a$.}
\end{figure*}

Since there is no coupling between the cloud of strings and the MCDF, their energy-momentum tensors can be linearly superposed in the Einstein field equations. This additivity enables us to construct the lapse function accordingly. Based on Eqs.~\eqref{metricofsc} and \eqref{metricofMCDF}, the lapse function in the presence of both cloud of strings and MCDF is given by
\begin{equation}
f(r) = 1 - a - \frac{2M}{r} - \frac{r^2}{3} \left( \frac{B}{1 + A} \right)^{\frac{1}{1 + \beta}} \mathcal{F}(r). \label{rightmetric}
\end{equation}
In what follows, the black holes considered in this study are the ones represented by Eq.~(\ref{rightmetric}). We perform a dimensional analysis of the parameters involved, adopting geometrized units where the gravitational constant $G$ and the speed of light $c$ are set to unity ($G=c=1$). In the EoS of MCDF, $p = A\rho - B/\rho^\beta$, consistency requires the parameters $A$ and $\beta$ to be dimensionless, while $[B] = L^{-2(1+\beta)}$. For the spacetime metric function in Eq.~(\ref{rightmetric}), the black hole mass $M$ has dimension $[M]=L$, and the cloud of strings parameter $a$ is dimensionless. Furthermore, mathematical consistency of the solution imposes a constraint on the dark fluid parameter $Q$, requiring its dimension to be $[Q] = L^{3 - 2/(1+A)}$, which explicitly depends on the dimensionless EoS parameter $A$.

We now analyze the asymptotic behavior of the energy density \(\rho(r)\) based on Eq.~\eqref{eq:rho}. In the small \(r\) regime (i.e., \(r \to 0\)), since \(\left( \frac{Q}{r^3} \right)^{\omega} \to \infty\), we can neglect the constant \(B\) and obtain
\begin{align}
\rho(r) &\approx \left[ \frac{1}{1 + A} \left( \frac{Q}{r^3} \right)^{\omega} \right]^{\frac{1}{1 + \beta}} \notag \\
&= \left( \frac{Q^\omega}{1 + A} \right)^{\frac{1}{1 + \beta}} r^{-3(1 + A)}.
\end{align}
Therefore, in the small-\(r\) region, the MCDF behaves like a fluid with energy density scaling as \(r^{-3(1 + A)}\). In the large \(r\) regime (i.e., \(r \to \infty\)), \(\left( \frac{Q}{r^3} \right)^{\omega} \to 0\), and the energy density tends to
\begin{equation}
\rho(r) \to \left( \frac{B}{1 + A} \right)^{\frac{1}{1 + \beta}}.
\end{equation}
This implies that the MCDF acts as a positive cosmological constant at large scales. This behavior is also confirmed by examining the asymptotic form of the lapse function \( f(r) \) in Eq.~(\ref{rightmetric}). As \( r \to \infty \), the function tends to
\begin{equation}
f(r) \to 1 - a - \frac{r^2}{3} \left( \frac{B}{1+A} \right)^{\frac{1}{1+\beta}}, \label{metricasymptotic}
\end{equation}
which reveals that the spacetime described by \( f(r) \) is asymptotically de Sitter, featuring both an event horizon \( r_h \) and a pseudo-cosmological horizon \( r_c \).

The effects of the parameters $A$, $B$, $\beta$, $Q$, and $a$ on the lapse function $f(r)$ are illustrated in Fig.~\ref{figoffr}. The following observations can be made: The parameter $Q$ predominantly affects the existence and position of the event horizon $r_h$, while the parameter $B$ governs the asymptotic behavior of $f(r)$ and thus controls the emergence and location of the pseudo-cosmological horizon $r_c$. The parameters $A$ and $\beta$ play a crucial role in modulating the strength and shape of the MCDF-induced deformation of spacetime geometry, thereby indirectly influencing both horizons. In particular, larger values of $\beta$ lead to more pronounced oscillatory features in $f(r)$, whereas increasing $A$ generally shifts the entire profile upward, delaying the occurrence of the event horizon. Moreover, the cloud of strings parameter $a$, representing an additional deviation from Schwarzschild geometry, uniformly lowers the $f(r)$ curve and reduces the range of the domain of outer communication, defined as the region between $r_h$ and $r_c$, where two observers can communicate without being causally disconnected by a horizon~\cite{Friedman:1993ty,Chrusciel:1994tr}.

\begin{figure*}[htbp]
\centering % \begin{center}/\end{center} takes some additional vertical space
\includegraphics[width=.48\textwidth]{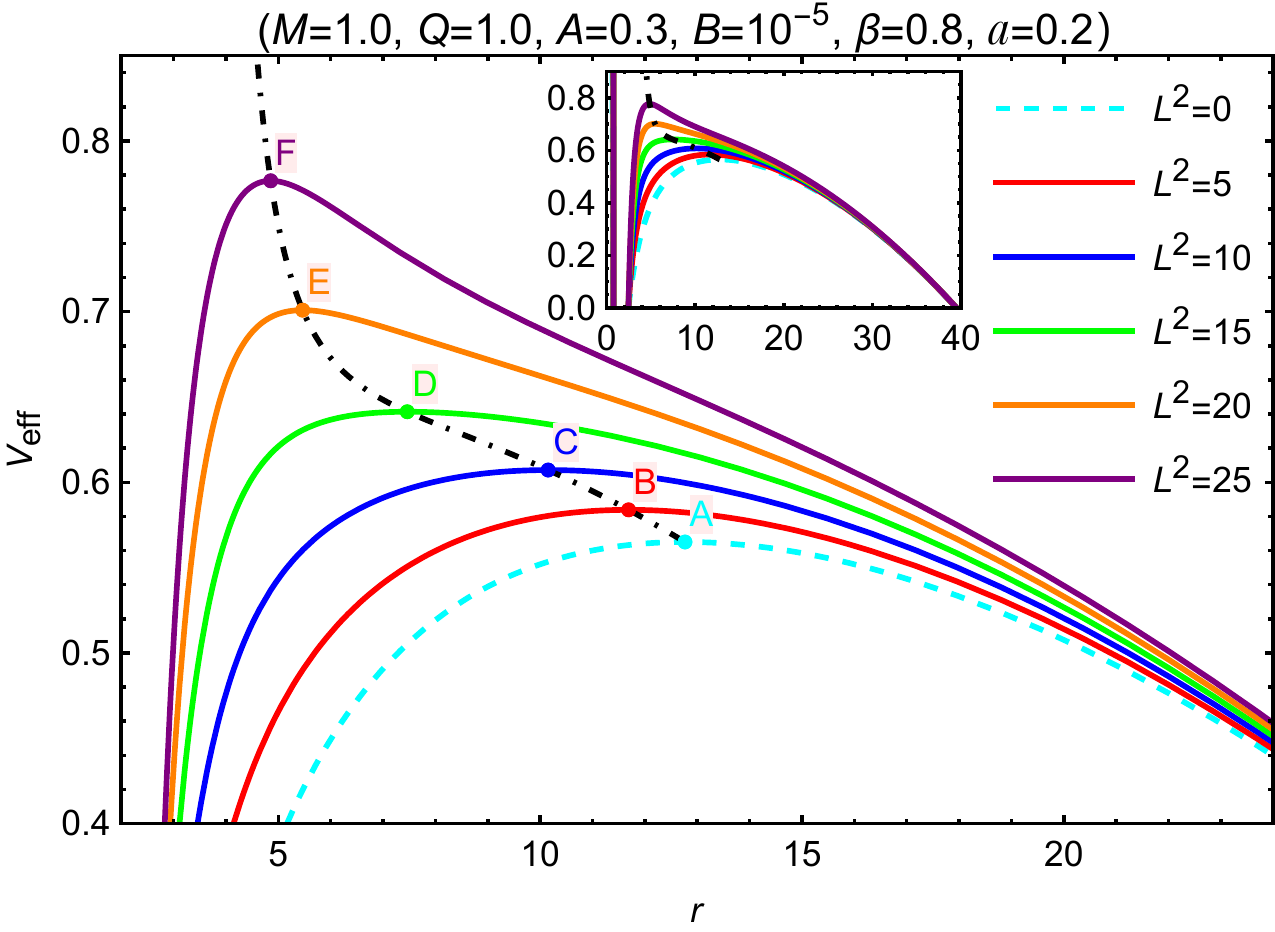}
\includegraphics[width=.49\textwidth]{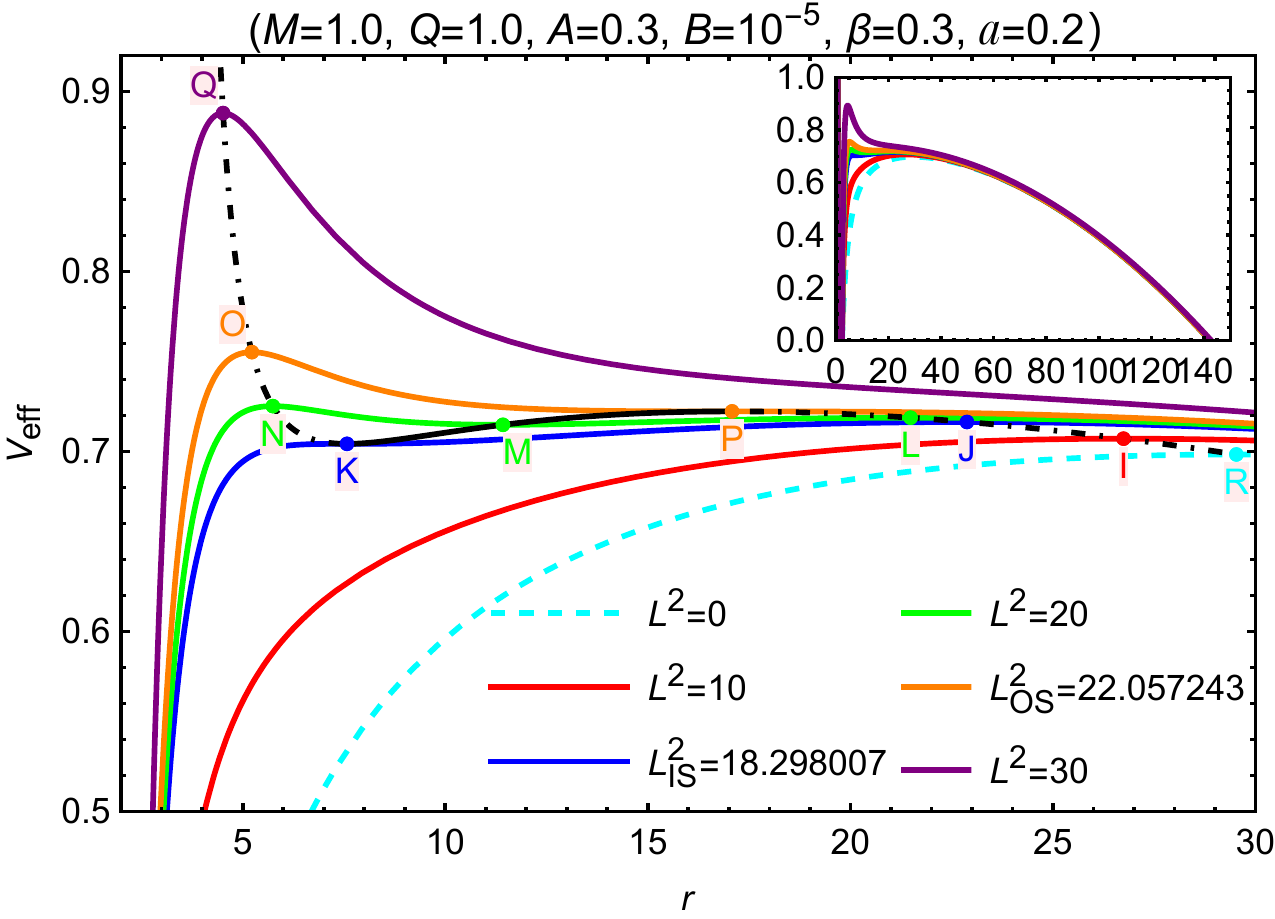}
% "\includegraphics" is very powerful; the graphicx package is already loaded
\caption{\label{FigUeff} The profile of the effective potential curves of timelike particles for black holes without (left) and with (right) SCOs. With varying $L^2$, the dot-dashed and solid black curves represents the unstable and stable circular orbit points, respectively. The cyan dashed curves correspond to $L^2=0$, which are unrealistic. The radial coordinates of points $K$, $A$, $J$, $P$ and $R$ are $r_{\rm K}=7.573887$, $r_{\rm A}=12.766328$, $r_{\rm J}=22.879906$, $r_{\rm P}=17.081879$ and $r_{\rm R}=29.532725$.}
\end{figure*}

\section{Geodesic Structure in the presence of MCDF and cloud of strings}
\label{secgeodesic}
In this section, we examine the existence and stability of circular orbits—both timelike and lightlike—in the black hole spacetime in the presence of MCDF and cloud of strings under Einstein gravity. The dynamics are governed by the Euler–Lagrange equation
\begin{equation}
\frac{d}{ds}\left(\frac{\partial \mathcal{L}}{\partial \dot{x}^{\mu}}\right) = \frac{\partial \mathcal{L}}{\partial x^{\mu}},
\label{eleq}
\end{equation}
where \( s \) denotes the affine parameter, and the overdot represents differentiation with respect to \( s \). The Lagrangian is given by
\begin{eqnarray}
\mathcal{L} &=& \frac{1}{2} g_{\mu\nu} \dot{x}^{\mu} \dot{x}^{\nu} \nonumber \\
&=& \frac{1}{2}\left(-f(r)\dot{t}^2 + \frac{\dot{r}^2}{f(r)} + r^2(\dot{\theta}^2 + \sin^2\theta\, \dot{\phi}^2)\right).
\label{laeq}
\end{eqnarray}

Assuming motion confined to the equatorial plane, we impose \( \theta = \pi/2 \) and \( \dot{\theta} = 0 \). Due to the spacetime’s time-translational and axial symmetries, the corresponding conserved quantities are
\begin{equation}
E = -\frac{\partial \mathcal{L}}{\partial \dot{t}}, \quad L = \frac{\partial \mathcal{L}}{\partial \dot{\phi}}.
\label{ELexpression}
\end{equation}

By combining Eqs.~(\ref{rightmetric}), (\ref{eleq}), and (\ref{laeq}), we derive the following equations of motion:
\begin{eqnarray}
\dot{t} &=& \frac{E}{f(r)}, \label{time} \\
\dot{\phi} &=& \frac{L}{r^2}, \label{psi} \\
\dot{r}^2 + \left(\delta + \frac{L^2}{r^2}\right)f(r) &=& E^2, \label{radial}
\end{eqnarray}
where
\begin{equation}
\delta = \begin{cases}
1, & \text{for timelike geodesics}; \\
0, & \text{for lightlike geodesics}.
\end{cases}
\end{equation}

The radial equation (\ref{radial}) can be recast as
\begin{equation}
\dot{r}^2 + V_{\rm eff}(r) = E^2,
\label{vbr}
\end{equation}
with the effective potential defined by
\begin{equation}
V_{\rm eff}(r) = \left(\delta + \frac{L^2}{r^2}\right)f(r).
\label{epotential}
\end{equation}

\subsection{The structure of timelike geodesics}
\begin{figure*}[htbp]
\centering % \begin{center}/\end{center} takes some additional vertical space
\includegraphics[width=.41\textwidth]{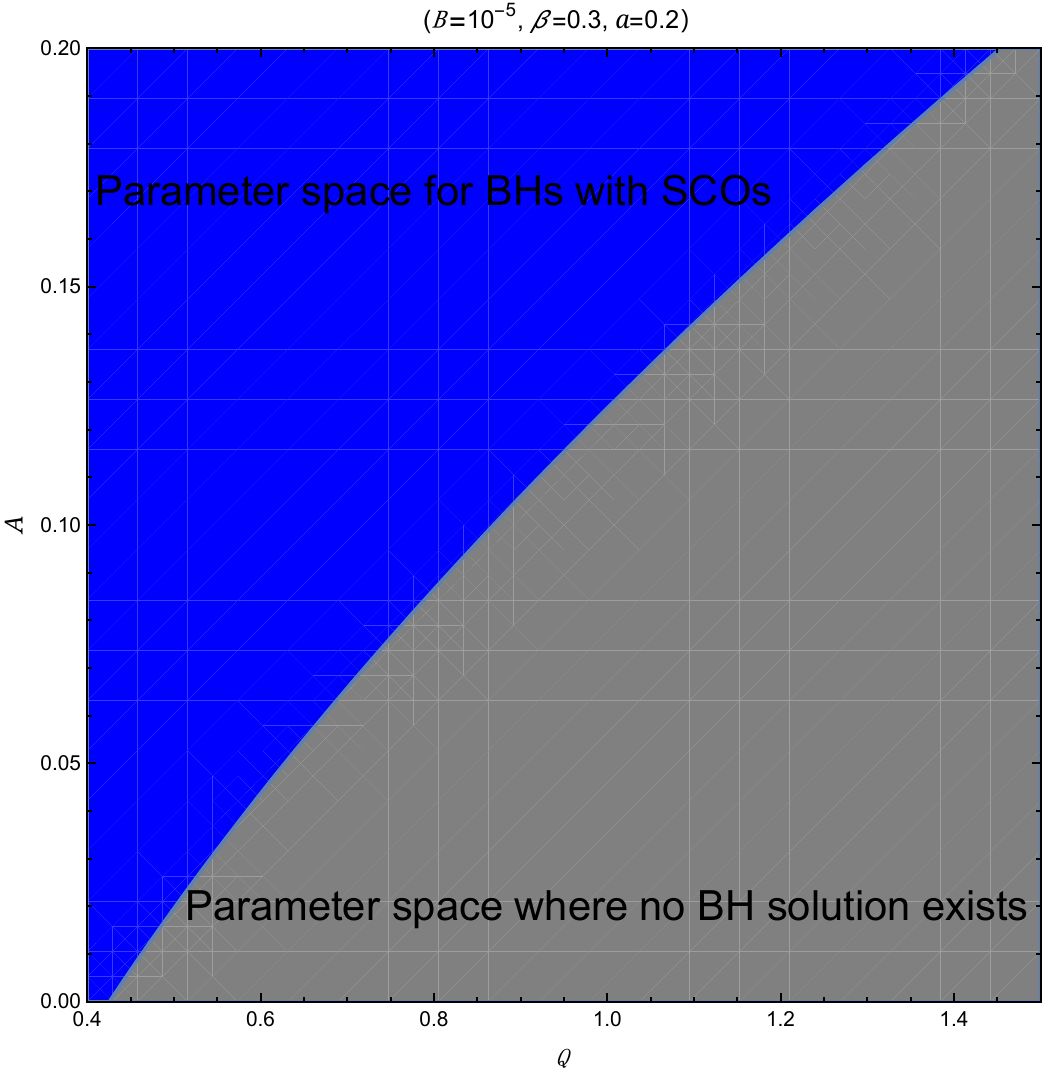}
\includegraphics[width=.41\textwidth]{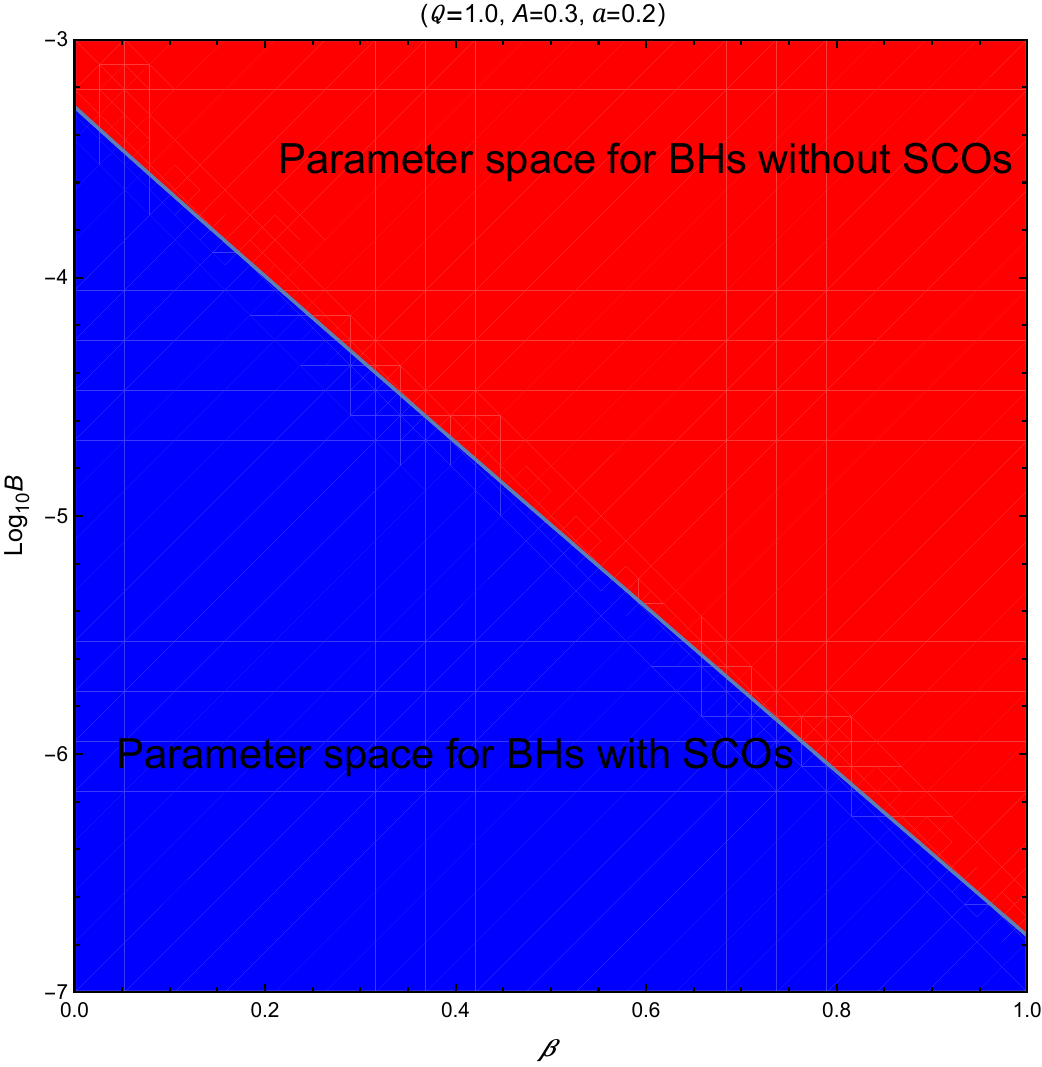}\\
\includegraphics[width=.41\textwidth]{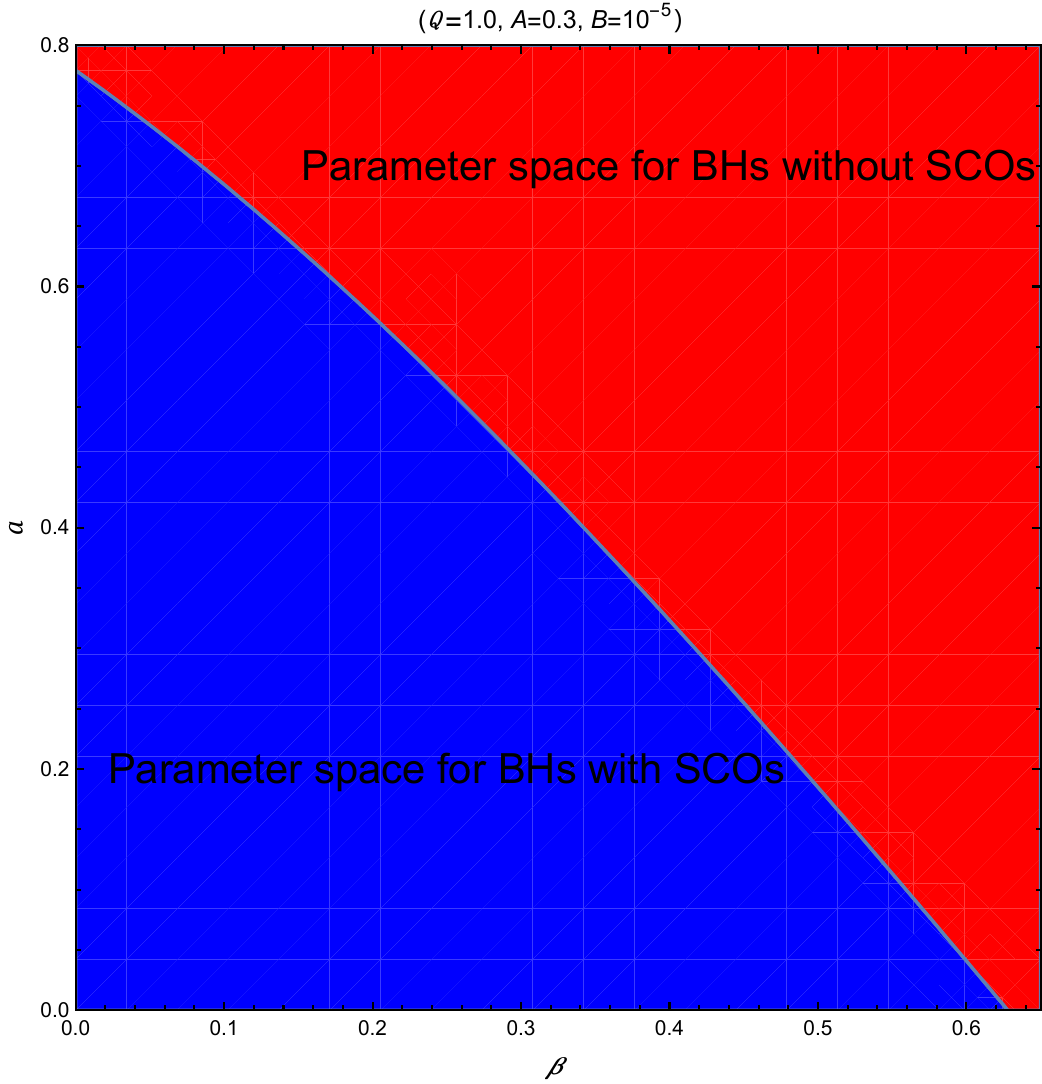}
\includegraphics[width=.41\textwidth]{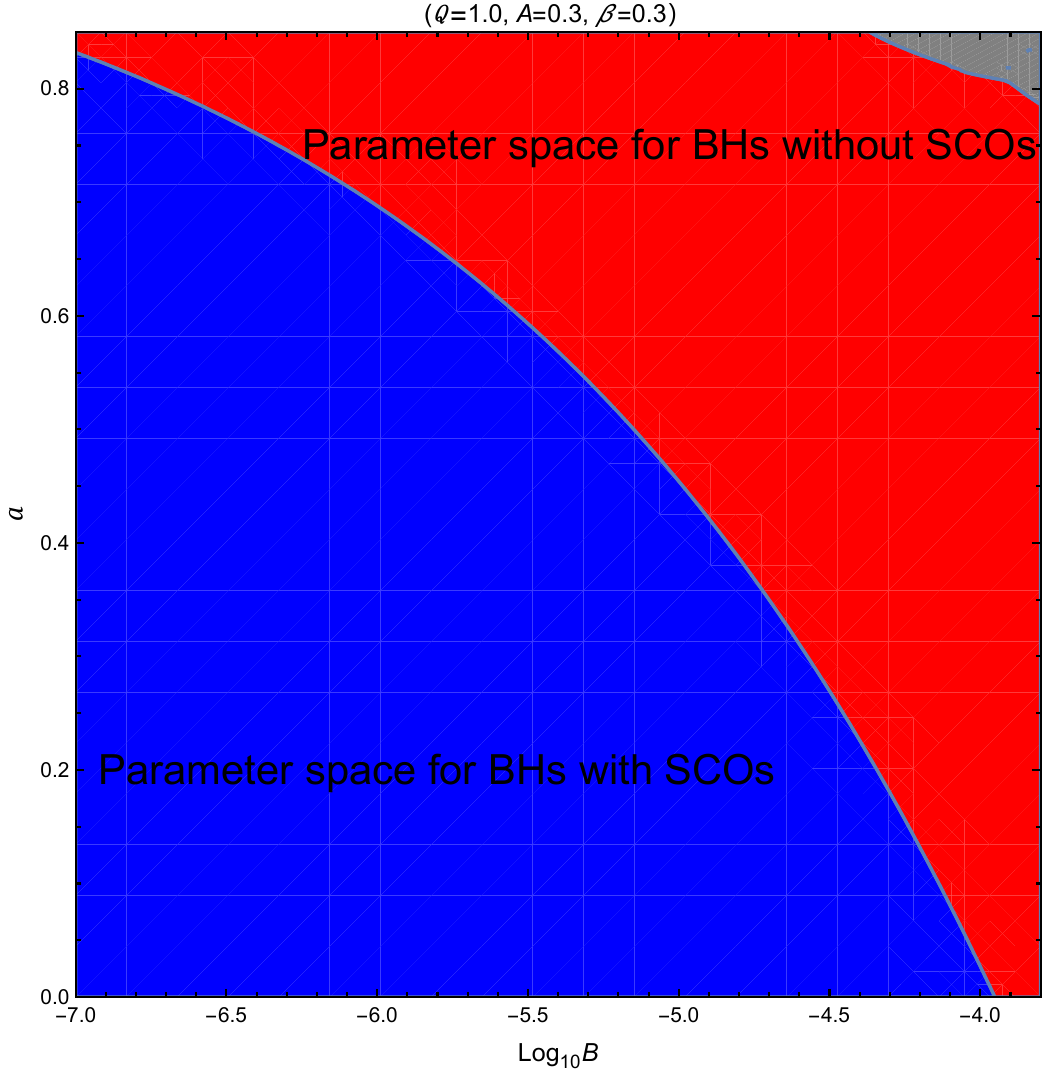}
% "\includegraphics" is very powerful; the graphicx package is already loaded
\caption{\label{Figpararegion} Parameter space of black holes without (red region) and with (blue region) SCOs. The gray region represents parameter space where no black hole solution exists. Here, we set $M=1$.}
\end{figure*}
Examining the motion of timelike particles is crucial, as it may significantly impact the profiles of accreting matter. According to Eq.~(\ref{epotential}), for timelike particles with \(\delta = 1\), the effective potential is written as
\begin{equation}\label{EquUeff}
V_{\rm eff}(r) = \left(1 + \frac{L^2}{r^2}\right) f(r).
\end{equation}
The effective potential depends on the mass \(M\), the angular momentum \(L\), the MCDF and cloud of strings parameters. By studying the effective potential profiles for different parameter values, one finds that, from the perspective of timelike orbits, black holes can be divided into two categories: those that allow for stable circular orbits (SCOs) and those that do not. We illustrate this classification through representative examples. The effective potential curves with various values of \(L^2\) for black holes without (left panel) and with (right panel) SCOs are shown in Fig.~\ref{FigUeff}.
\begin{figure*}[htbp]
  \centering
  \begin{minipage}{0.32\textwidth}
    \includegraphics[width=\linewidth]{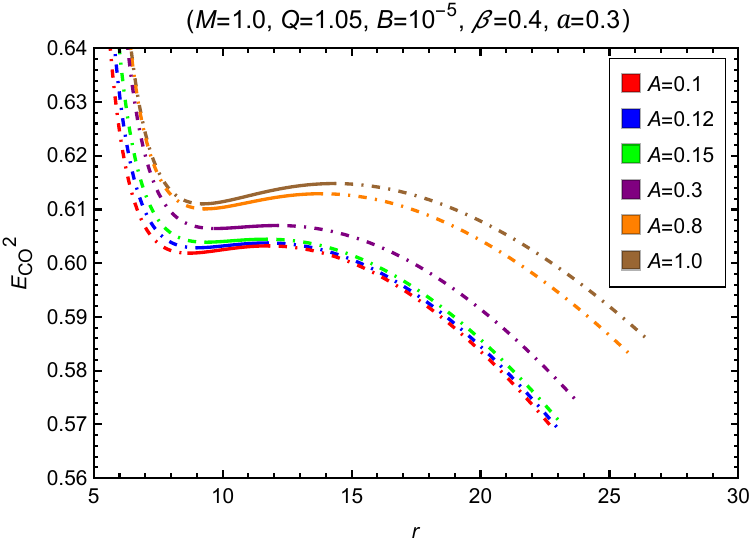}
  \end{minipage}
  \begin{minipage}{0.32\textwidth}
    \includegraphics[width=\linewidth]{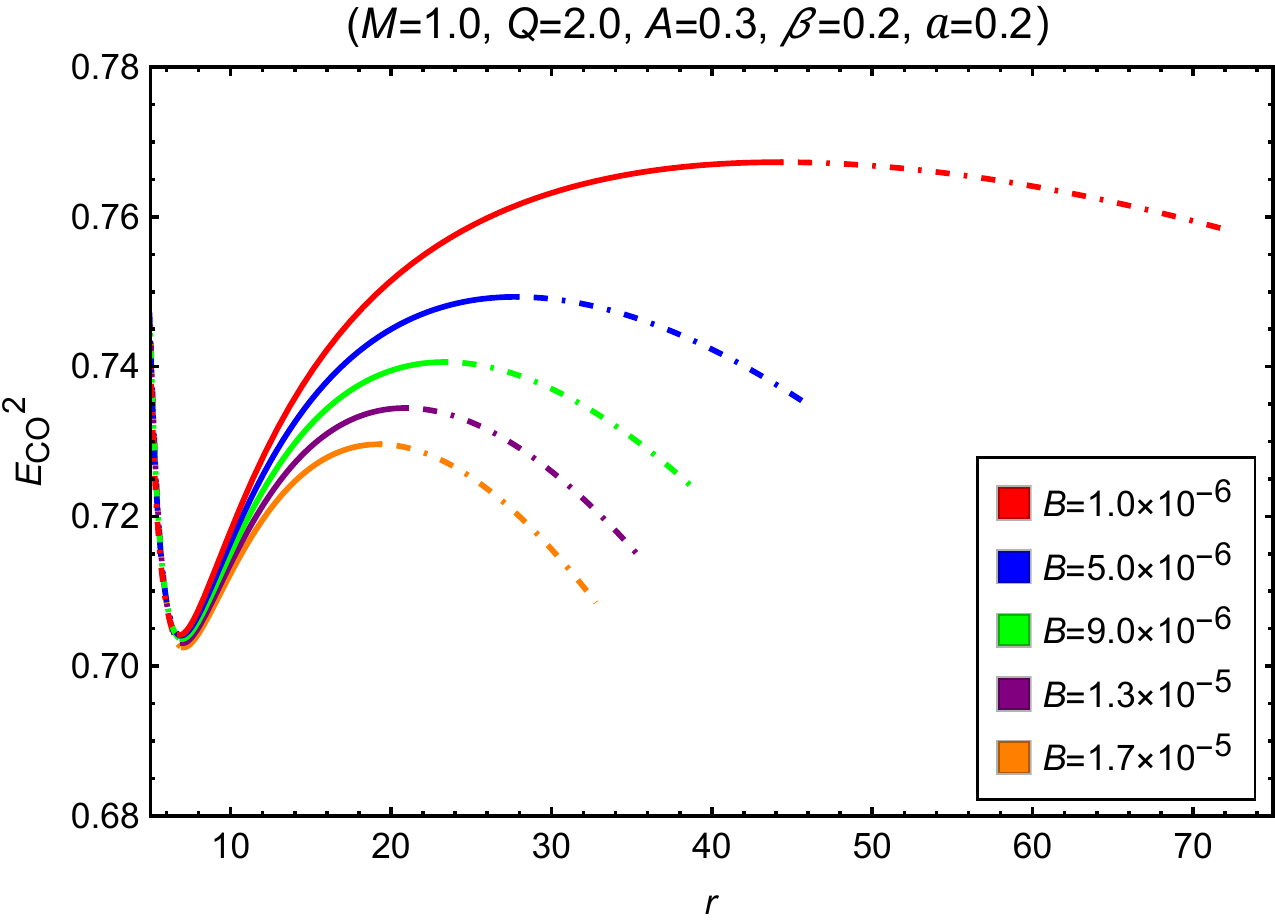}
  \end{minipage}
 \begin{minipage}{0.32\textwidth}
    \includegraphics[width=\linewidth]{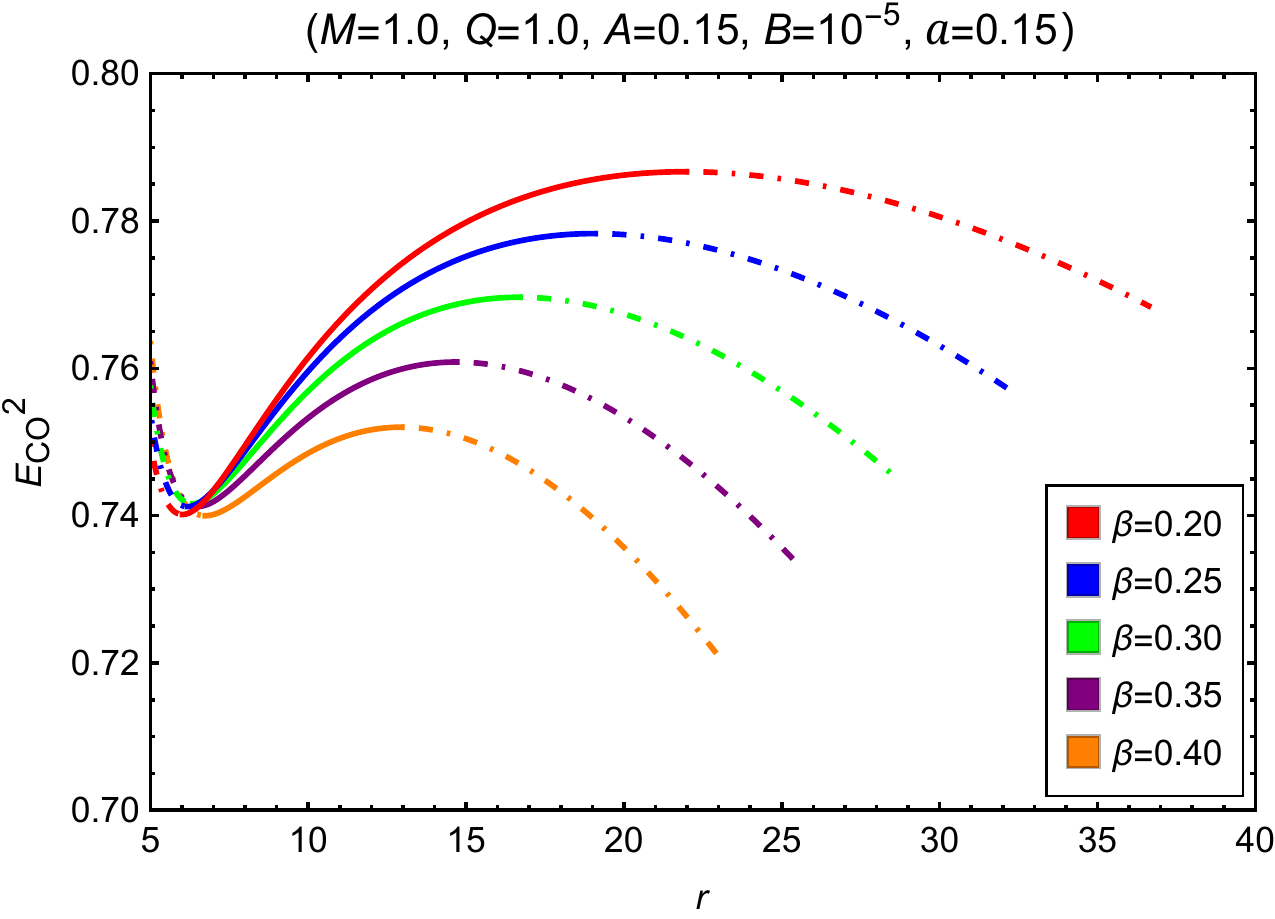}
  \end{minipage}
  \begin{minipage}{0.32\textwidth}
    \includegraphics[width=\linewidth]{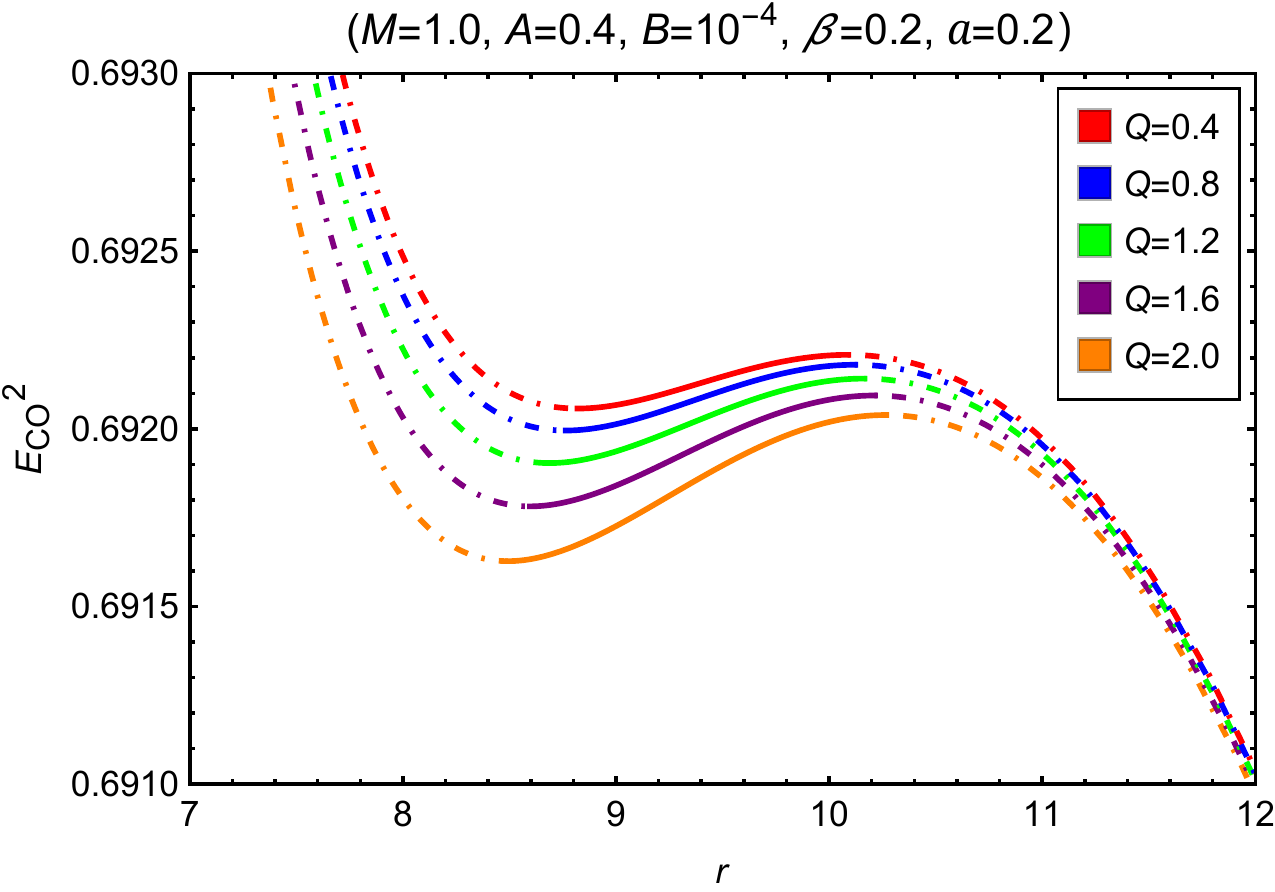}
  \end{minipage}
    \begin{minipage}{0.32\textwidth}
    \includegraphics[width=\linewidth]{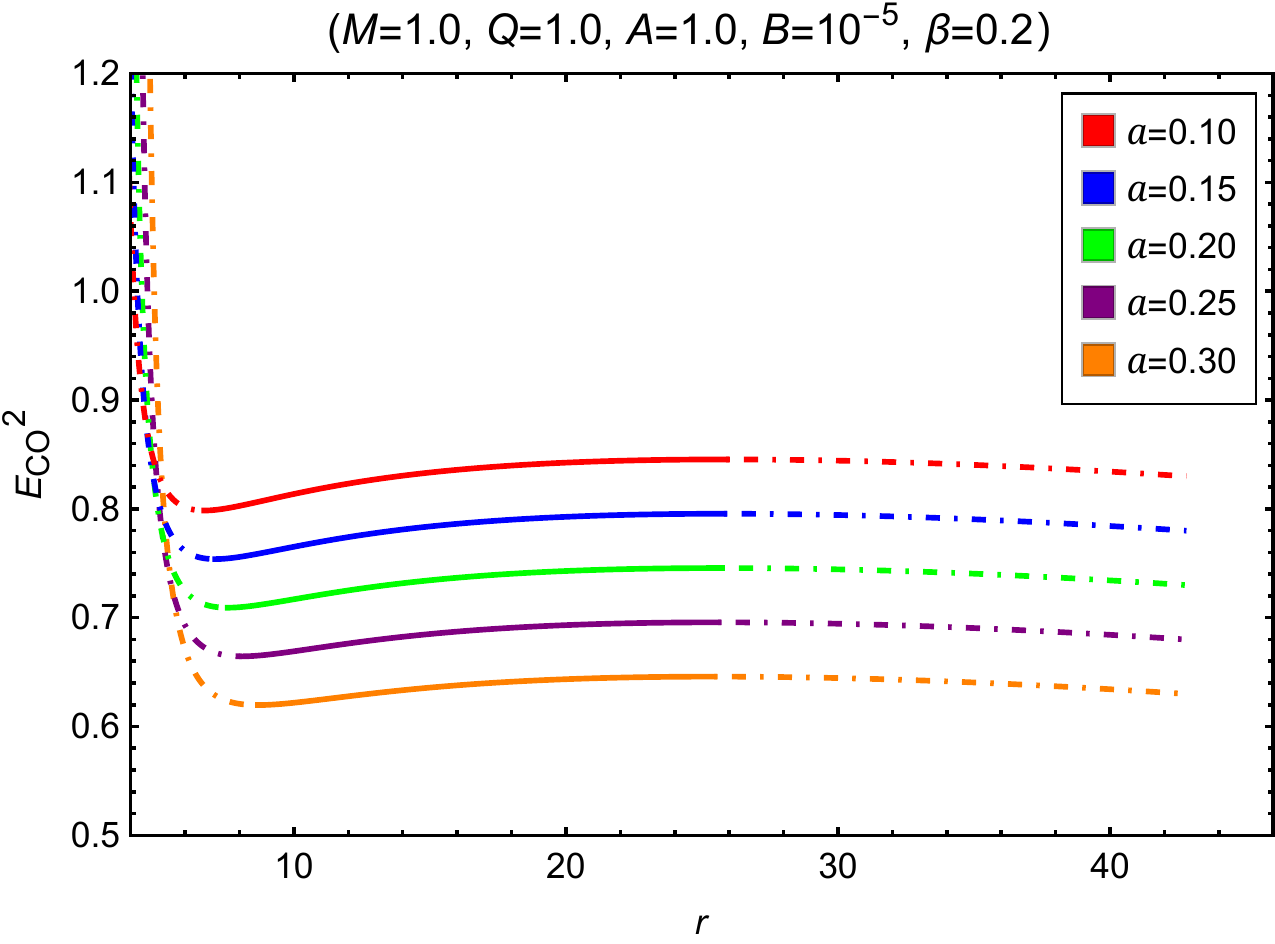}
  \end{minipage}
  \caption{\label{FigEE} The radial profiles of $E_{\rm CO}^2$ with varying $A$, $B$, $\beta$, $Q$ and $a$. The dotted-dashed and solid segments of a curve depict the positions of unstable and stable circular orbits, respectively.}
\end{figure*}
\begin{figure*}[htbp]
  \centering
  \begin{minipage}{0.32\textwidth}
    \includegraphics[width=\linewidth]{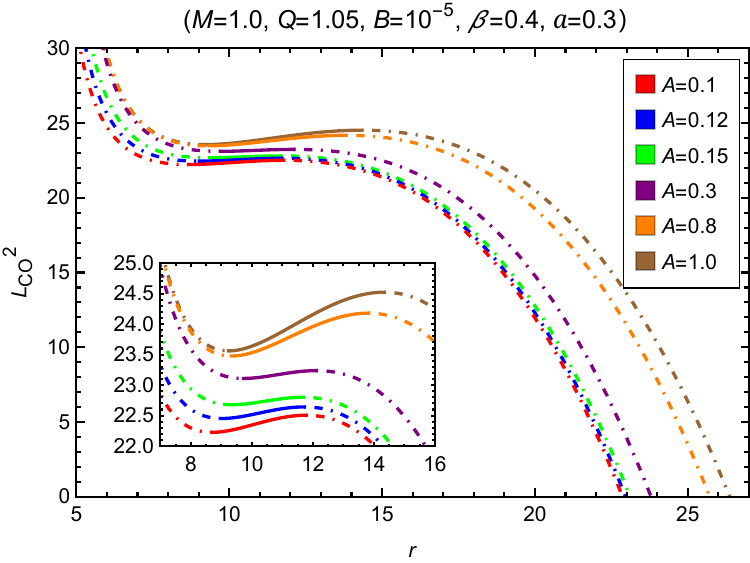}
  \end{minipage}
  \begin{minipage}{0.32\textwidth}
    \includegraphics[width=\linewidth]{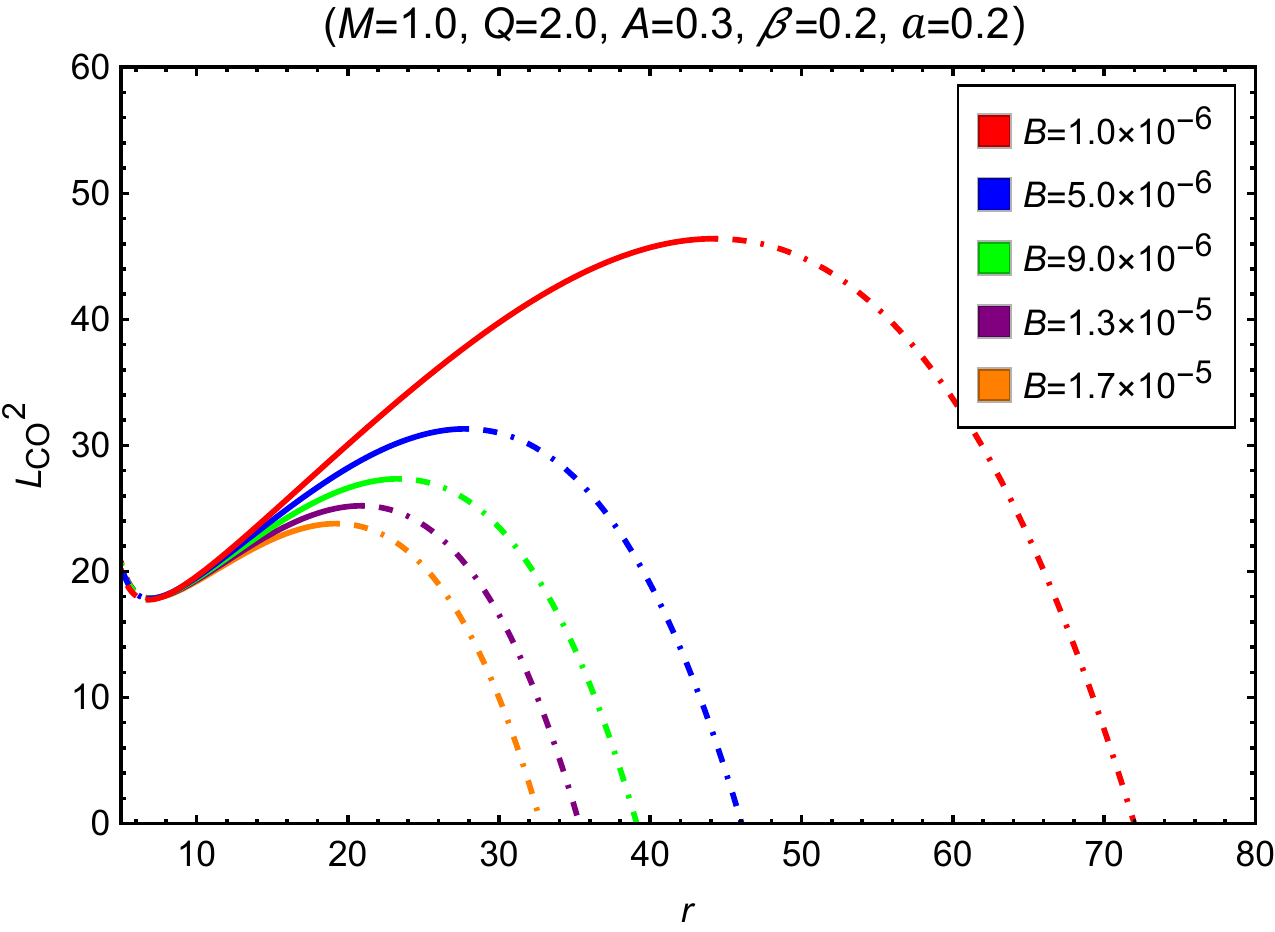}
  \end{minipage}
 \begin{minipage}{0.32\textwidth}
    \includegraphics[width=\linewidth]{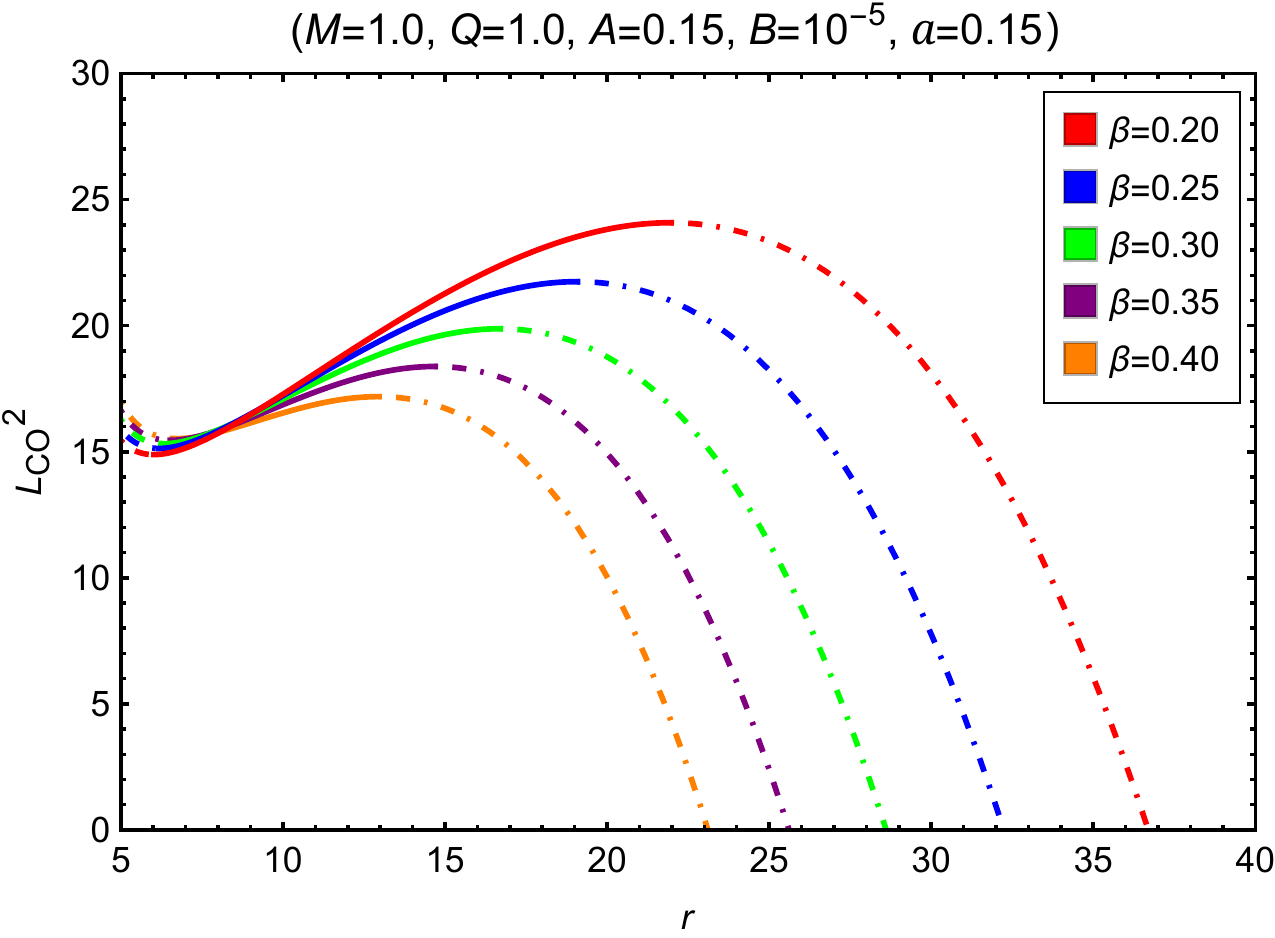}
  \end{minipage}
  \begin{minipage}{0.32\textwidth}
    \includegraphics[width=\linewidth]{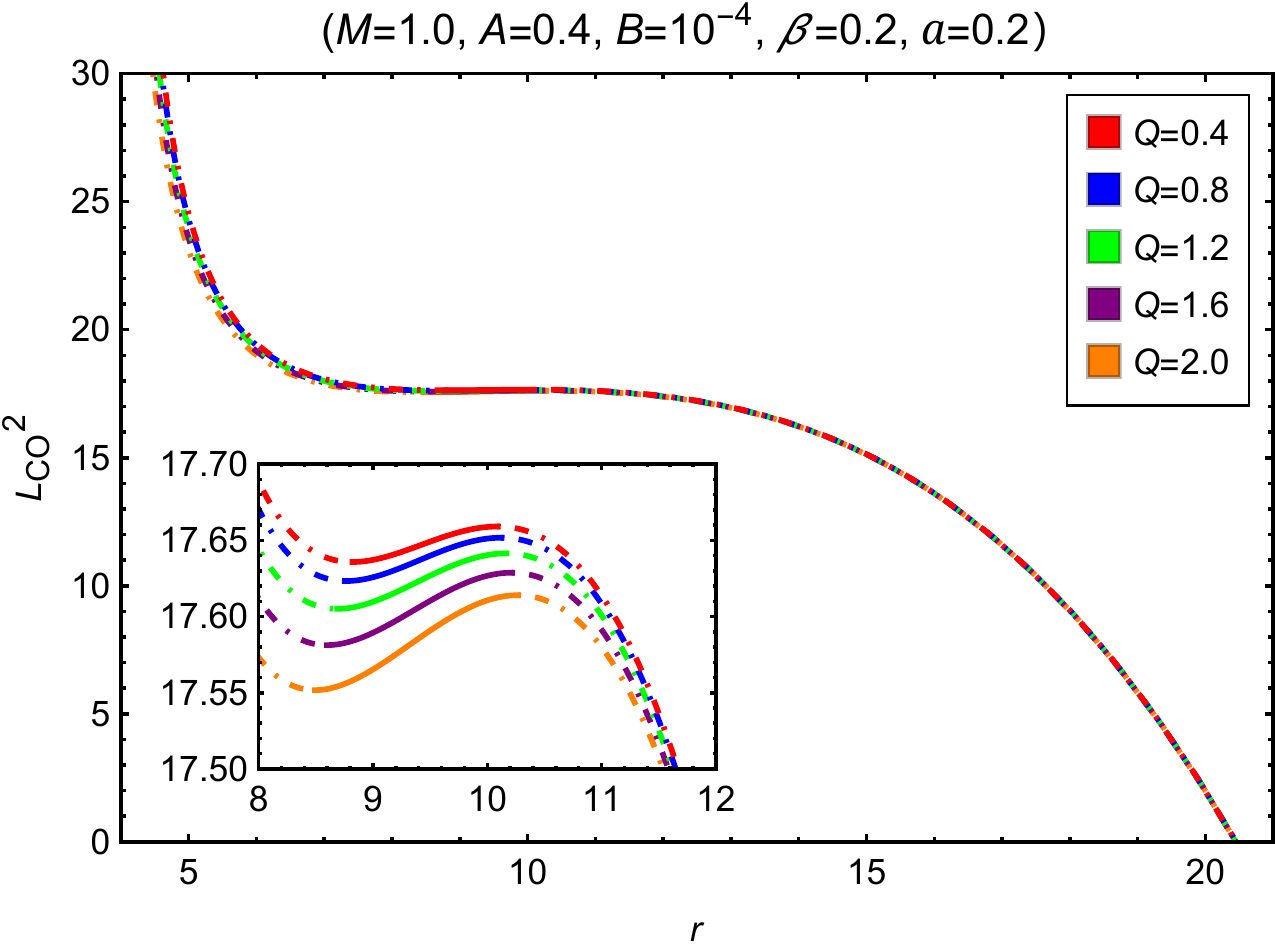}
  \end{minipage}
    \begin{minipage}{0.32\textwidth}
    \includegraphics[width=\linewidth]{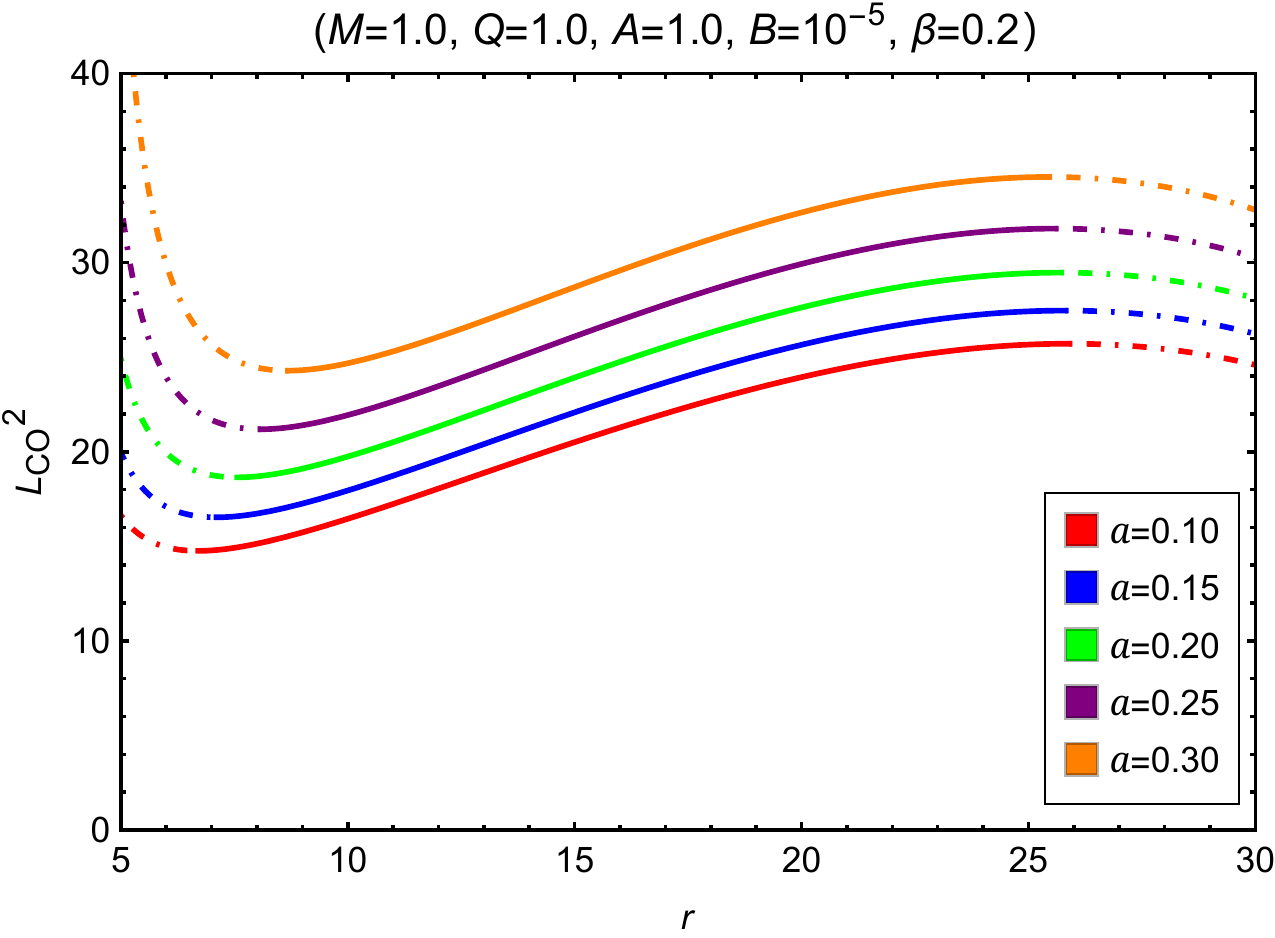}
  \end{minipage}
  \caption{\label{FigLL} The radial profiles of $L_{\rm CO}^2$ with varying $A$, $B$, $\beta$, $Q$ and $a$. The dotted-dashed and solid segments of a curve depict the positions of unstable and stable circular orbits, respectively. }
\end{figure*}
\begin{figure*}[htbp]
  \centering
  \begin{minipage}{0.32\textwidth}
    \includegraphics[width=\linewidth]{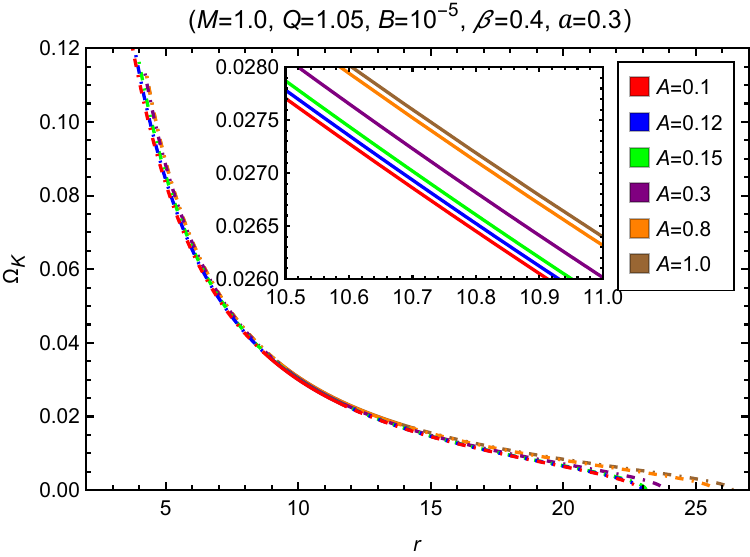}
  \end{minipage}
  \begin{minipage}{0.32\textwidth}
    \includegraphics[width=\linewidth]{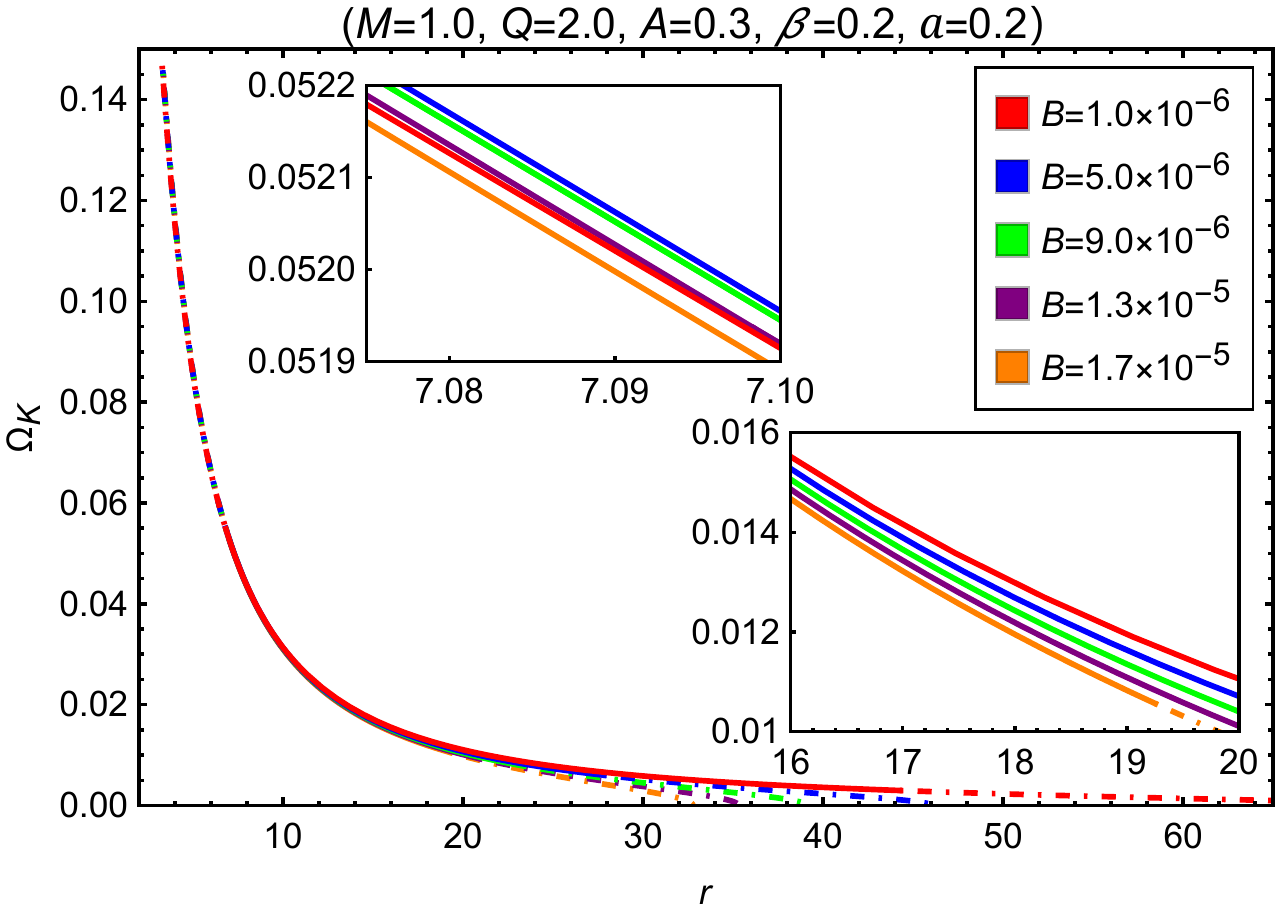}
  \end{minipage}
 \begin{minipage}{0.32\textwidth}
    \includegraphics[width=\linewidth]{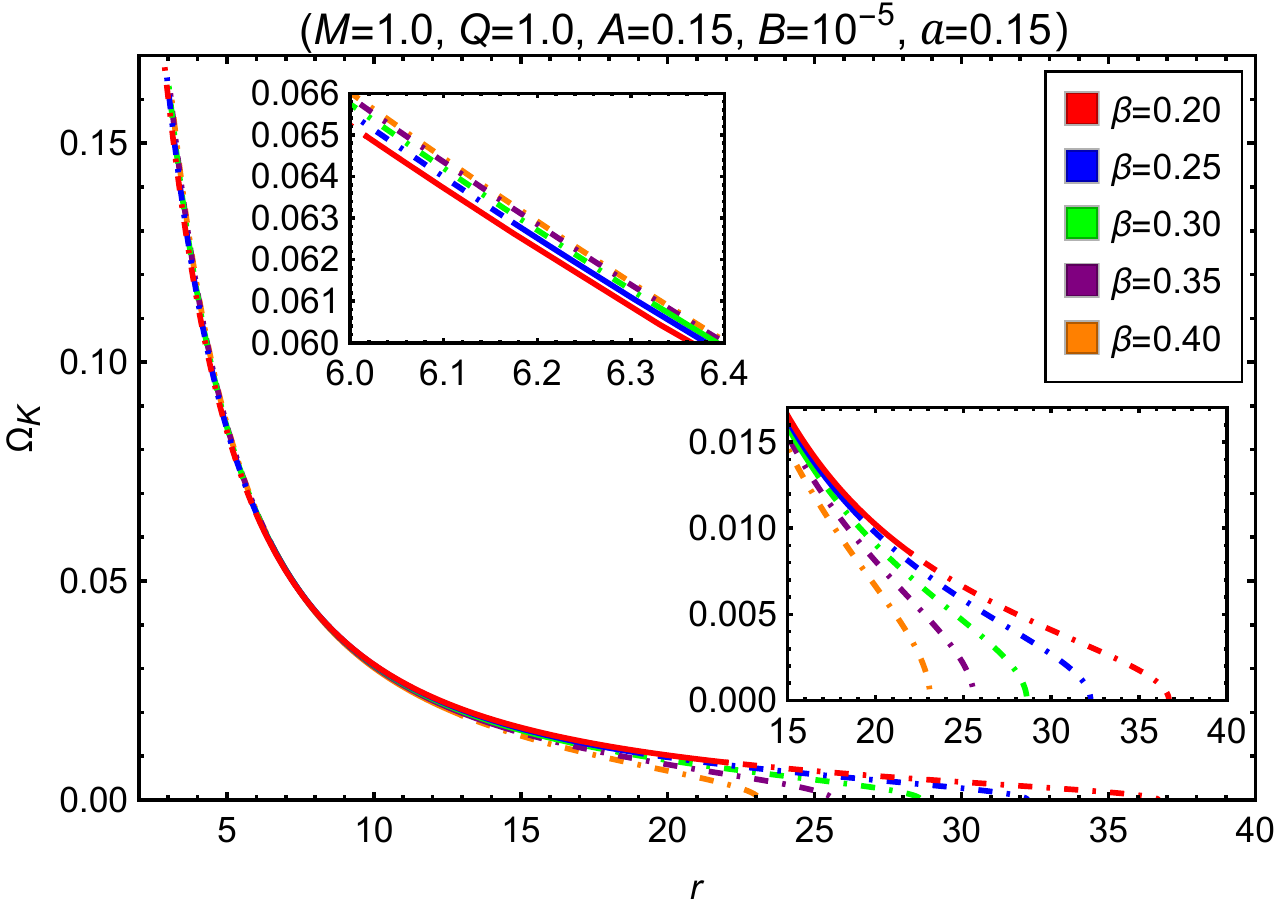}
  \end{minipage}
  \begin{minipage}{0.32\textwidth}
    \includegraphics[width=\linewidth]{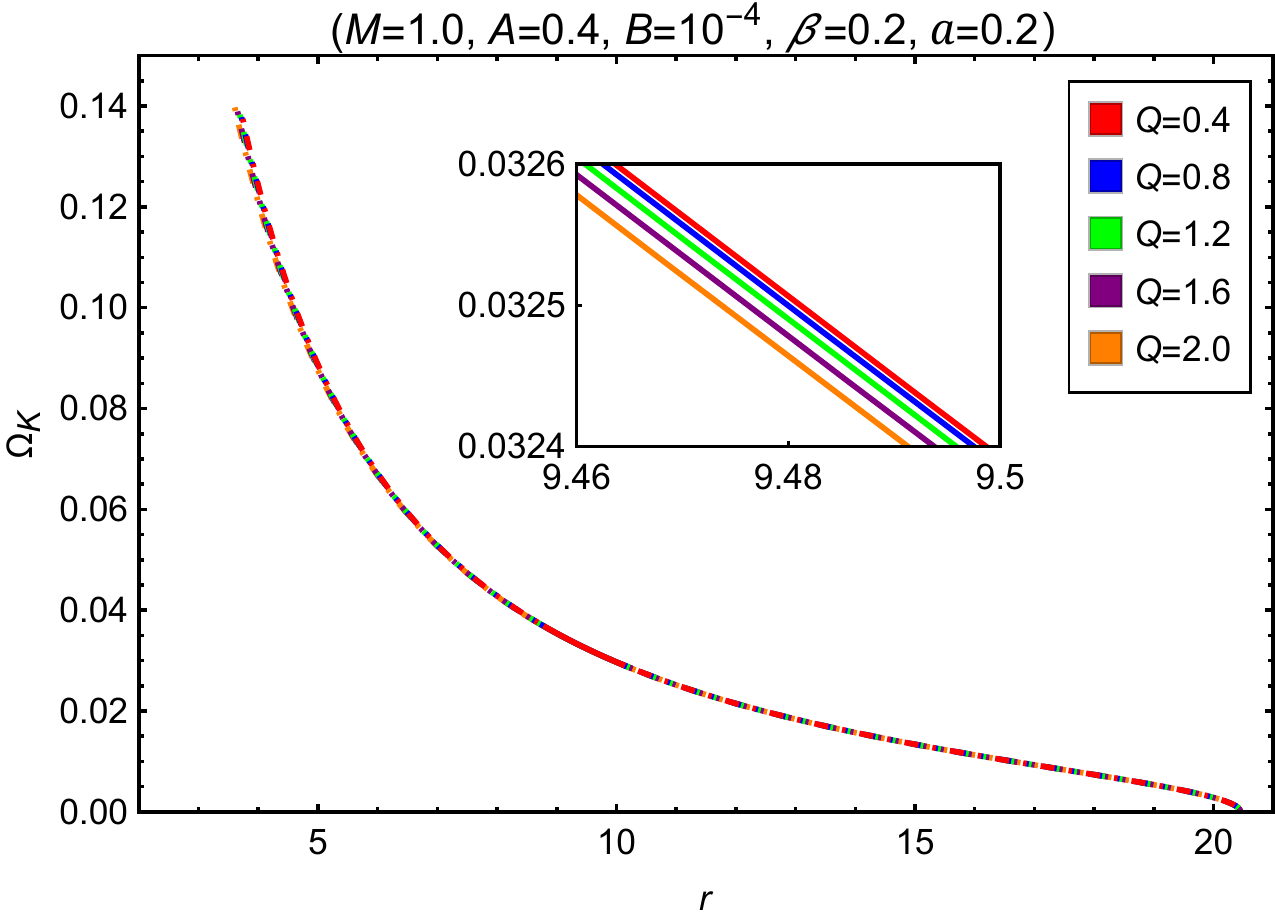}
  \end{minipage}
    \begin{minipage}{0.32\textwidth}
    \includegraphics[width=\linewidth]{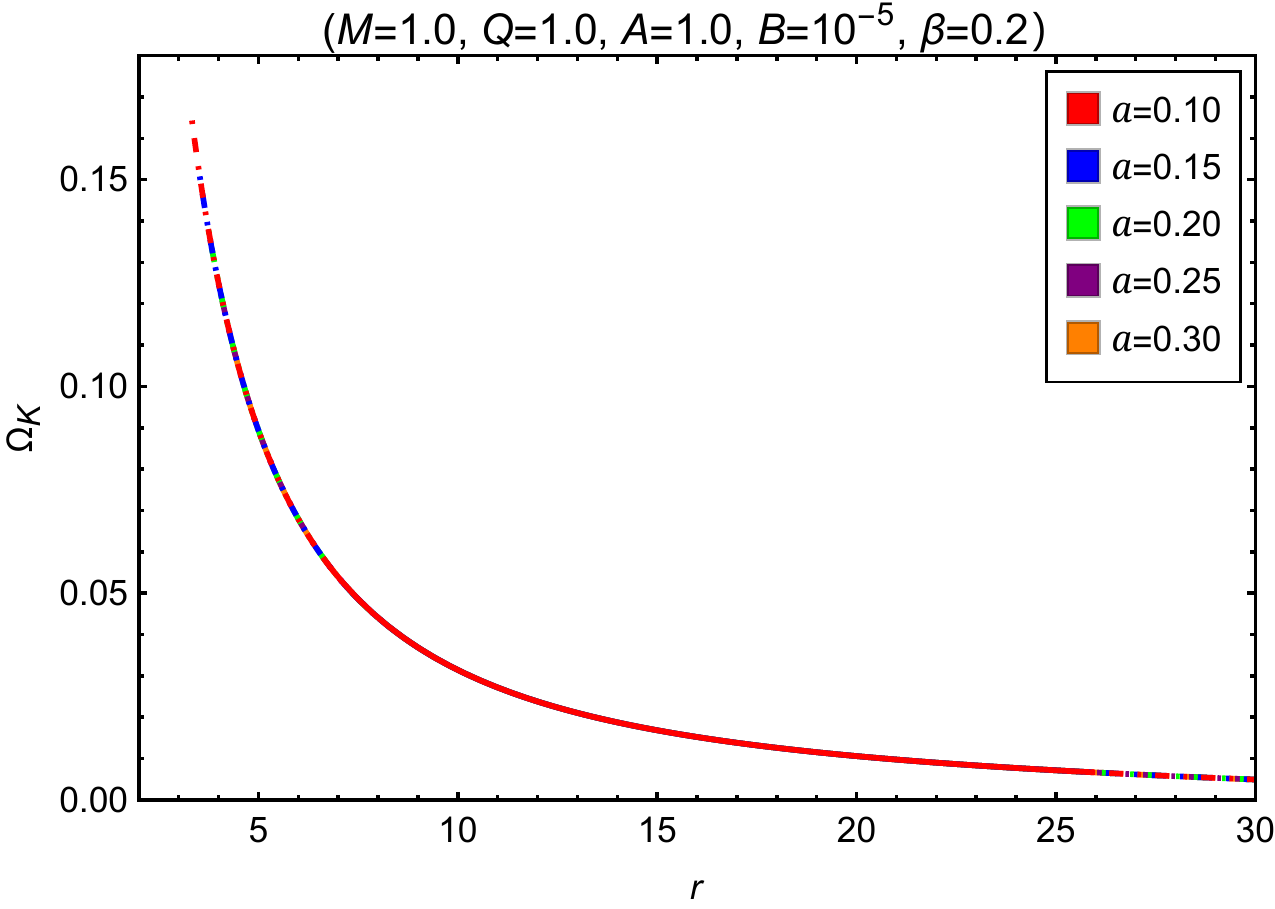}
  \end{minipage}
  \caption{\label{figOmigaK} Radial dependence of Keplerian frequencies of test particles around a black hole for different values of $A$, $B$, $\beta$, $Q$ and $a$. The dotted-dashed and solid segments of a curve depict the positions of unstable and stable circular orbits, respectively.}
\end{figure*}
For the case without SCOs, the effective potential curves corresponding to different values of \(L^2\) each exhibit only a single maximum. This implies that particles in this case can have only one unstable circular orbit (when \(E^2\) equals the peak of the potential curve). In contrast, the case with SCOs presents a more complex structure, and the classification of angular momentum values can be summarized as follows:
\begin{enumerate}
  \item When \( 0 < L^2 < L^2_{\rm IS} \), the potential exhibits only one maximum, corresponding to a single unstable circular orbit (e.g., point I).
  \item When \( L^2 = L^2_{\rm IS} \), the particle can occupy both an unstable circular orbit (e.g., point J) and the ISCO (e.g., point K).
  \item When \( L^2_{\rm IS} < L^2 < L^2_{\rm OS} \), the potential shows two maxima and one minimum, indicating one stable circular orbit (e.g., point M) and two unstable circular orbits (e.g., points N and L).
  \item When \( L^2 = L^2_{\rm OS} \), the particle possesses an unstable circular orbit (e.g., point O) and the OSCO (e.g., point P).
  \item When \( L^2 > L^2_{\rm OS} \), the potential once again features only one maximum, indicating a single unstable orbit (e.g., point Q).
\end{enumerate}
For black holes concerned in this work, the existence of SCOs is primarily influenced by the parameters \( B \), \( \beta \), and \( a \), whereas \( Q \) and \( A \) are more crucial for determining the presence of the event horizon and thus the existence of the black hole solution. Fig.~\ref{Figpararegion} illustrates the dependence of black hole and SCO existence on different MCDF and cloud of strings parameters for selected fixed values.

The requirement $V_{\rm{eff}}(r)=V'_{\rm{eff}}(r)=0$ for circular orbits yields
\begin{equation}\label{ELexpressions}
E_{\rm CO}^2=\frac{2f(r)^2}{2f(r)-rf'(r)},~~\quad~~L_{\rm CO}^2=\frac{r^3f'(r)}{2f(r)-rf'(r)}.
\end{equation}
We show, respectively, the radial profiles of $E_{\rm CO}^2$ and $L_{\rm CO}^2$ of the circular orbits with varying MCDF and cloud of strings parameters in Figs.~\ref{FigEE} and \ref{FigLL}. To summarize the effects, we observe that increasing the parameters $B$, $\beta$, and $Q$ leads to a decrease in both $E_{\rm CO}^2$ and $L_{\rm CO}^2$, indicating a weakening of the gravitational potential and a reduction in the required energy and angular momentum for circular motion. The parameter $A$ demonstrates an opposite influence, leading to an increase in both the energy and angular momentum of the circular orbits. In contrast, the string cloud parameter $a$ shows distinct behavior: increasing $a$ decreases $E_{\rm CO}^2$ while increasing $L_{\rm CO}^2$. This suggests that although circular orbits become energetically more favorable in the presence of a stronger string cloud background, they require higher angular momentum to remain stable. When the circular orbit radius approaches the photon sphere radius $r_{\rm ph}$, as defined later in Eq.~(\ref{Eqrph}), both $E_{\rm CO}^2$ and $L_{\rm CO}^2$ tend to infinity. Therefore, $r_{\rm ph}$ can be considered as the minimum cutoff radius for timelike circular orbits. As the circular orbit radius increases continuously, $L_{\rm CO}^2$ approaches 0, indicating the maximum cutoff radius for circular orbits. The SCOs satisfy the equations $V'_{\rm{eff}}(r)=0$  and $V''_{\rm{eff}}(r)=0$. Formally, the radii of SCOs yields
\begin{equation}\label{RISCO}
r_{\rm{SCO}}=-\frac{3f(r_{\rm{SCO}})f'(r_{\rm{SCO}})}{f(r_{\rm{SCO}})f''(r_{\rm{SCO}})-2f'(r_{\rm{SCO}})^2}.
\end{equation}
It can be observed from Figs.~\ref{FigEE} and \ref{FigLL} that, $r_{\rm SCO}$ is a monotonic function of $E_{\rm CO}^2$, as well as $L_{\rm CO}^2$, thus the minimum and maximum values of $E_{\rm CO}^2$ (and $L_{\rm CO}^2$) for SCOs correspond to the ISCO and OSCO, respectively. If Eq.~(\ref{RISCO}) has no solution, it indicates the absence of SCO. Based on the data in Table~\ref{tab:physical_quantities}, we briefly summarize the effects of the MCDF and cloud of strings parameters on the SCO boundaries. Increasing the parameter $Q$ decreases the ISCO radius while increasing the OSCO radius, thereby expanding the radial span of stable orbits. Conversely, increasing parameters $B$, $\beta$, or $a$ leads to an increase in $r_{\text{ISCO}}$ and a decrease in $r_{\text{OSCO}}$, consequently shrinking the region of stable circular orbits. Notably, the parameter $A$ exhibits a more complex, non-monotonic influence on both $r_{\text{ISCO}}$ and $r_{\text{OSCO}}$ within the parameter range examined.

The angular velocity of a particle orbiting a black hole measured by an observer located at infinity, known as the Keplerian frequency, is defined by
\begin{equation}\label{Omigaexpression}
\Omega_{\rm{K}}=\frac{d\phi}{dt}\equiv\frac{\dot{\phi}}{\dot{t}}\Rightarrow\Omega_{\rm{K}}^2=\frac{f'(r)}{2r}.
\end{equation}
The radial dependence of the Keplerian frequencies for test particles around a black hole is shown in Fig.~\ref{figOmigaK}. From Fig.~\ref{figOmigaK}, one can observe that $\Omega_{\rm K}(r)$ is a monotonically decreasing function of $r$. One can observe the figures to understand the influence of each parameter on $\Omega_{\rm K}$. It is worth noting that parameter B, as well as $\beta$, have opposite effects on $\Omega_{\rm K}$ near the minimum cutoff end and near the maximum cutoff end of the circular orbits. The parameter $a$, however, does not affect the Keplerian frequency of circular orbits, which is understandable considering the definition of $\Omega_{\rm{K}}$ in Eq.~(\ref{Omigaexpression}) and the lapse function in Eq.~(\ref{rightmetric}).
\begin{figure*}[htbp]
\centering % \begin{center}/\end{center} takes some additional vertical space
\includegraphics[width=.48\textwidth]{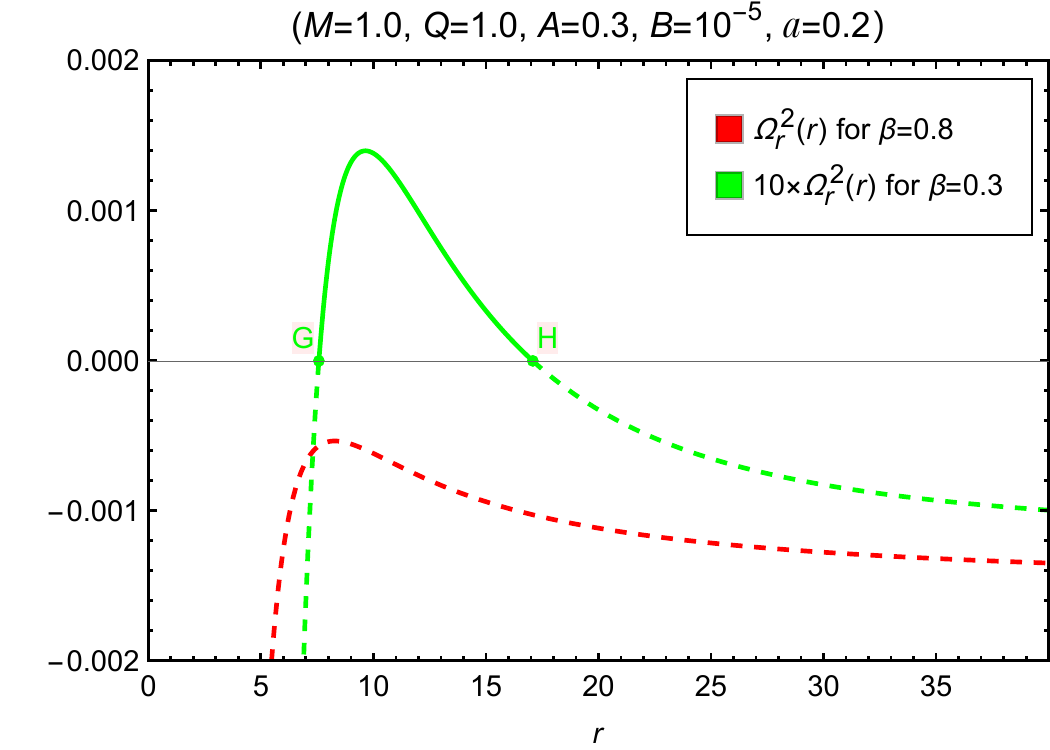}
\includegraphics[width=.49\textwidth]{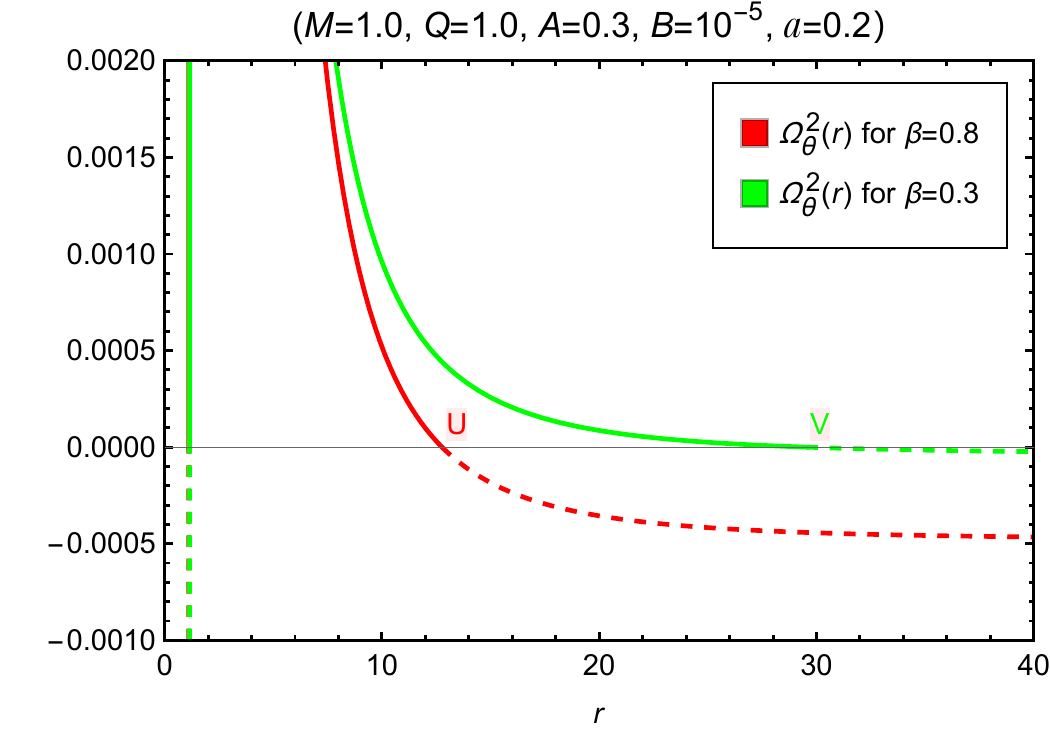}
% "\includegraphics" is very powerful; the graphicx package is already loaded
\caption{\label{Figomiga2}  Radial (left) and vertical (right) epicyclic frequencies for black holes with ($\beta=0.3$) and without ($\beta=0.8$) SCOs. The radial coordinates of points $\mathrm{G}$, $\mathrm{H}$, $\mathrm{U}$ and $\mathrm{V}$ are $r_\mathrm{G}=7.573887$, $r_\mathrm{H}=17.081879$, $r_\mathrm{U}=12.766328$ and $r_\mathrm{V}=29.532725$.}
\end{figure*}

The orbit properties of timelike particles can also be studied by examining the oscillatory motions and the epicyclic frequencies. In order to dertermine the radial locations where circular equatorial motion is either stable or unstable in the radial or vertical directions, one can calculate the radial and vertical epicyclic frequencies $\Omega_r$ and $\Omega_{\theta}$ as follows. According to Eqs.~(\ref{time}) and (\ref{radial}), radial and vertical motions around a circular equatorial orbit are governed by the equations
{\small
\begin{eqnarray}
&&\frac{1}{2}\left(\frac{dr}{dt}\right)^2=-\frac{1}{2}\frac{f(r)^3}{E^2}\left[1-\frac{E^2}{f(r)}+\frac{L^2}{r^2\sin^2{\theta}}\right]\equiv V_{\rm{eff}}^{(r)},  \label{radialeom} \\
&&\frac{1}{2}\left(\frac{d\theta}{dt}\right)^2=-\frac{1}{2}\frac{f(r)^2}{r^2E^2}\left[1-\frac{E^2}{f(r)}+\frac{L^2}{r^2\sin^2{\theta}}\right]\equiv V_{\rm{eff}}^{(\theta)},
 \label{verticaleom}
\end{eqnarray}
}where the factor $\sin^2{\theta}$ is recovered to study the orbital perturbations. We then introduce small perturbations $\delta r$ and $\delta \theta$, and take the coordinate time derivative of Eqs.~(\ref{radialeom}) and (\ref{verticaleom}), which yields
\begin{equation}\label{Eqperturbation}
\frac{d^2(\delta r)}{dt^2}=\frac{d^2V_{\rm{eff}}^{(r)}}{dr^2}\delta r,~~\quad~~\frac{d^2(\delta \theta)}{dt^2}=\frac{d^2V_{\rm{eff}}^{(\theta)}}{d\theta^2}\delta \theta.
\end{equation}
The radial and vertical epicyclic frequencies can be derived as
\begin{equation}\label{epifrequncies}
\Omega_r^2=\frac{d^2V_{\rm{eff}}^{(r)}}{dr^2},~~\quad~~\Omega_\theta^2=\frac{d^2V_{\rm{eff}}^{(\theta)}}{d\theta^2}.
\end{equation}
By combining Eqs.~(\ref{ELexpressions}), (\ref{radialeom}) and (\ref{verticaleom}), one can obtain the explicit expressions for epicyclic frequencies
\begin{equation}\label{Omigaexpressions}
\Omega_r^2=-f'(r)^2+\frac{f(r)}{2r}\left[3f'(r)+rf''(r)\right],~~\quad~~\Omega_\theta^2=\frac{f'(r)}{2r}.
\end{equation}
Fig.~\ref{Figomiga2} illustrates the radial profiles of the radial epicyclic frequency $\Omega_r^2(r)$ (left panel) and the vertical epicyclic frequency $\Omega_\theta^2(r)$ (right panel) for test particles orbiting a black hole without SCOs and a black hole with SCOs. For the black hole without SCOs, $\Omega_r^2(r)$ remains negative across all radii, indicating the absence of stable circular orbits—consistent with the conclusion drawn from the left panel of Fig.~\ref{FigUeff}. The vertical epicyclic frequency $\Omega_\theta^2(r)$ drops to zero at $r_\mathrm{U}$, which coincides with the radius $r_\mathrm{A}$ in the left panel of Fig.~\ref{FigUeff}, where the angular momentum $L$ vanishes. In contrast, for the black hole with SCOs, $\Omega_r^2(r)$ becomes positive in the range $r_\mathrm{G}<r<r_\mathrm{H}$, suggesting radial stability of equatorial circular orbits within this interval. The right panel further shows that these orbits are also vertically stable, as $\Omega_\theta^2(r)$ remains positive over the same range, becoming zero only at $r_\mathrm{V}>r_\mathrm{H}$. This $r_\mathrm{V}$ corresponds to the radius $r_\mathrm{R}$ in the right panel of Fig.~\ref{FigUeff}, where the angular momentum $L$ also vanishes. Notably, $r_\mathrm{G}=r_\mathrm{K}$ and $r_\mathrm{H}=r_\mathrm{P}$, identifying points G and H as the ISCO and OSCO, respectively.

\subsection{The structure of lightlike geodesics}
\begin{table*}[htbp]
\caption{The radii of the ISCO and OSCO, the radius $r_{\rm{ph}}$ and impact parameter $b_{\rm{ph}}$ of the photon sphere, the event horizon $r_{\rm{h}}$ and pseudo-cosmological horizon $r_\mathrm{c}$, as well as the rays classification parameters $b_m^{\pm}$ with varying $Q$, $A$, $B$, $\beta$ and $a$ for $M=1.0$.}
\label{tab:physical_quantities}
\vspace{2mm}
\centering
\scriptsize
\renewcommand{\arraystretch}{1.5}
\newcolumntype{Y}{>{\centering\arraybackslash}X} % 居中对齐的自动伸缩列
\begin{tabularx}{\textwidth}{c|c|Y|Y|Y|Y|Y|Y|Y|Y|Y|Y|Y}
\toprule[\lightrulewidth]
\hline
\textbf{Fixed} & \textbf{Variable} & $r_{\mathrm{h}}$ & $r_{\mathrm{c}}$ & $r_{\rm ISCO}$ & $r_{\rm OSCO}$ & $r_{\mathrm{ph}}$ & $b_{\mathrm{ph}}$ & $b_1^-$ & $b_2^-$ & $b_2^+$ & $b_3^-$ & $b_3^+$ \\
\hline
\multirow{3}{*}{$\renewcommand{\arraystretch}{0.98}\begin{array}{c}
Q=1.05\\B=10^{-5}\\\beta=0.4\\a=0.3\end{array}$} & A=0.1 & 2.26732 & 90.07571 & 8.69919 & 11.76843& 3.75397 & 8.33683 & 3.45915 & 7.35818 & 12.36752 & 8.20790 & 8.61584 \\
& A=0.15 & 2.54482 & 91.54112 & 9.36178 & 11.67575 & 3.99203 & 8.59743 & 3.85441 & 7.82149 & 12.48722 & 8.51560 & 8.82638 \\
& A=1.0 & 2.85834 &111.87080 &9.23875 & 14.33245 & 4.28558 & 8.89129 & 4.27697 & 8.29310 & 12.74755 & 8.84333 & 9.06851 \\
\hline
\multirow{3}{*}{$\renewcommand{\arraystretch}{0.98}\begin{array}{c}
Q=2.0\\A=0.3\\\beta=0.2\\a=0.2\end{array}$} & $B=10^{-4}$ & 2.06708 & 78.93323 &8.14395 &10.07712 & 3.41944 & 6.97848 & 3.12700 & 6.35341 & 9.23575 & 6.91452 & 7.11420 \\
& $B=0.5\times10^{-4}$ & 2.03214 &105.79988 &7.40045 & 13.53791 & 3.39488 & 6.94228 & 3.06052 & 6.28799 & 9.36265 & 6.87328 & 7.08580 \\
& $B=10^{-6}$ & 1.83055 & 545.23543 &6.83086 & 44.26347 & 3.29520 & 6.81586 & 2.74856 & 5.98564 & 9.74406 & 6.71442 & 6.98476 \\
\hline
\multirow{3}{*}{$\renewcommand{\arraystretch}{0.98}\begin{array}{c}
Q=1.0\\A=0.15\\B=10^{-5}\\a=0.15\end{array}$} & $\beta=0.3$ & 1.74743 & 139.96221 &6.33758 &16.68805 & 3.01939 & 6.13287 & 2.62472 & 5.49560 & 8.22275 & 6.06048 & 6.26408 \\
& $\beta=0.4$ & 1.88695 & 101.28807 &6.73993 &12.94359 & 3.12036 & 6.25400 & 2.82861 & 5.71129 & 8.19369 & 6.19947 & 6.37058 \\
& $\beta=0.5$ & 1.97661 & 76.45065 &7.52082 &9.81981 & 3.20106 & 6.35455 & 2.96907 & 5.86329 & 8.14638 & 6.30907 & 6.45946 \\
\hline
\multirow{3}{*}{$\renewcommand{\arraystretch}{0.98}\begin{array}{c}
A=0.4\\B=10^{-4}\\\beta=0.2\\a=0.2 \end{array}$} & Q=0.1 & 2.50106 & 81.44970 &8.84294 &10.07227 & 3.74908 & 7.28353 & 3.72032 & 6.90541 & 9.33042 & 7.25871 & 7.37223 \\
& Q=2.0 & 2.31756 & 81.44970 &8.49246 &10.26470 & 3.60602 & 7.15721 & 3.47076 & 6.68531 & 9.30284 & 7.11930 & 7.26596 \\
& Q=3.0 & 1.91828 & 81.44970 &8.18079 &10.40254 & 3.43682 & 7.01502 & 2.92762 & 6.24640 & 9.27527 & 6.92923 & 7.15244 \\
\hline
\multirow{3}{*}{$\renewcommand{\arraystretch}{0.98}\begin{array}{c}
Q=1.0\\A=1.0\\B=10^{-5}\\\beta=0.2\end{array}$} & $a=0.0$ & 1.99475 & 279.10105 &6.02473 &26.02853 & 2.99764 & 5.19572 & 2.85662 & 5.01576 & 6.11544 & 5.18737 & 5.22656 \\
& $a=0.1$ & 2.21906 & 264.61416 &6.70534 &25.89194 & 3.33196 & 6.08644 & 3.21079 & 5.82398 & 7.55001 & 6.07201 & 6.14030\\
& $a=0.2$ & 2.49834 & 249.27535 &7.56325 &25.70801 & 3.74927 & 7.26384 & 3.65443 & 6.86943 & 9.73024 & 7.23779 & 7.36221 \\
\hline
\bottomrule[\lightrulewidth] % 添加粗线作为下边框
\end{tabularx}
\end{table*}
\begin{figure*}[htbp]
\centering % \begin{center}/\end{center} takes some additional vertical space
\includegraphics[width=.48\textwidth]{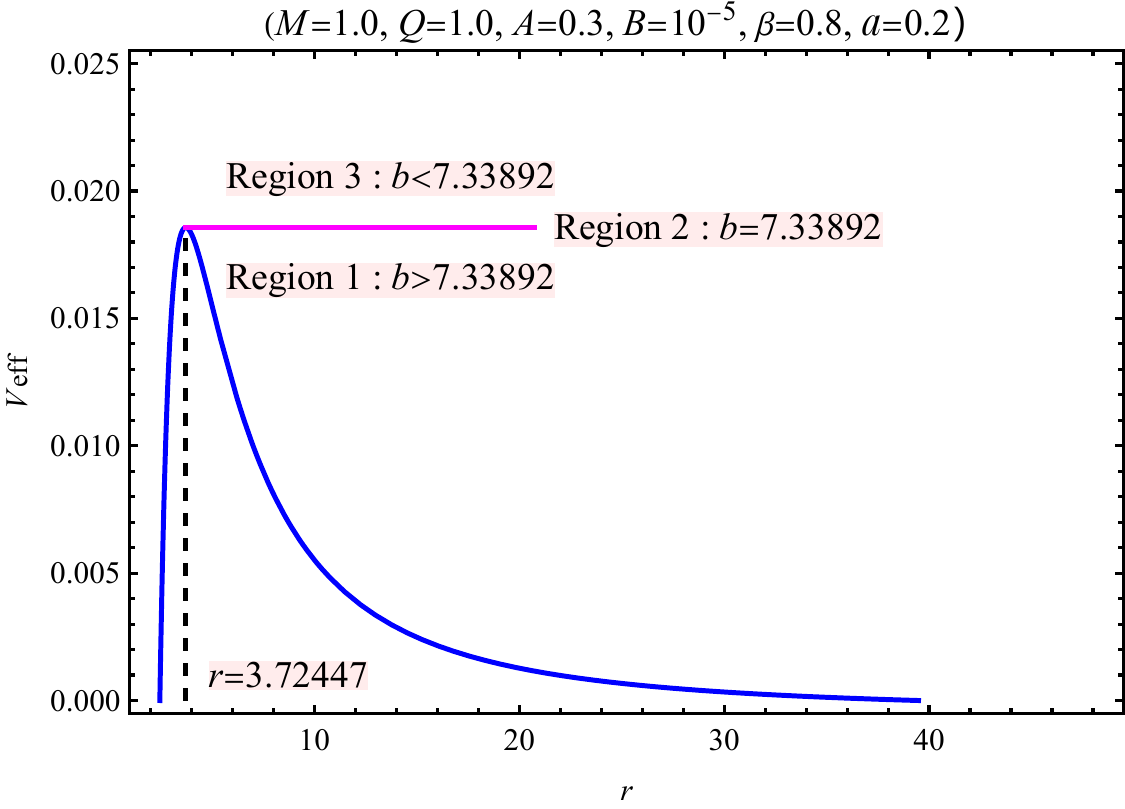}
\includegraphics[width=.48\textwidth]{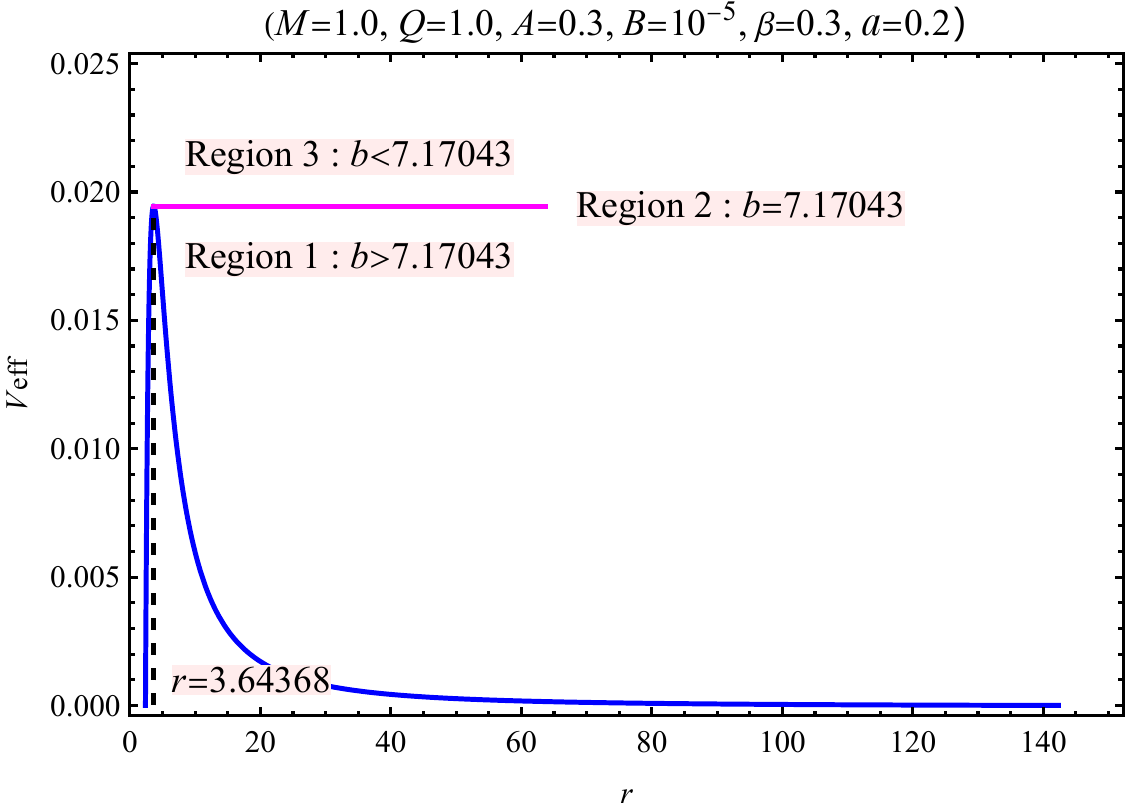}
% "\includegraphics" is very powerful; the graphicx package is already loaded
\caption{\label{FigVeff}   The profile of the effective potential (blue lines) for black holes without (left) and with (right) SCOs. The dashed lines indicate the radii of the photon sphere $r_{\rm{ph}}$. Region 2 (pink lines) correspond to $V_{\rm{eff}}(r)=E_{\rm{ph}}^2$ ($b=b_{\rm{ph}}$), while  Region  1  and  Region 3 correspond to $V_{\rm{eff}}(r)<E_{\rm{ph}}^2$ ($b>b_{\rm{ph}}$) and  $V_{\rm{eff}}(r)>E_{\rm{ph}}^2$  ($b<b_{\rm{ph}}$), respectively. }
\end{figure*}
According to Eq.~(\ref{epotential}), for lightlike particles, $\delta=0$, the effective potential is written as
\begin{equation}\label{EquUefflight}
V_{\rm eff}(r)=\frac{L^2}{r^2}f(r).
\end{equation}
At the photon sphere, the motion of the light ray satisfies $\dot{r}=0$ and $\ddot{r}=0$, implying
\begin{eqnarray}
V_{\rm{eff}}(r)=E_{\rm{ph}}^2, ~~~ V_{\rm{eff}}^{'}(r) = 0,
\label{condition1}
\end{eqnarray}
where the prime $'$ denotes the derivative with respect to the radial coordinate $r$. Considering Eq.~(\ref{EquUeff}), the radial coordinate of the photon sphere is determined by
\begin{equation}\label{Eqrph}
f(r_{\rm{ph}})-\frac{1}{2}r_{\rm{ph}}f'(r_{\rm{ph}})=0.
\end{equation}
Based on Eq.~(\ref{Eqrph}) and with the help of Eq.~(\ref{ELexpressions}), we can obtain the radius $r_{\rm{ph}}$ and impact parameter $b_{\rm{ph}}={|L_{\rm{ph}}|}/{E_{\rm{ph}}}$ of the photon sphere for different values of model parameters. Generally, obtaining analytic results for the radius and impact parameter is challenging, so we resort to numerical methods for their determination. The numerical results of radii and impact parameters of the photon sphere, as well as the event horizons and pseudo-cosmological horizons, are listed in Table \ref{tab:physical_quantities}. From Table~\ref{tab:physical_quantities}, it can be observed that as the parameter $A$ increases, all of $r_{\rm{h}}$, $r_{\rm{c}}$, $r_{\rm{ph}}$, and $b_{\rm{ph}}$ increase, with a particularly notable expansion in the outer communication region, defined as the radial distance between the pseudo-cosmological horizon and the event horizon. In contrast, increasing the values of $B$, $\beta$, or $a$ leads to an increase in $r_{\rm{h}}$, $r_{\rm{ph}}$, and $b_{\rm{ph}}$, while $r_{\rm{c}}$ decreases. As a result, the outer communication region becomes narrower with increasing $B$, $\beta$, or $a$. On the other hand, as the charge parameter $Q$ increases, $r_{\rm{h}}$, $r_{\rm{ph}}$, and $b_{\rm{ph}}$ all decrease, whereas $r_{\rm{c}}$ exhibits a slight increase. Consequently, the radial extent of the outer communication region is mildly broadened with increasing $Q$.

The particle’s motion, governed by Eq.~(\ref{vbr}), depends on the impact parameter and the effective potential. As shown in Fig.\ref{FigVeff}, the effective potential vanishes at the event horizon, increases to a maximum at the photon sphere $r_{\rm{ph}}$, and then decreases toward the pseudo-cosmological horizon. For lightlike geodesics, the potential features a single peak corresponding to an unstable circular orbit. Photon trajectories are classified into three regions depending on the impact parameter $b$
\begin{itemize}
    \item \textbf{Region 1} ($b > b_{\rm{ph}}$): photons originating from $r > r_{\rm{ph}}$ are reflected outward by the potential barrier, while those from $r < r_{\rm{ph}}$ fall into the singularity.
    \item \textbf{Region 2} ($b = b_{\rm{ph}}$): photons asymptotically orbit the black hole at $r_{\rm{ph}}$.
    \item \textbf{Region 3} ($b < b_{\rm{ph}}$): photons plunge directly into the black hole without encountering a turning point.
\end{itemize}
The path of the light ray can be illustrated based on the equation of motion. Combining Eqs.~(\ref{psi}) and (\ref{radial}), we obtain
\begin{equation}
\frac{dr}{d\phi}=\pm r^2 \sqrt{\frac{1}{b^2}-\frac{1}{r^2}f(r)}. \label{drp}
\end{equation}
To facilitate integration, we introduce the variable $u=1/r$. Thus, Eq.~(\ref{drp}) transforms into
\begin{equation}
\frac{du}{d\phi}=\mp \sqrt{\frac{1}{b^2}-u^2f\left(\frac{1}{u}\right)}\equiv \Phi(u).\label{gu}
\end{equation}
with
{\small
\begin{equation}
f\left(\frac{1}{u}\right) = 1 - a - 2Mu - \frac{1}{3u^2} \left( \frac{B}{1+A} \right)^{\frac{1}{1+\beta}} \mathcal{F}\left(\frac{1}{u}\right).
\end{equation}
}

The geometry of the geodesics is determined by the roots of the equation $ \Phi(u) =0$. Specifically, for $ b>b_{\rm{ph}}$, light will be deflected at the radial position $ u_i$ that satisfies $ \Phi(u_i) =0$. Therefore, finding the radial position $u_i$ is crucial for determining the trajectory of the light ray. Additionally, the observer's location is significant. While observers are typically situated at an infinite boundary for asymptotically flat spacetime, in our black hole model, the pseudo-cosmological horizon is present. Physically, the observer should be positioned within the domain of outer communication, which lies between the event horizon and the pseudo-cosmological horizon, similar to de Sitter spacetime. Here, to study the trajectory of the light ray, we place the observer near the pseudo-cosmological horizon.

\begin{figure}[H]
  \centering
    \includegraphics[width=0.7\linewidth]{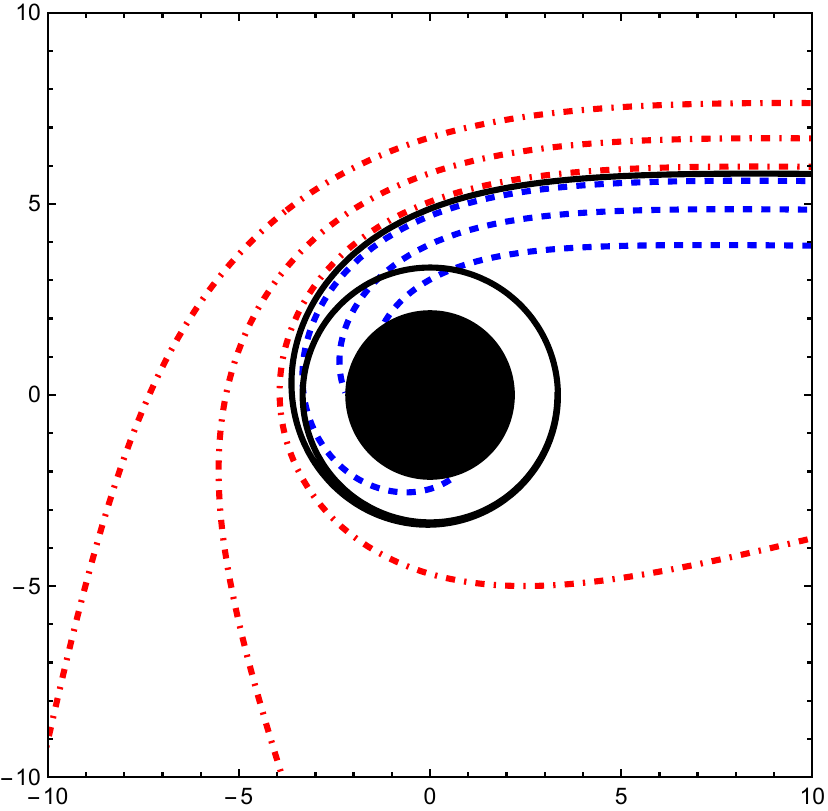}
  \caption{\label{Figtrajectory} Trajectories of light rays around the black hole for varying impact parameters: $b < b_{\rm ph}$ (blue, dashed), $b = b_{\rm ph}$ (black, solid), and $b > b_{\rm ph}$ (red, dot-dashed). The black hole parameters are set to $A = 1.0$, $B = 10^{-5}$, $\beta = 0.2$, $Q = 1.0$, and $a = 0.1$.
}
\end{figure}

Utilizing Eq.~\eqref{gu}, we determine the trajectories of light rays, as depicted in Fig.~\ref{Figtrajectory}. All light rays approach the black hole from the right. For impact parameters $b < b_{\rm ph}$, the light rays (blue, dashed lines) are entirely captured by the black hole. When $b > b_{\rm ph}$, the light rays (red, dot-dashed lines) are deflected and do not enter the black hole; notably, some rays passing close to the black hole can be reflected back toward the side from which they originated. At the critical impact parameter $b = b_{\rm ph}$, the light rays (black, solid lines) asymptotically orbit the black hole, corresponding to the photon sphere. This behavior aligns with the analysis of the effective potential discussed earlier. The deflection of light rays contributes to the formation of the black hole shadow. Since the observed light originates from accreting matter, the profiles of the accretion matter play a crucial role in determining the characteristics of black hole shadows, which will be discussed in Section~\ref{disk}.
\subsection{Shadow and observation constraints}
\begin{figure*}[htbp]
  \centering
  \begin{minipage}{0.32\textwidth}
    \includegraphics[width=\linewidth]{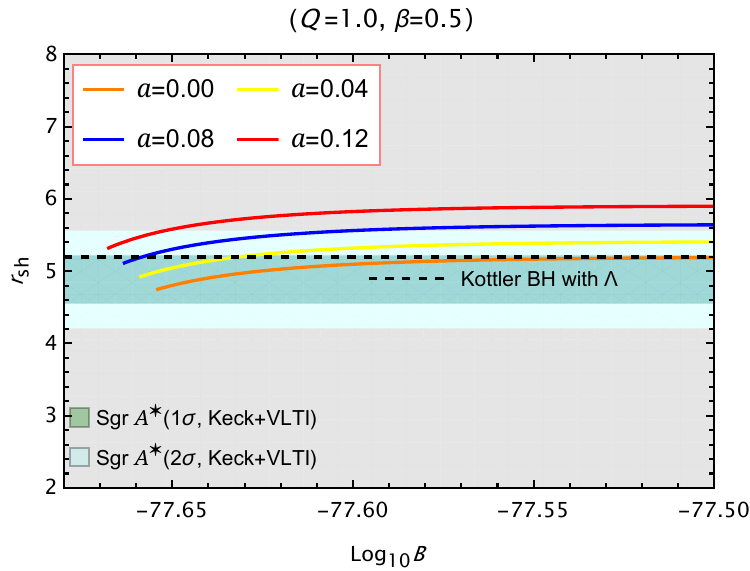}
  \end{minipage}
  \begin{minipage}{0.32\textwidth}
    \includegraphics[width=\linewidth]{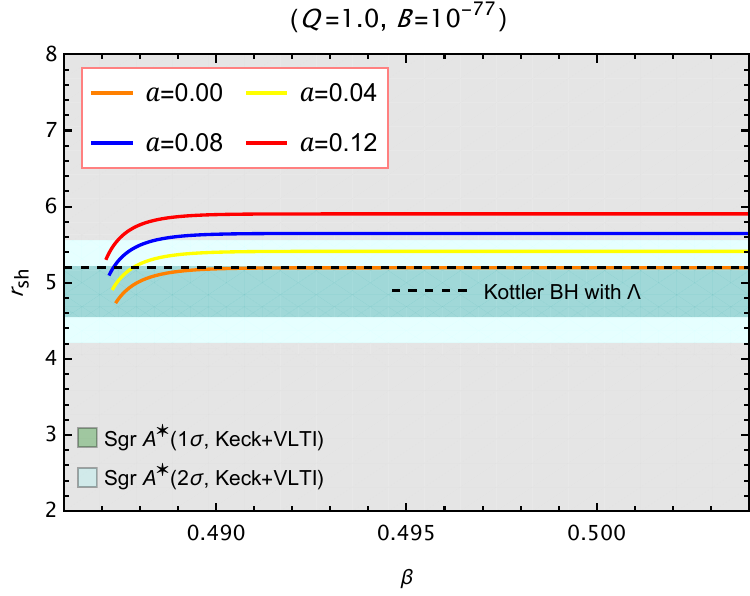}
  \end{minipage}
 \begin{minipage}{0.315\textwidth}
    \includegraphics[width=\linewidth]{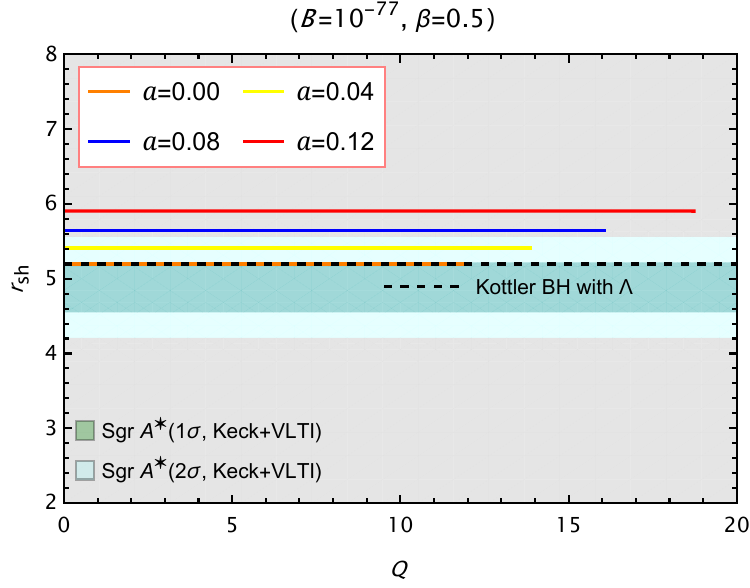}
  \end{minipage}
  \caption{\label{Figconstraints1} Shadow radius in $M$ unit as a function of $\mathrm{Log}_{10}B$ (left), $\beta$ (middle) and $Q$ (right). The dark-green and light-green shaded regions represent the regions of $1\sigma$ and $2\sigma$ confidence intervals, respectively, with respect to the Sgr ${\rm A}^{\ast}$ observations. Here we set $r_{\rm O}\sim26996\ {\rm ly}\backsimeq2.55\times10^{20}\mathrm{m}$. The left endpoints of the plots in the left and middle panels, as well as the right endpoints in the right panel, indicate the presence of extremal black hole solutions, where the event horizon coincides with the inner horizon.
}
\end{figure*}
\begin{figure*}[htbp]
  \centering
  \begin{minipage}{0.32\textwidth}
    \includegraphics[width=\linewidth]{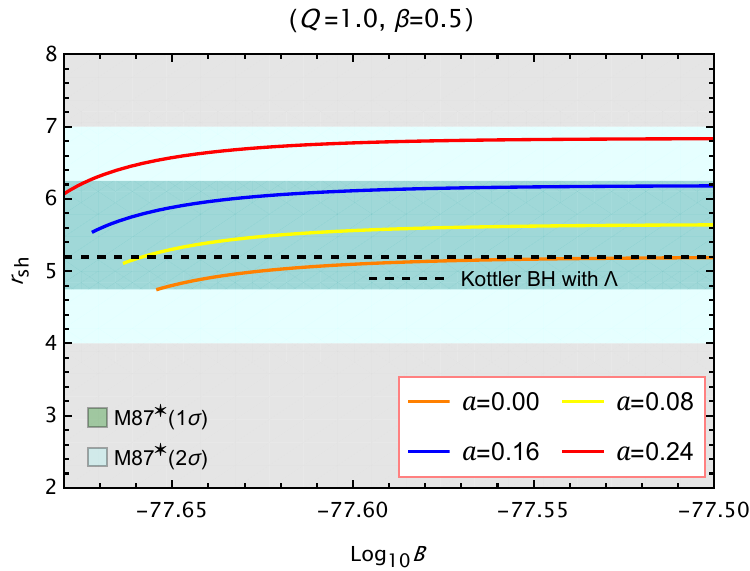}
  \end{minipage}
  \begin{minipage}{0.32\textwidth}
    \includegraphics[width=\linewidth]{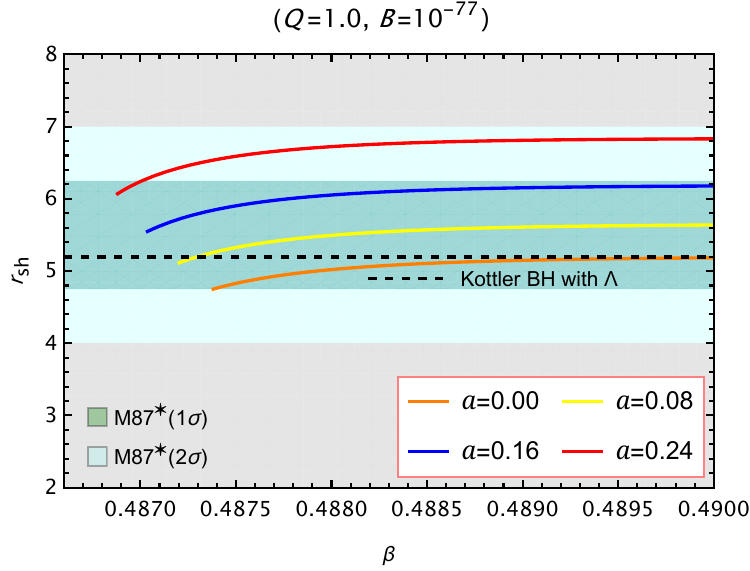}
  \end{minipage}
 \begin{minipage}{0.315\textwidth}
    \includegraphics[width=\linewidth]{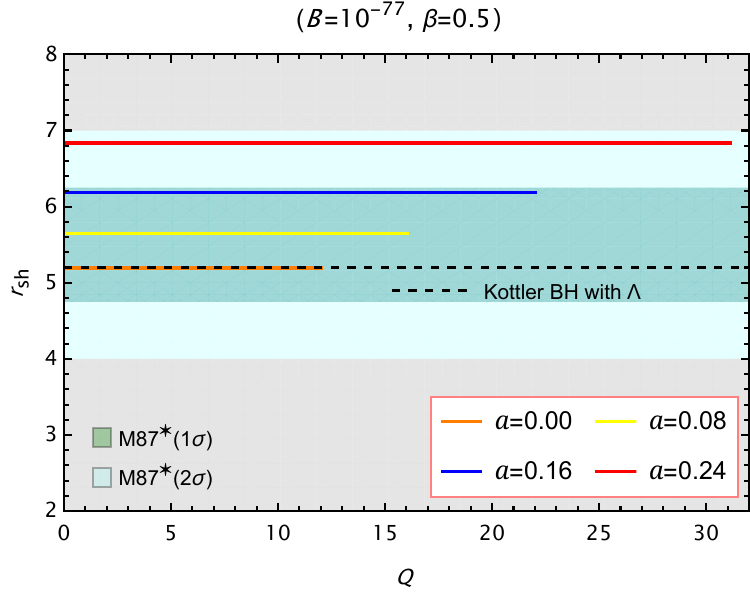}
  \end{minipage}
  \caption{\label{Figconstraints2} Shadow radius in $M$ unit as a function of $\mathrm{Log}_{10}B$ (left), $\beta$ (middle) and $Q$ (right). The dark-green and light-green shaded regions represent the regions of $1\sigma$ and $2\sigma$ confidence intervals, respectively, with respect to the M$87^{\ast}$ observations. Here we set $r_{\rm O}\sim16.4 \ {\rm Mpc}\backsimeq5.06\times10^{23}\mathrm{m}$. The left endpoints of the plots in the left and middle panels, as well as the right endpoints in the right panel, indicate the presence of extremal black hole solutions, where the event horizon coincides with the inner horizon.
}
\end{figure*}
For a black hole spacetime with a (pseudo-)cosmological horizon, such as a Kottler black hole (also known as Schwarzshild de Sitter black hole), a quintessence-black hole or a CDF-black hole~\cite{Li:2024abk}, the size of the black hole shadow can explicitly depend on the radial coordinate of the observer, and on whether the observer is static or comoving. For a static observer located at a distance $r_{\rm O}$, the angular size of the black hole shadow $\alpha_{\rm{sh}}$ is given by (see \cite{Perlick:2021aok})
\begin{equation}\label{Eqsinrsh}
{\rm sin}^2 \alpha_{{\rm sh}}=\frac{r^2_{{\rm ph}}}{f(r_{\rm ph})}\frac{f(r_{\rm O})}{r^2_{\rm O}}.
\end{equation}
It is evident that the observed size of a black hole shadow is dependent on the location of the observer. Therefore, specifying the size of a black hole shadow within a metric that includes a pseudo-cosmological horizon is conditional.

In the physically relevant small-angle approximation, it is easy to see that the shadow size is given by (see \cite{Perlick:2021aok} for more detailed discussions)
\begin{equation}\label{Eqrsh}
r_{{\rm sh}}=r_{{\rm ph}}\sqrt{\frac{f(r_{\rm O})}{f(r_{\rm ph})}}.
\end{equation}
Note that the black hole shadow radius is affected by the MCDF and cloud of strings parameters. The explicit dependence of the shadow size on the observer's position is clear in Eq.~(\ref{Eqrsh}). We assume that the MCDF is responsible for the accelerated expansion of the universe. Considering the asymptotic behavior of $f(r)$ at spatial infinity as shown in Eq.~(\ref{metricasymptotic}), we identify $\left( \frac{B}{1+A} \right)^{\frac{1}{1+\beta}}$ with the cosmological constant $\Lambda$. This serves as a prerequisite for constraining the model parameters using astronomical observations. That is,
\begin{equation}\label{Acondition}
\left( \frac{B}{1+A} \right)^{\frac{1}{1+\beta}} = \Lambda, \quad \text{with} \quad \Lambda = 1.47\times10^{-52}~\mathrm{m}^{-2}.
\end{equation}
In terms of observation, the black hole shadow diameter $r_{\rm sh}$ can be measured with the EHT data. The bounds for the black hole shadow radius in $M$ unit can be deduced as
\begin{align}
 \begin{array}{c}
                            \text{Sgr A$^{\ast}$\textsuperscript{\cite{Vagnozzi:2022moj}}} \\
                            \text{\footnotesize Keck+VLTI}
                          \end{array} &\left\{\begin{array}{c}
                            \text{$4.55\lesssim r_{\rm sh}/M \lesssim 5.22$~~($1\sigma$)} \\
                            \text{$4.21\lesssim r_{\rm sh}/M \lesssim 5.56$~~($2\sigma$)}
                          \end{array}\right., & ~\\
\text{M$87^{\ast}$\textsuperscript{\cite{Bambi:2019tjh,Allahyari:2019jqz}}} &\left\{\begin{array}{c}
                            \text{$4.75\lesssim r_{\rm sh}/M \lesssim 6.25$~~($1\sigma$)} \\
                            \text{$4\lesssim r_{\rm sh}/M \lesssim 7$~~($2\sigma$)}
                          \end{array}\right..
\end{align}
\begin{table}[htbp]
\centering
\caption{Allowed parameter space of $B$ and $A$ constrained by Sgr A* and M87* data, for two values of $a$ with fixed $Q=1.0$, $\beta=0.5$.}
\label{tab:paramsB}
\renewcommand{\arraystretch}{1.4}
\resizebox{\columnwidth}{!}{
\begin{tabular}{|c|c|c|c|c|c|}
\hline
\multicolumn{1}{|c|}{} & \multicolumn{1}{c|}{} & \multicolumn{4}{c|}{\textbf{(lower bound, upper bound)}} \\ \cline{3-6}
\textbf{Date} & \textbf{Params}& \multicolumn{2}{c|}{$a=0.00$} & \multicolumn{2}{c|}{$a=0.08$} \\ \cline{3-6}
\multicolumn{1}{|c|}{} & \multicolumn{1}{c|}{} & $1\sigma$ & $2\sigma$ & $1\sigma$ & $2\sigma$ \\
\hline
\multirow{2}{*}{Sgr A*} & $\text{Log}_{10} B$ & ($-77.654,\text{-}$) & ($-77.654,\text{-}$) & ($-77.663,-77.657$) & ($-77.663,-77.601$) \\ \cline{2-6}
                        & $A$                & ($0.245,\text{-}$)   & ($0.245,\text{-}$)   & ($0.218, 0.236$)   & ($0.218, 0.406$)   \\ \hline
\multirow{2}{*}{M87*}  & $\text{Log}_{10} B$ & ($-77.654,\text{-}$) & ($-77.654,\text{-}$) & ($-77.663,\text{-}$)      & ($-77.663,\text{-}$)      \\ \cline{2-6}
                       & $A$                & ($0.245,\text{-}$)   & ($0.245,\text{-}$)   & ($0.218,\text{-}$)        & ($0.218,\text{-}$)        \\ \hline
\end{tabular}
}
\end{table}
\begin{table}[htbp]
\centering
\caption{Allowed parameter space of $\beta$ and $A$ constrained by Sgr A* and M87* data, for two values of $a$ with fixed $Q=1.0$, $B=10^{-77}$.}
\label{tab:paramsbeita}
\renewcommand{\arraystretch}{1.4}
\resizebox{\columnwidth}{!}{
\begin{tabular}{|c|c|c|c|c|c|}
\hline
\multicolumn{1}{|c|}{} & \multicolumn{1}{c|}{} & \multicolumn{4}{c|}{\textbf{(lower bound, upper bound)}} \\ \cline{3-6}
\textbf{Date} & \textbf{Params}& \multicolumn{2}{c|}{$a=0.00$} & \multicolumn{2}{c|}{$a=0.08$} \\ \cline{3-6}
\multicolumn{1}{|c|}{} & \multicolumn{1}{c|}{} & $1\sigma$ & $2\sigma$ & $1\sigma$ & $2\sigma$ \\
\hline
\multirow{2}{*}{Sgr A*} & $\beta$ & ($0.4874,\text{-}$) & ($0.4874,\text{-}$) & ($0.4872,0.4873$) & ($0.4872,0.4884$) \\ \cline{2-6}
                        & $A$                & ($0.244,\text{-}$)   & ($0.244,\text{-}$)   & ($0.218, 0.236$)   & ($0.218, 0.406$)   \\ \hline
\multirow{2}{*}{M87*}  & $\beta$ & ($0.4874,\text{-}$) & ($0.4874,\text{-}$) & ($0.4872,\text{-}$)      & ($0.4872,\text{-}$)      \\ \cline{2-6}
                       & $A$                & ($0.244,\text{-}$)   & ($0.244,\text{-}$)    & ($0.218,\text{-}$)        & ($0.218,\text{-}$)        \\ \hline
\end{tabular}
}
\end{table}
For different values of the parameter $a$, we fix two parameters at a time
and compute the influence of the varying parameters $B$, $\beta$, and $Q$
on the shadow radius $r_{\rm sh}$ according to Eq.~(\ref{Eqrsh}). The results are shown in Figs.~\ref{Figconstraints1} and \ref{Figconstraints2},
together with the observational constraints on the black hole shadow radius
from Sgr~${\rm A}^{\ast}$ and M$87^{\ast}$. It is worth noting that the parameter $A$ is not explicitly shown in the figures,
as it is treated as a derived quantity from $B$ and $\beta$
using Eq.~(\ref{Acondition}). To comprehensively consider the impact of MCDF and cloud of strings, we also take into account the Kottler black hole in the figures, which has the following metric lapse function
\begin{equation}\label{Kottlermetric}
f_{\rm Kottler}(r)=1-\frac{2M}{r}-\frac{\Lambda}{3}r^2.
\end{equation}
In the left panels of Fig.~\ref{Figconstraints1} and Fig.~\ref{Figconstraints2}, for fixed values of $\beta$ and $Q$,
the shadow radius increases monotonically with $\log_{10}B$.
This trend becomes more evident for larger values of $a$.
According to Eq.~(\ref{Acondition}), the parameter $A$ is positively correlated with $B$.
Therefore, one can infer that $A$ also enlarges the shadow size in this case.

In the middle panels, fixing $B$ and $Q$, the shadow radius again grows with increasing $\beta$,
and the effect is amplified by larger $a$.
Similarly, due to the positive correlation between $A$ and $\beta$,
it follows that $A$ contributes to an increase in $r_{\rm sh}$ as well.

In the right panels, where $B$ and $\beta$ are held constant,
we find that $Q$ has negligible influence on the shadow radius,
as all curves appear nearly flat regardless of the value of $a$.

These results suggest that the shadow radius is sensitive to the parameters $B$, $\beta$ and $A$,
as well as the string cloud parameter $a$,
but largely insensitive to the electric charge $Q$ under the considered parameter range. As shown in all three panels, the curve of $r_{\rm sh}$ with $a=0$
approaches that of the Kottler black hole with $\Lambda$,
serving as a limiting case of our model. Compared to the Kottler case, the black hole with MCDG and cloud of strings
offers more flexibility in fitting observational data,
covering a broader range of shadow radii consistent with observations.
As a result, using the $1\sigma$ and $2\sigma$ confidence intervals of the $r_{\rm sh}$ for Sgr A$^{\ast}$ and M$87^{\ast}$, lower and upper bounds on the parameter are presented in Tables~\ref{tab:paramsB} and \ref{tab:paramsbeita}. Note that when determining the lower bounds of the parameters, we also took into account the constraints imposed by the existence of the black hole solution.

As we shall see later, not only is the position of observer important for determining the shadow radius, it is also crucial for the observation of the optical images of a black hole. In a given black hole spacetime, when the observer is close to the black hole, light rays with a large impact parameter will fail to reach the observer. To avoid such a situation, in our following study on the optical images of the black hole, we will adopt the convention that placing the observer as far away as possible from the black hole, specifically inside the domain of outer communication and near the pseudo-cosmological horizon.
\section{Optical appearance with thin disk accretion }
\label{disk}
In this section, we consider an optically and geometrically thin accretion disk viewed face-on. As discussed in~\cite{Gralla:2019xty}, a distinctive feature of such a setup is the formation of lensing and photon rings around the black hole shadow. The lensing ring consists of light rays that intersect the disk plane twice outside the horizon, while the photon ring is formed by rays that intersect it three or more times. Thus, analyzing photon trajectories in our model is essential to differentiate between these two structures.

\subsection{Number of orbits of the deflected light trajectories}
\begin{figure*}[htbp]
\centering % \begin{center}/\end{center} takes some additional vertical space
\includegraphics[width=.328\textwidth]{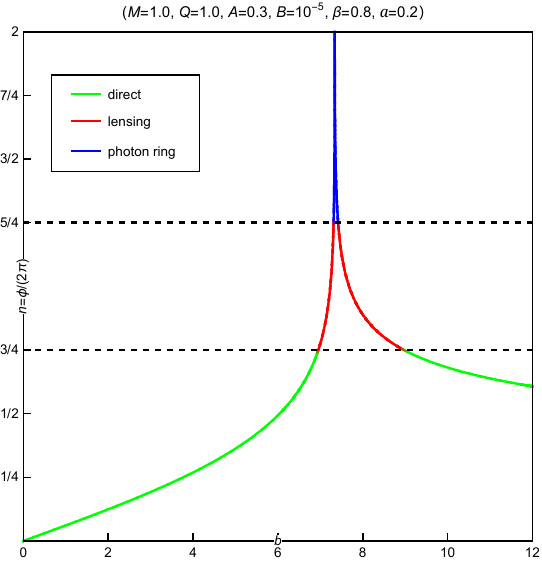}
\includegraphics[width=.328\textwidth]{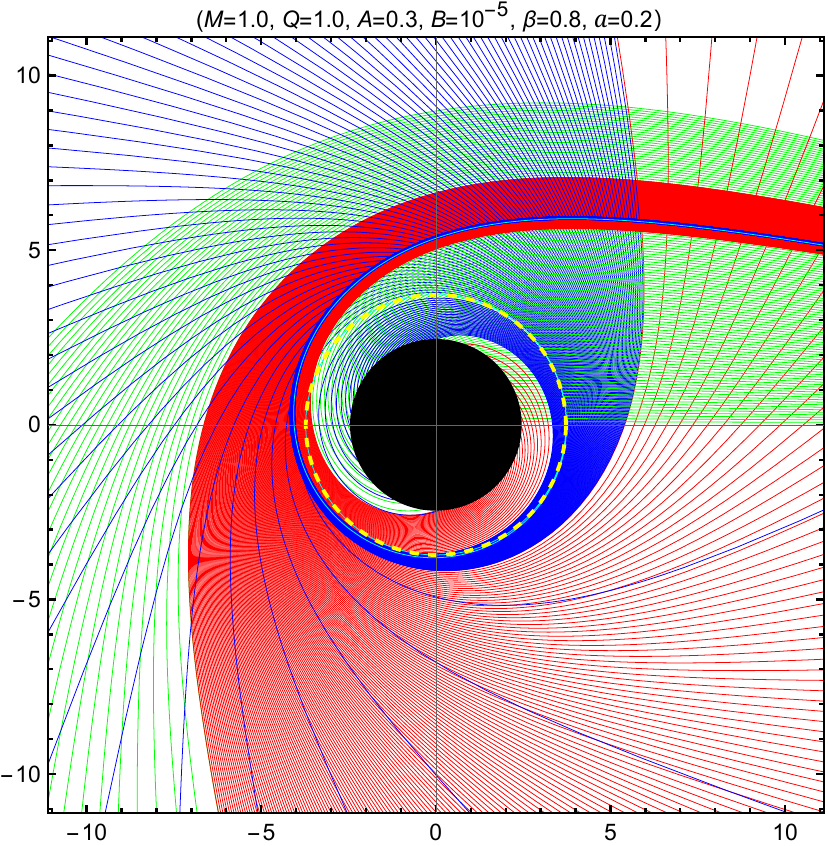}
\includegraphics[width=.328\textwidth]{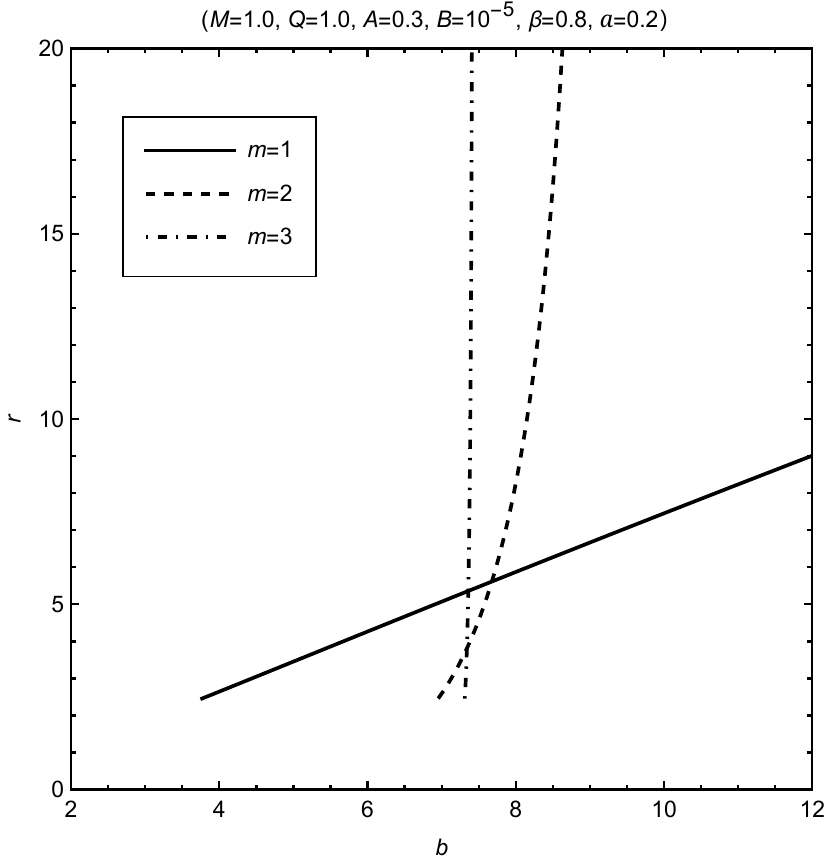}\\
\includegraphics[width=.328\textwidth]{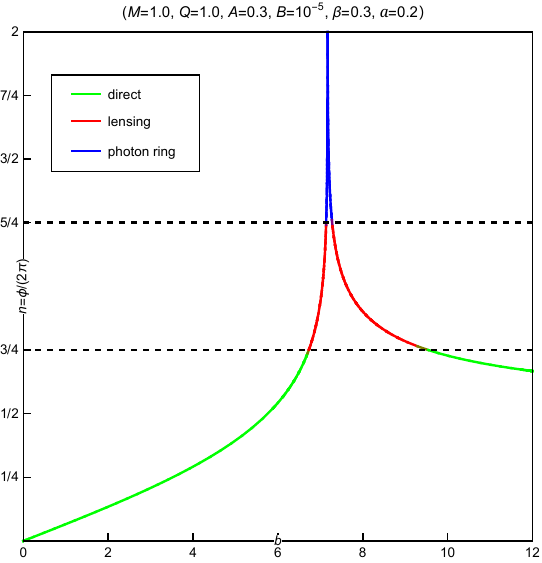}
\includegraphics[width=.328\textwidth]{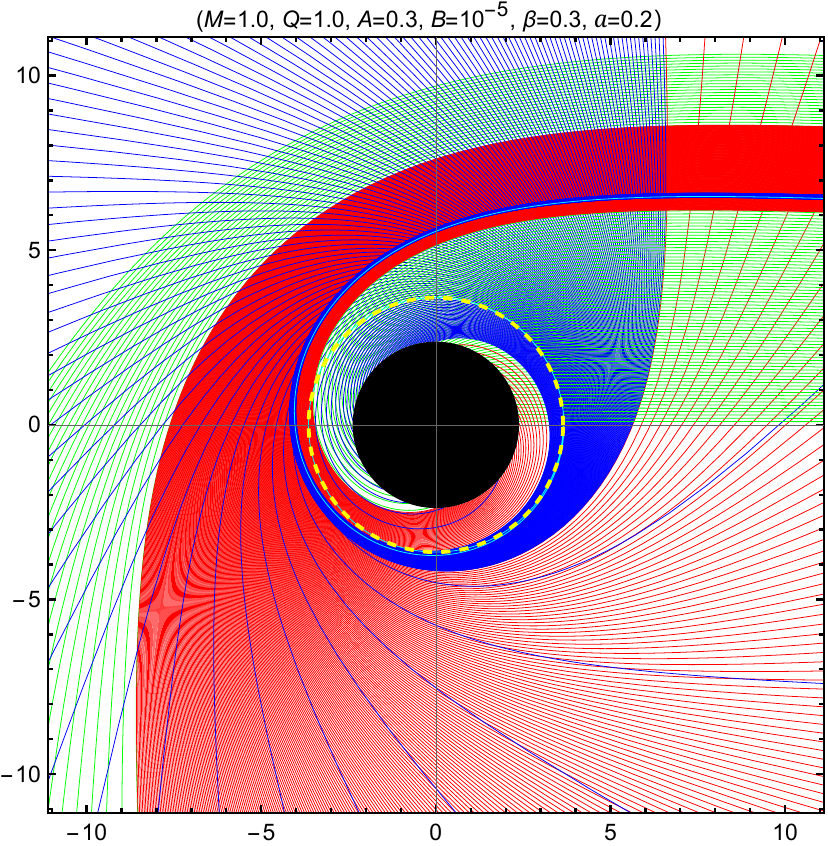}
\includegraphics[width=.328\textwidth]{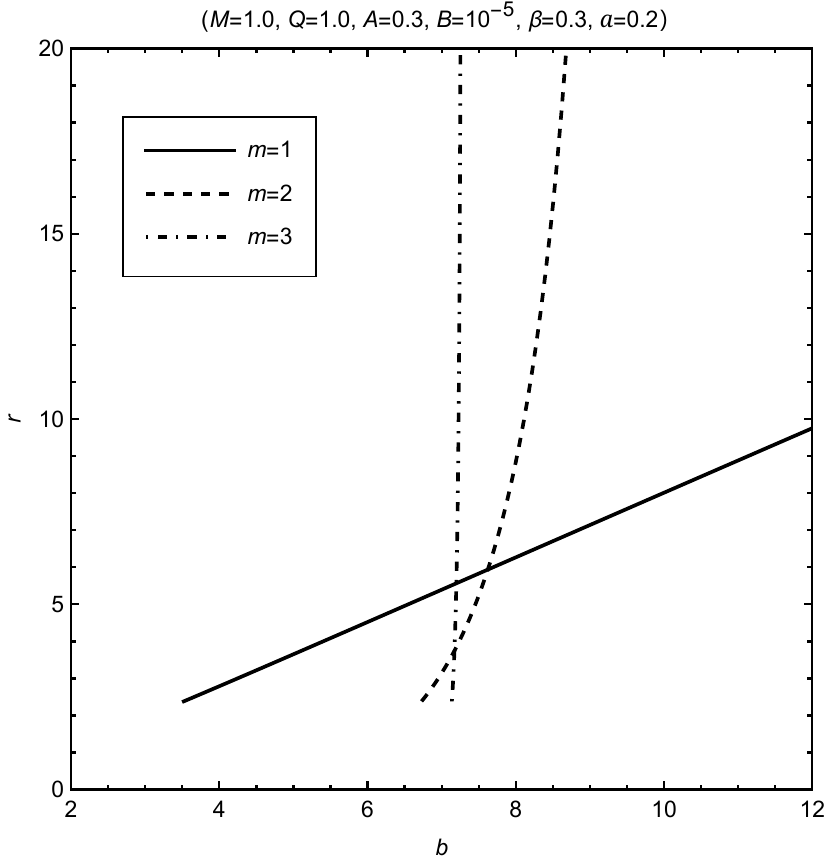}
\caption{\label{Figtrajectory2} Behavior of photons as a function of impact parameter $b$ in the black hole spacetime without (top row) and with (bottom row) SCOs. \textbf{Left column:} The fractional number of orbits $n = \phi / 2\pi$ is shown, where $\phi$ is the total azimuthal angle accumulated outside the horizon. The direct, lensing, and photon ring regions correspond to $n < 3/4$ (green), $3/4 < n < 5/4$ (red), and $n > 5/4$ (blue), respectively. \textbf{Middle column:} Selected photon trajectories are displayed in polar coordinates $(r, \phi)$. The spacings in impact parameter are $1/10$, $1/100$, and $1/1000$ for the direct (green), lensing (red), and photon ring (blue) bands, respectively. The black holes are represented as solid black disks. \textbf{Right column:} The first three transfer functions $r_m(b)$ are shown for a face-on thin disk. The solid, dashed, and dot-dashed lines indicate the radial positions of the first, second, and third photon-disk intersections, respectively.}
\end{figure*}
The trajectories of light rays are shown in the middle column of Fig.~\ref{Figtrajectory2}. In these diagrams, green lines represent direct emission, red lines represent lensing rings, and blue lines correspond to photon rings. These labels follow the definitions outlined in \cite{Gralla:2019xty}, where light rays intersect the disk plane once, twice, or more than twice, respectively.

Another approach to distinguish the trajectories of light rays is by considering the total number of orbits, denoted as \( n = \frac{\phi}{2\pi} \) \cite{Gralla:2019xty}. The total number of orbits is displayed in the left column of Fig.~\ref{Figtrajectory2}. The green, red, and blue lines maintain their designations for direct emission, lensing rings, and photon rings, respectively. According to the definitions of these light rings, it is clear that direct emissions correspond to \( n < 3/4 \), lensing rings to \( 3/4 < n < 5/4 \), and photon rings to \( n > 5/4 \). The parameter ranges of \( b \) for direct emission, photon rings, and lensing rings are provided in the last five columns of Table~\ref{tab:physical_quantities} for varying values of \( A \), \( B \), \( \beta \), \( Q \), and \( a \).
\subsection{Observed specific intensities and transfer functions}
We now turn to the analysis of the observed specific intensity arising from the accretion of a thin disk. Following the framework of \cite{Gralla:2019xty}, we consider isotropic emission in the rest frame of static worldlines within the thin disk, which lies in the equatorial plane of the black hole. A static observer is assumed to be located at the North pole. Let \( I_{\mathrm{e}}(r) \) and \( \nu_{\mathrm{e}} \) denote the emitted specific intensity and frequency, respectively, while \( I_{\mathrm{O}}(r) \) and \( \nu_{\mathrm{O}} \) denote their observed counterparts. According to Liouville’s theorem, which ensures the conservation of \( I_{\mathrm{e}}/\nu_{\mathrm{e}}^3 \) along a light ray, the observed specific intensity can be written as
\begin{equation}
I_{\mathrm{O}}(r) = \left[\frac{f(r)}{f(r_{\mathrm{O}})}\right]^{3/2} I_{\mathrm{e}}(r).
\end{equation}

The total observed intensity is obtained by integrating the specific intensity over frequency:
{\small
\begin{eqnarray}
I_{\mathrm{obs}}(r) &&= \int I_{\mathrm{O}}(r) \, d\nu_{\mathrm{O}} \nonumber \\
&&= \int \left[\frac{f(r)}{f(r_{\mathrm{O}})}\right]^2 I_{\mathrm{e}}(r) \, d\nu_{\mathrm{e}} = \left[\frac{f(r)}{f(r_{\mathrm{O}})}\right]^2 I_{\mathrm{em}}(r),
\end{eqnarray}
}
where \( I_{\mathrm{em}}(r) = \int I_{\mathrm{e}}(r) \, d\nu_{\mathrm{e}} \) represents the total emitted specific intensity near the disk.

When a light ray is traced backward from the observer and intersects the disk, it collects radiation from the disk at the point of intersection. For rays with \( 3/4 < n < 5/4 \), the red trajectories curve around the black hole and strike the far side of the disk (see the red lines in the middle column of Fig.~\ref{Figtrajectory2}), resulting in additional brightness from a second encounter. For \( n > 5/4 \), the blue rays bend around the black hole more extensively and strike the near side of the disk again (see the blue lines in Fig.~\ref{Figtrajectory2}), gaining brightness from a third intersection. Therefore, the total observed intensity is the sum of contributions from each disk intersection:
\begin{equation}
I_{\mathrm{obs}}(b) = \sum_{m} \left[\frac{f(r)}{f(r_{\mathrm{O}})}\right]^2 I_{\mathrm{em}} \bigg|_{r = r_m(b)}, \label{lir}
\end{equation}
where \( r_m(b) \) denotes the radial coordinate of the \( m \)-th intersection of the light ray with the disk plane outside the horizon, commonly referred to as the transfer function. For simplicity, we neglect any absorption in the thin disk, which could otherwise reduce the observed intensity from multiple crossings.

From Eq.~(\ref{lir}), it is evident that the observed intensity depends on the observer's location. Specifically, the intensity is relatively high when \( r_{\mathrm{O}} \) is small or close to \( r_c \), and significantly lower when the observer is located far from both the event horizon and the quasi-cosmological horizon.

The transfer function characterizes the mapping between the impact parameter \( b \) and the disk coordinate \( r \), and its slope \( dr/db \) acts as a demagnification factor. The right column of Fig.~\ref{Figtrajectory2} presents the transfer functions as functions of \( b \), where the solid, dashed, and dot-dashed curves correspond to the first (\( m = 1 \)), second (\( m = 2 \)), and third (\( m = 3 \)) disk crossings, respectively. The first transfer function essentially reflects the redshifted brightness profile of the disk. The second transfer function is associated with the lensing ring, including contributions from photon rings, and provides a highly demagnified image of the disk's far side. The third transfer function corresponds to the photon ring and yields an extremely demagnified image of the near side of the disk, as its slope approaches infinity. As a result, this contribution is negligible in the total observed brightness.
\begin{figure*}[htbp]
\centering % \begin{center}/\end{center} takes some additional vertical space
\includegraphics[width=.34\textwidth]{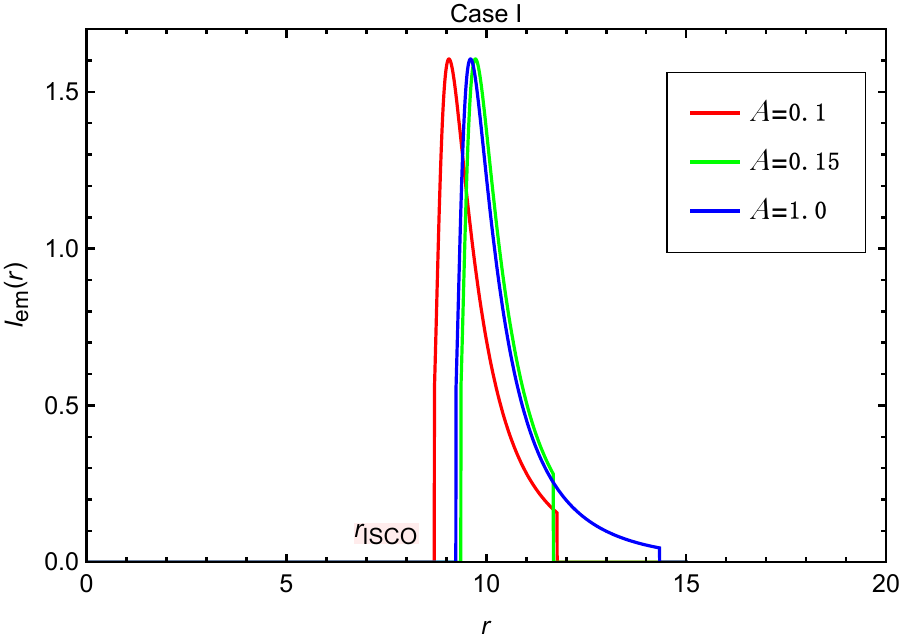}
\includegraphics[width=.34\textwidth]{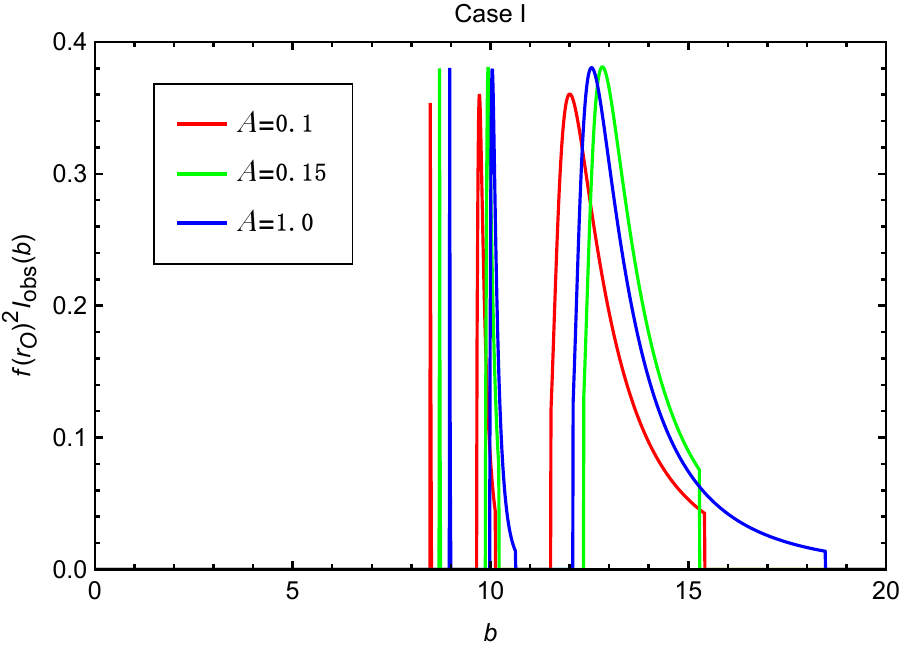}
\includegraphics[width=.28\textwidth]{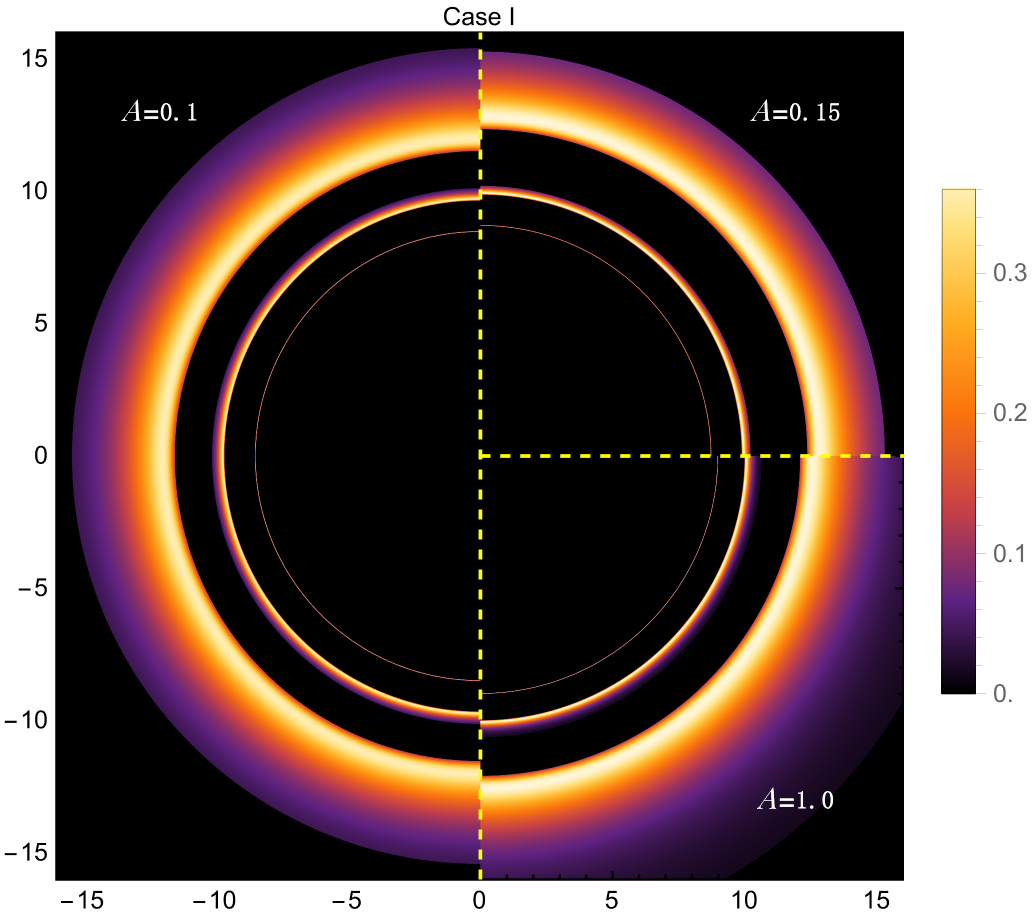}
\includegraphics[width=.34\textwidth]{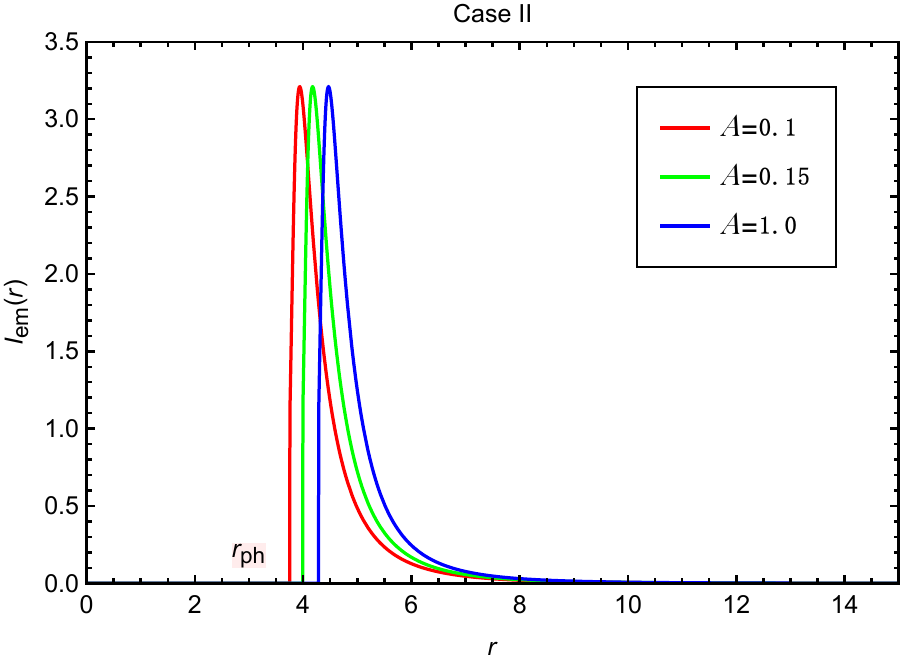}
\includegraphics[width=.34\textwidth]{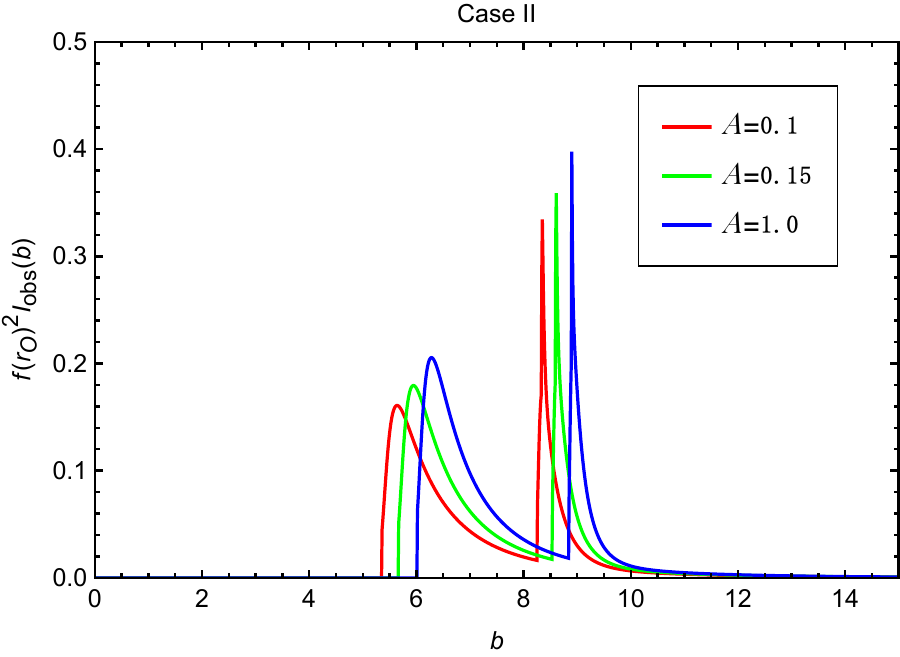}
\includegraphics[width=.28\textwidth]{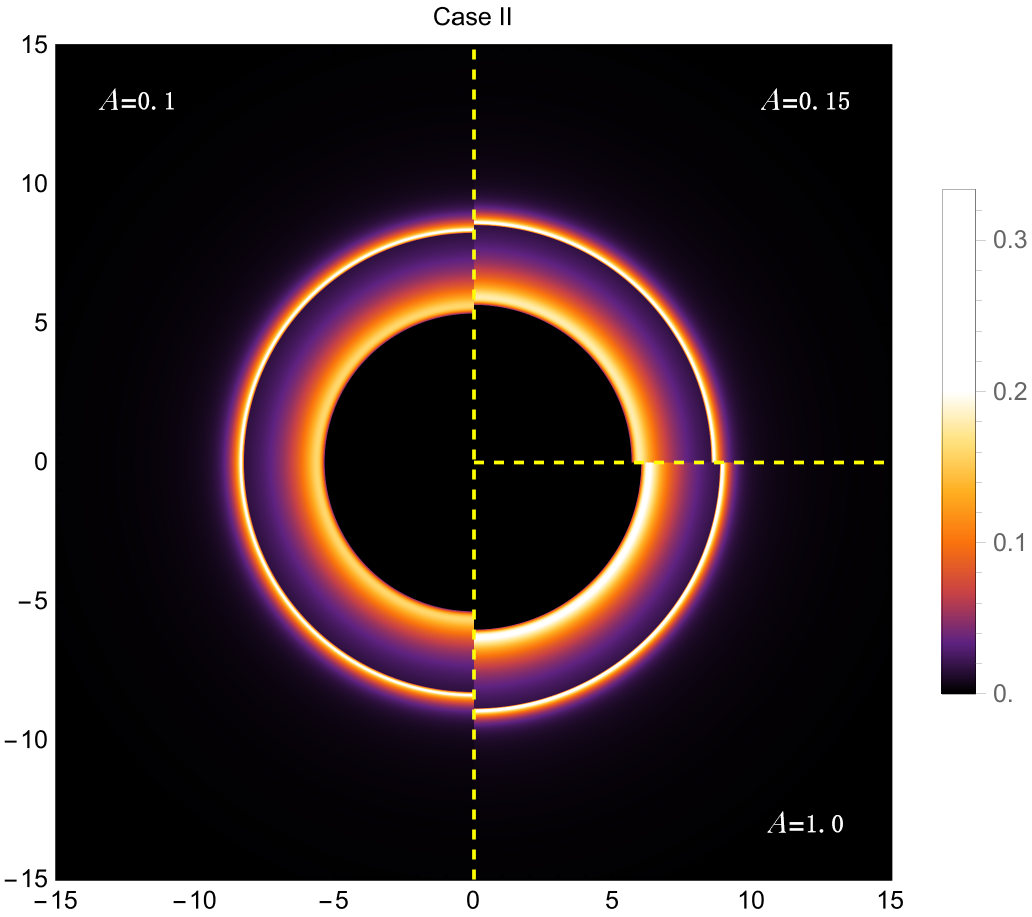}
\includegraphics[width=.34\textwidth]{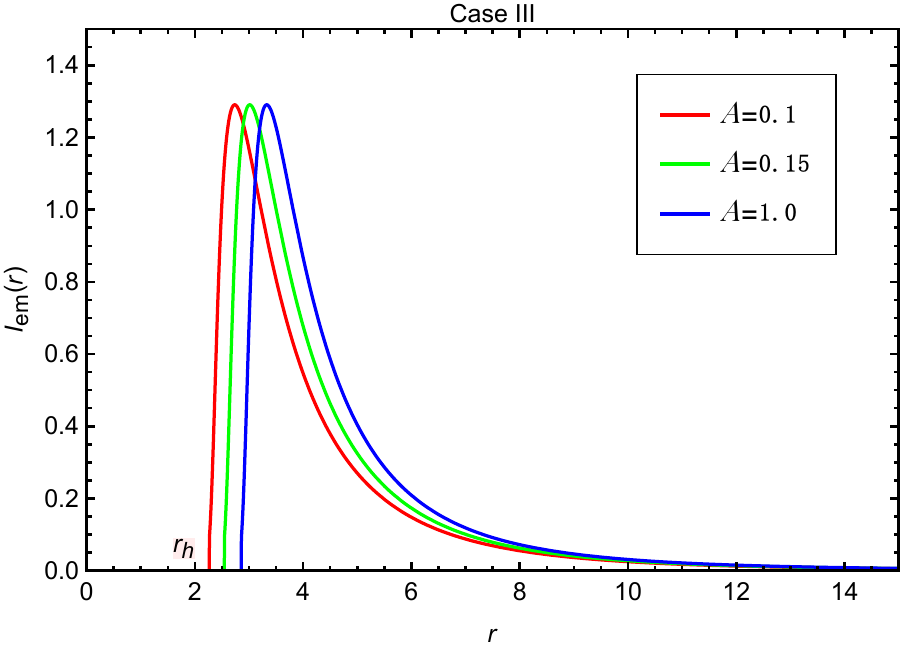}
\includegraphics[width=.34\textwidth]{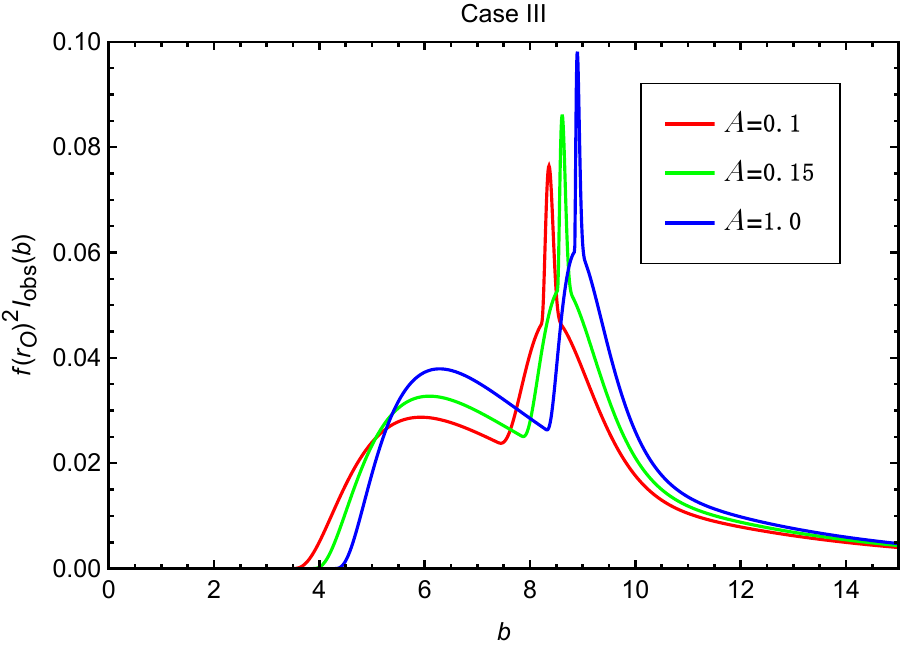}
\includegraphics[width=.28\textwidth]{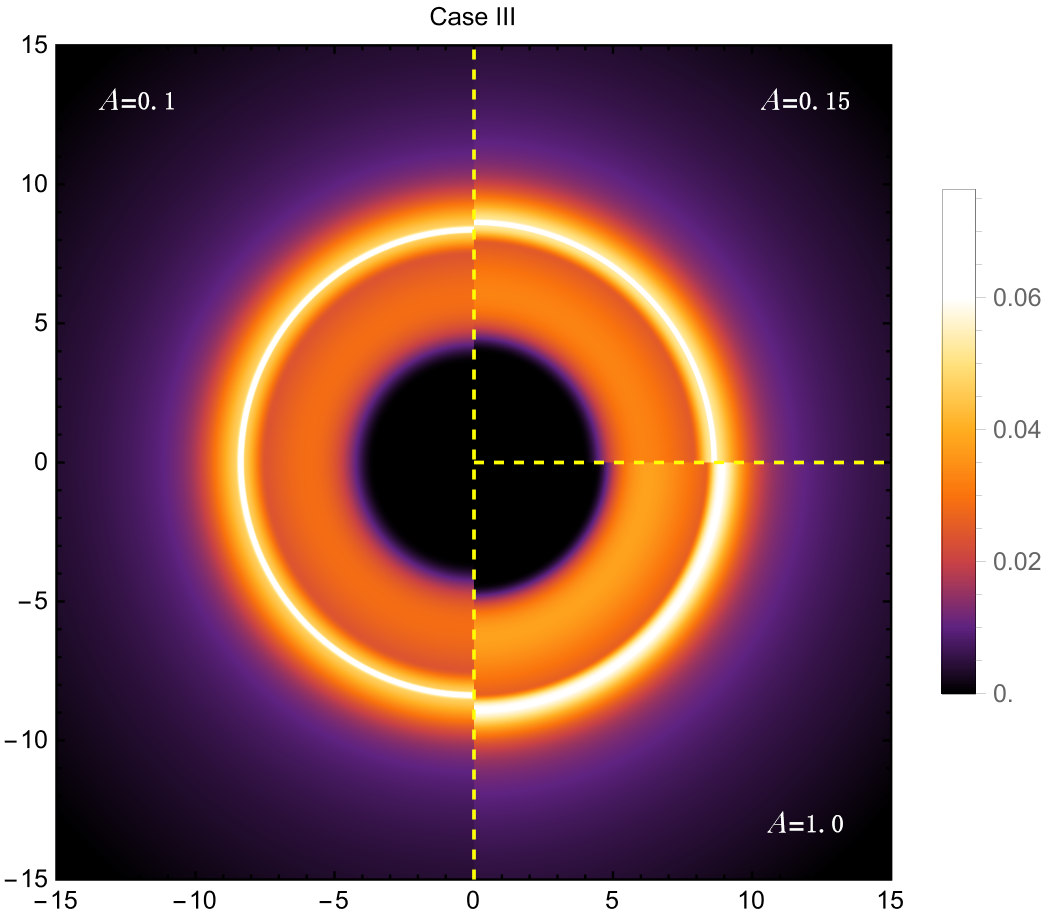}
% "\includegraphics" is very powerful; the graphicx package is already loaded
\caption{\label{FigvaryAA}Observational appearances of a geometrically and optically thin disk with different emissivity profiles near black hole with $M=1$, $Q=1.05$, $B=10^{-5}$, $\beta=0.4$ and $a=0.3$, viewed from a face-on orientation. The left column shows the profiles of various emissions $I_{\mathrm{em}}(r)$. The middle column exhibits the observed intensities $I_{\mathrm{obs}}(b)$ as a function of the impact parameter $b$. The red, green and blue curves correspond to $A=0.1$, $A=0.15$ and $A=1.0$, respectively. The right column shows the 2-dim density plots of the observed intensities $I_{\mathrm{obs}}(b)$.}
\end{figure*}
\begin{figure*}[htbp]
\centering % \begin{center}/\end{center} takes some additional vertical space
\includegraphics[width=.34\textwidth]{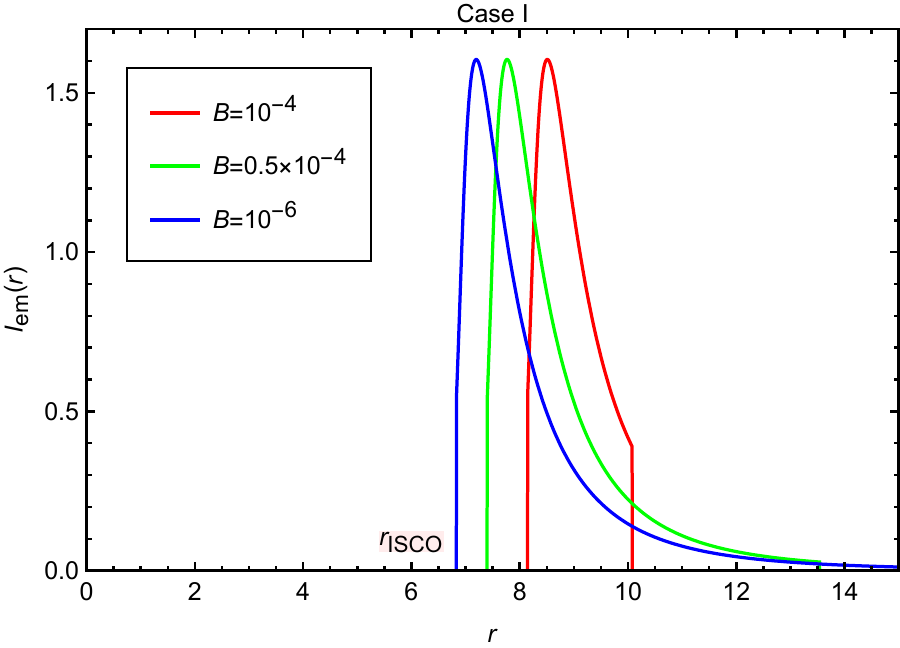}
\includegraphics[width=.34\textwidth]{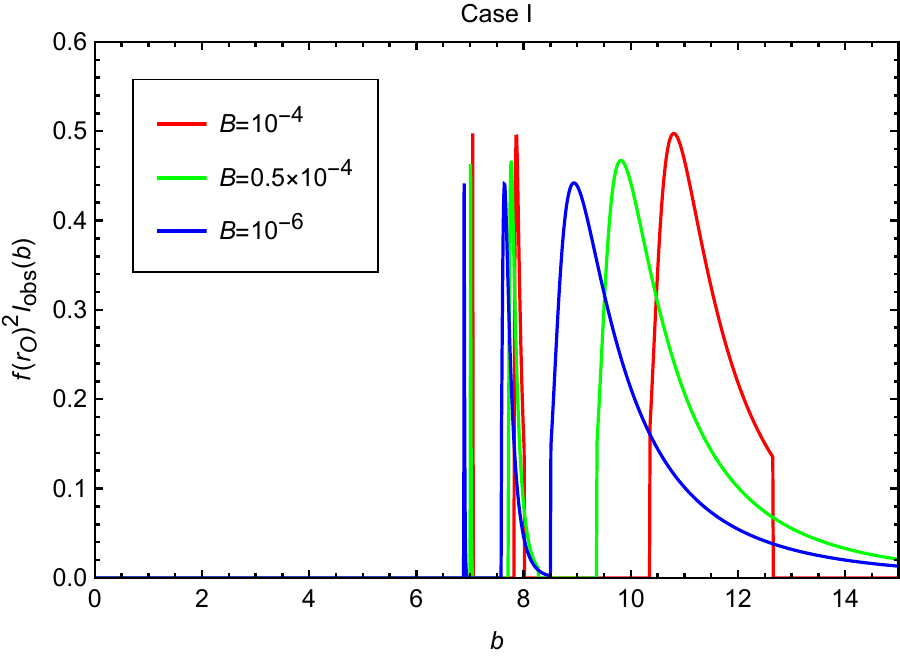}
\includegraphics[width=.28\textwidth]{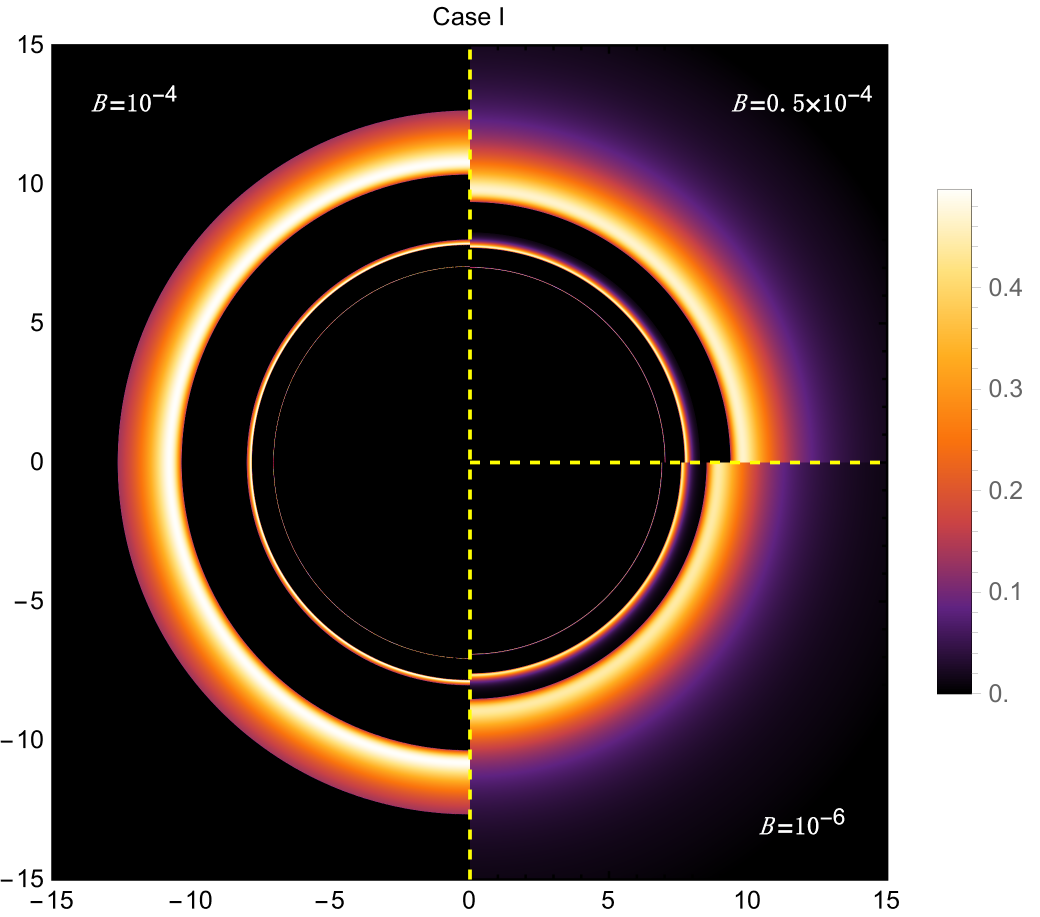}
\includegraphics[width=.34\textwidth]{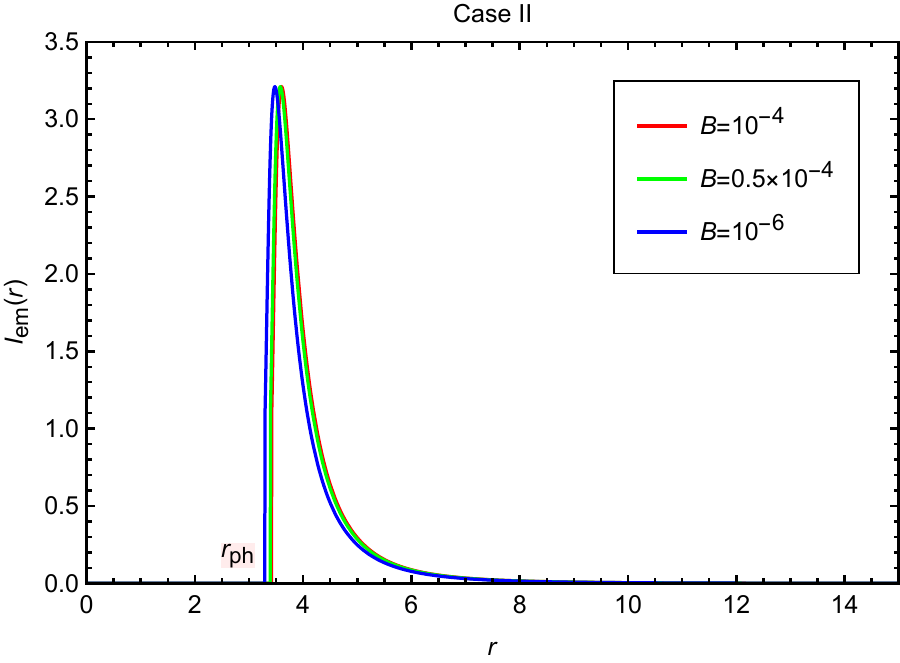}
\includegraphics[width=.34\textwidth]{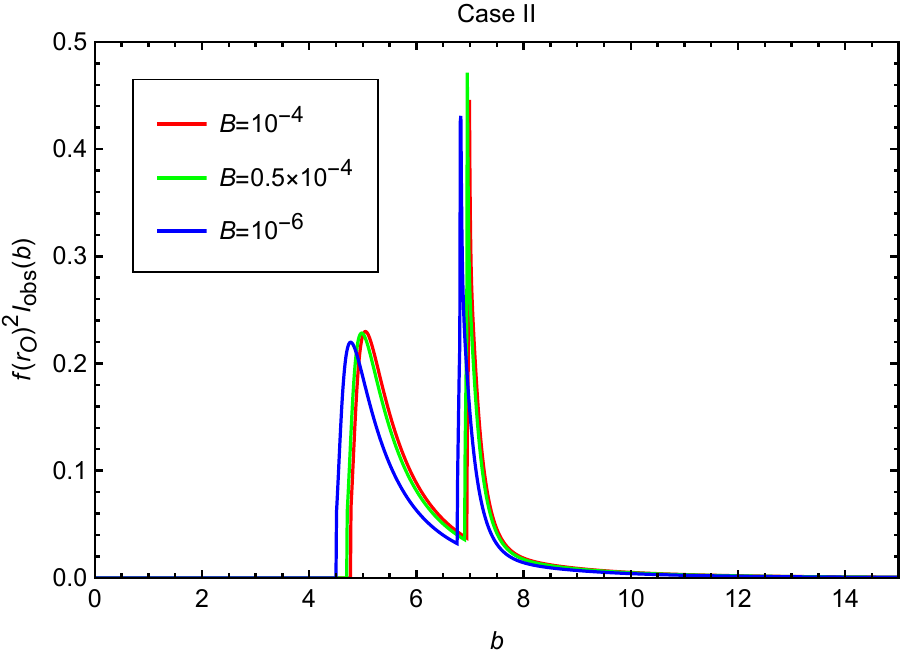}
\includegraphics[width=.28\textwidth]{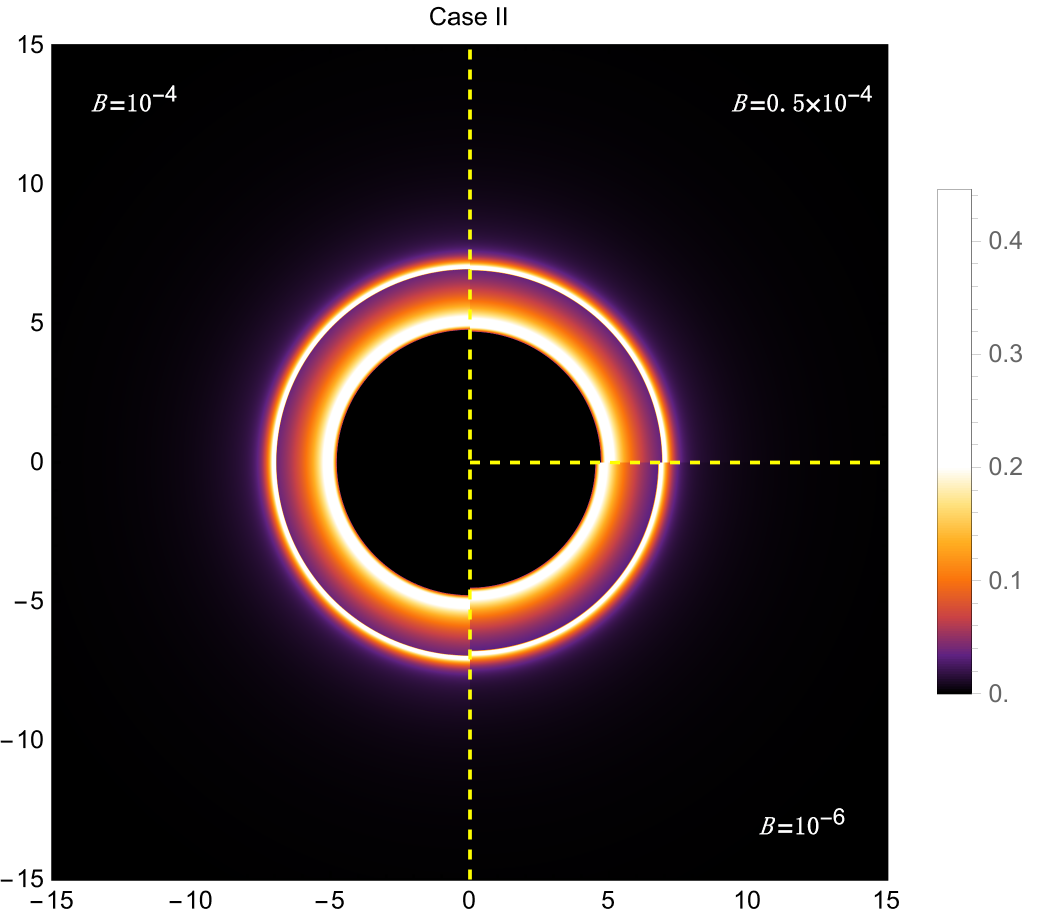}
\includegraphics[width=.34\textwidth]{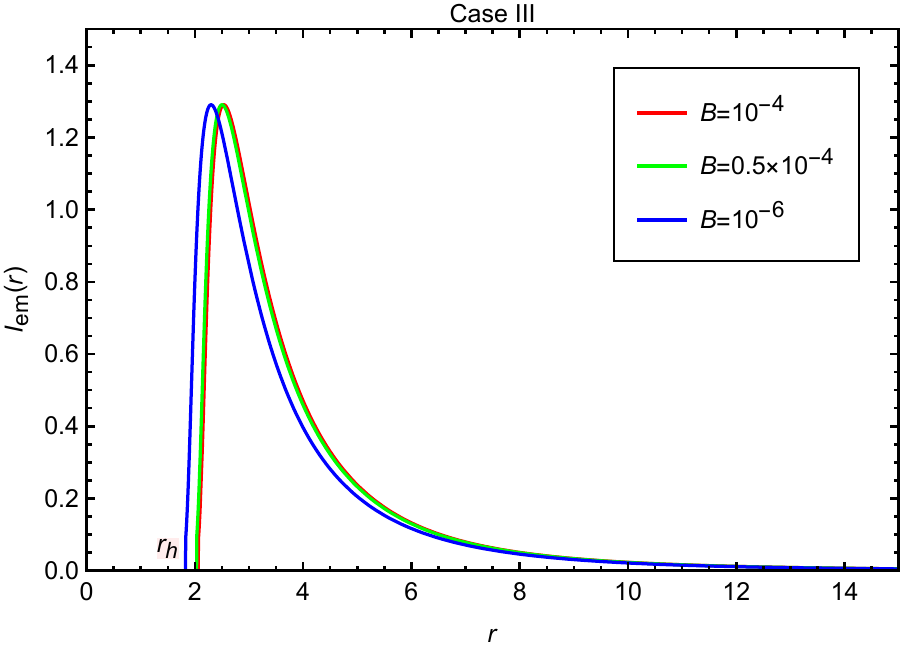}
\includegraphics[width=.34\textwidth]{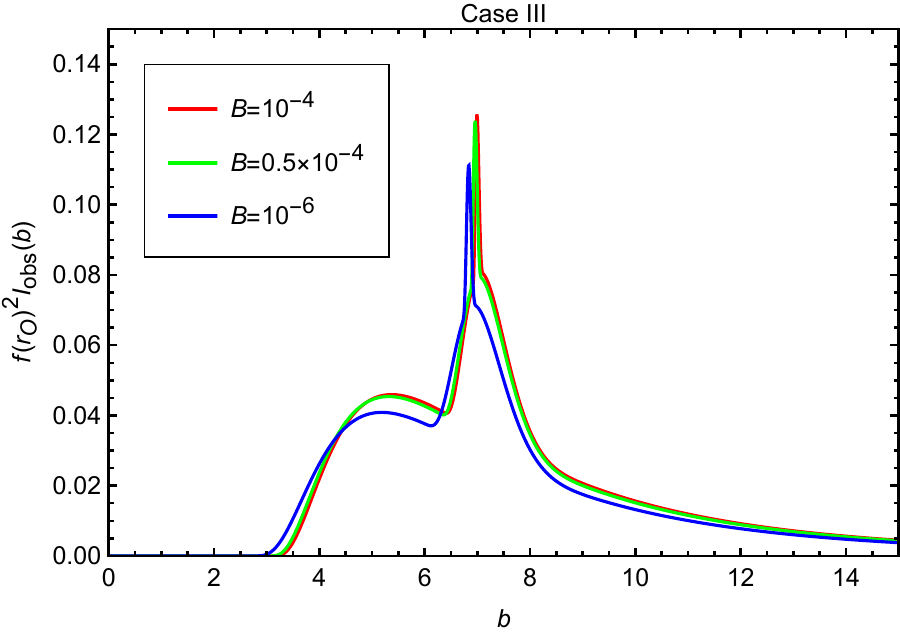}
\includegraphics[width=.28\textwidth]{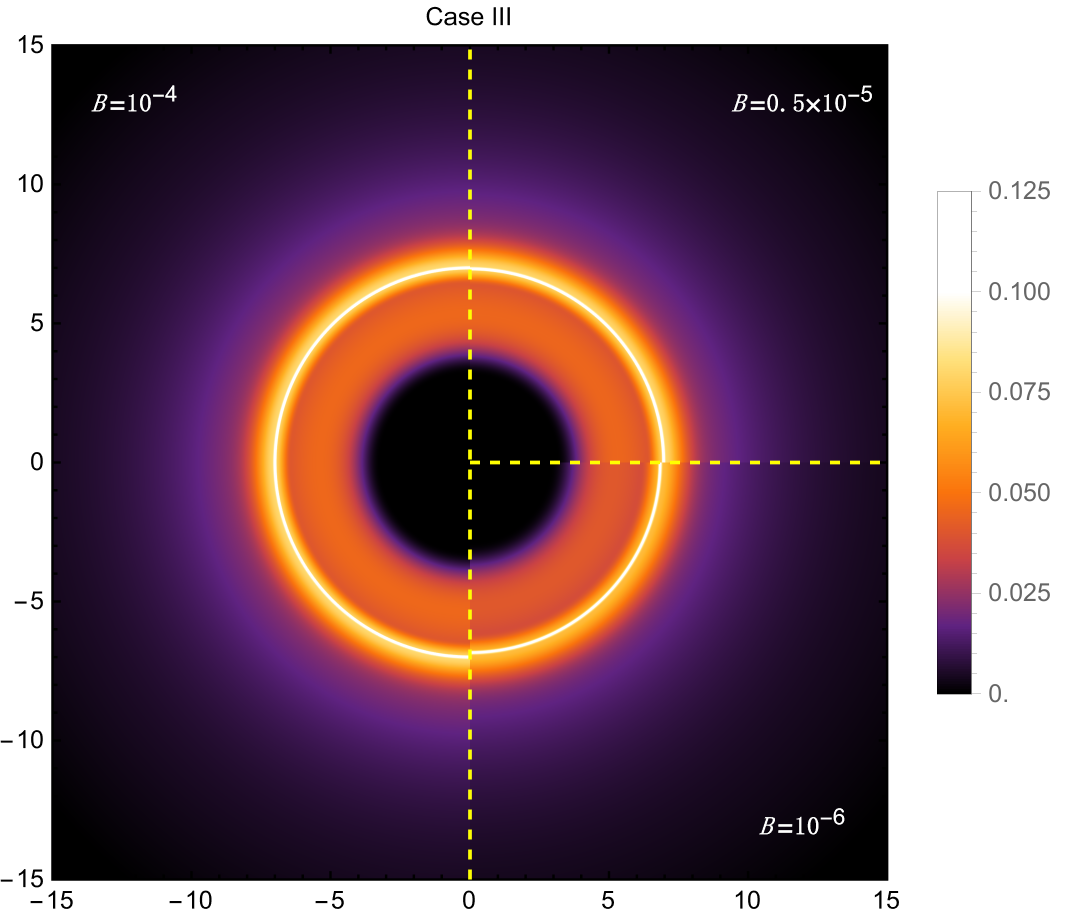}
% "\includegraphics" is very powerful; the graphicx package is already loaded
\caption{\label{FigvaryB}Observational appearances of a geometrically and optically thin disk with different emissivity profiles near black hole with $M=1$, $Q=2.0$, $A=0.3$, $\beta=0.2$ and $a=0.2$, viewed from a face-on orientation. The left column shows the profiles of various emissions $I_{\mathrm{em}}(r)$. The middle column exhibits the observed intensities $I_{\mathrm{obs}}(b)$ as a function of the impact parameter $b$. The red, green and blue curves correspond to $B=10^{-4}$, $B=0.5\times10^{-4}$ and $B=10^{-6}$, respectively. The right column shows the 2-dim density plots of the observed intensities $I_{\mathrm{obs}}(b)$.}
\end{figure*}
\begin{figure*}[htbp]
\centering % \begin{center}/\end{center} takes some additional vertical space
\includegraphics[width=.34\textwidth]{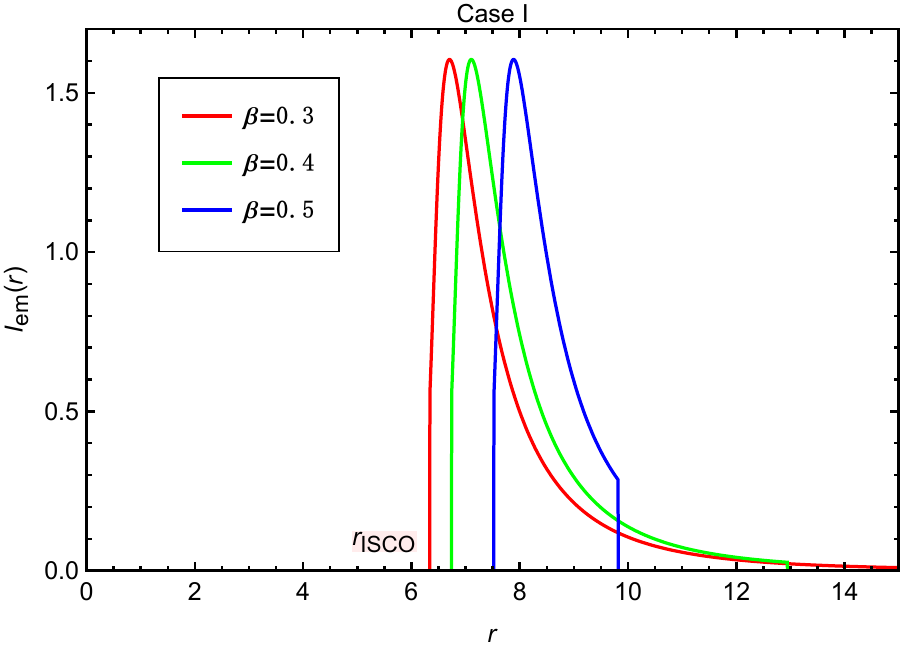}
\includegraphics[width=.34\textwidth]{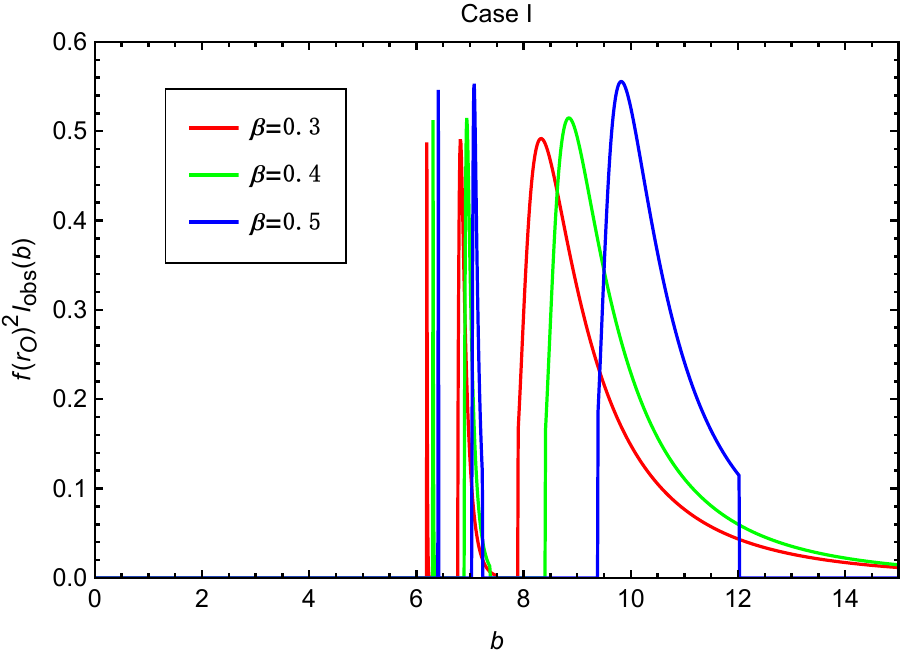}
\includegraphics[width=.28\textwidth]{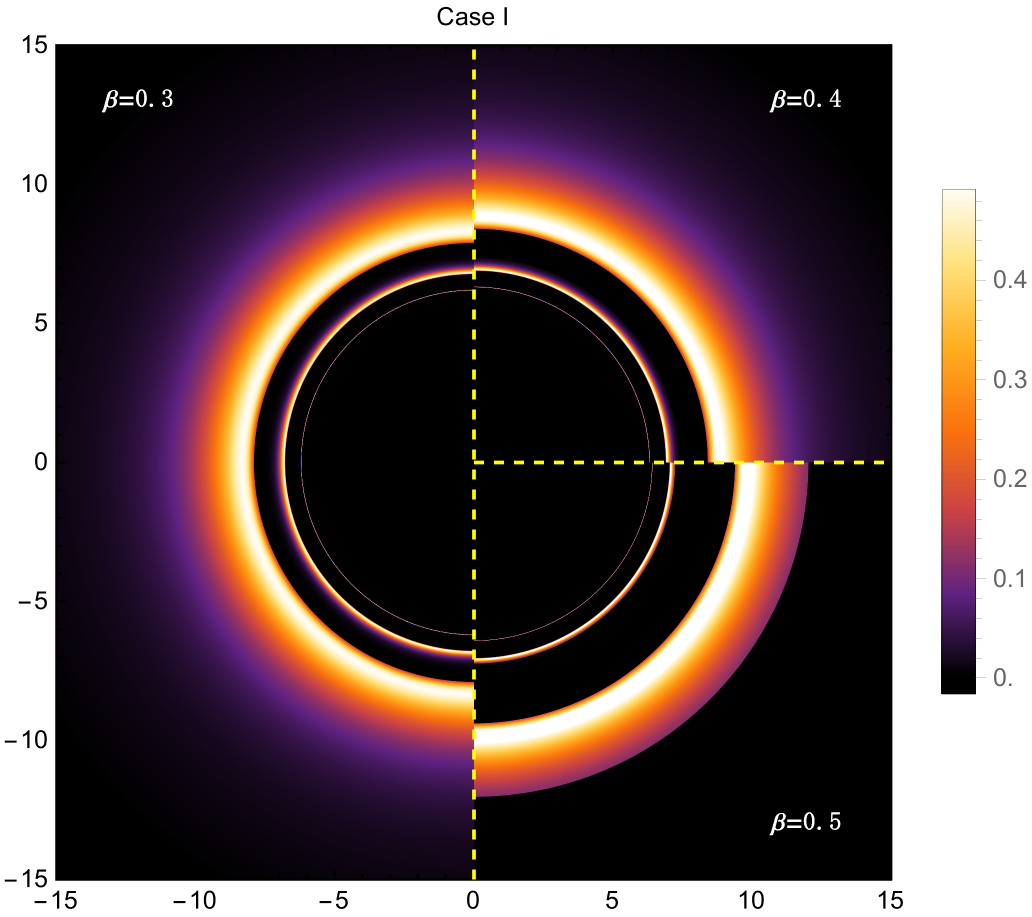}
\includegraphics[width=.34\textwidth]{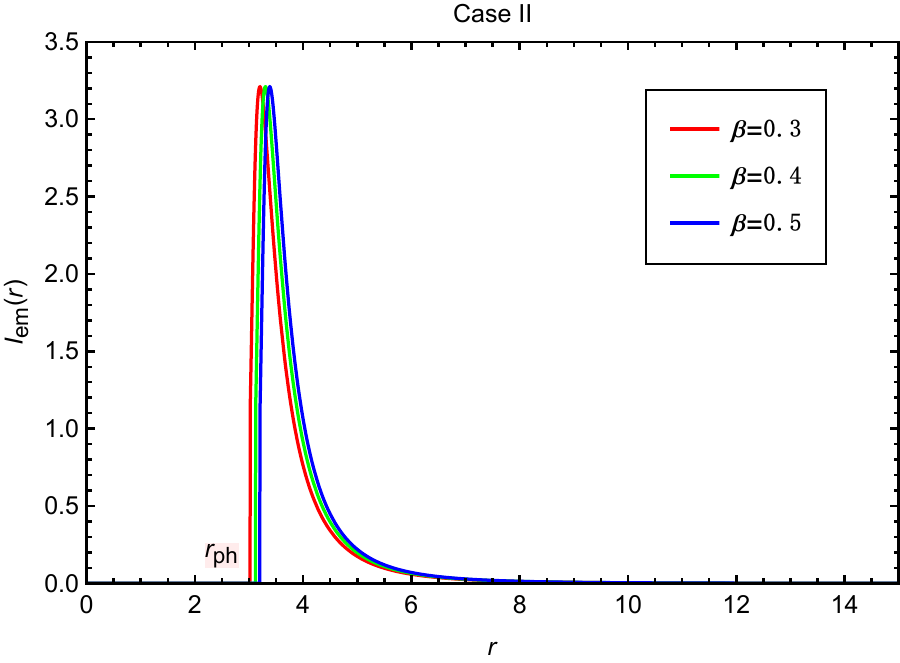}
\includegraphics[width=.34\textwidth]{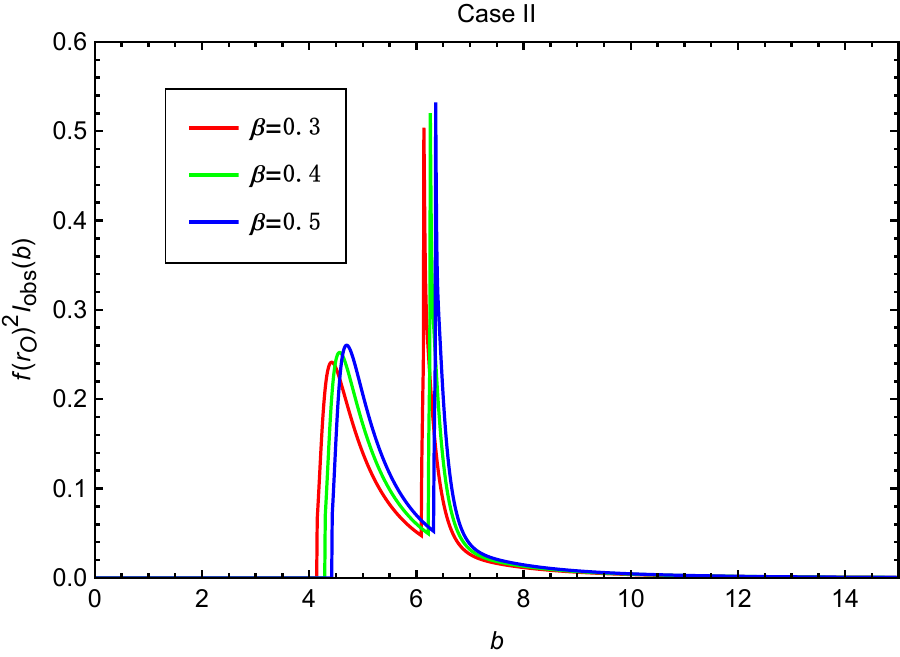}
\includegraphics[width=.28\textwidth]{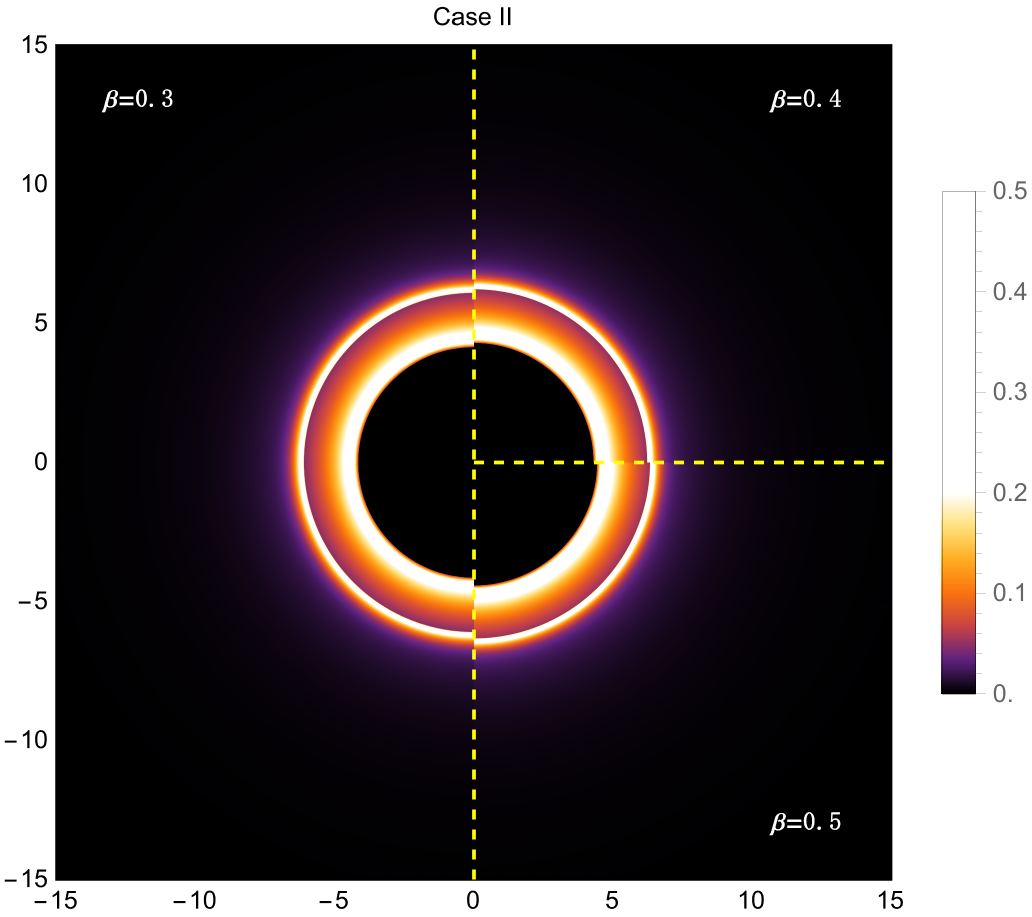}
\includegraphics[width=.34\textwidth]{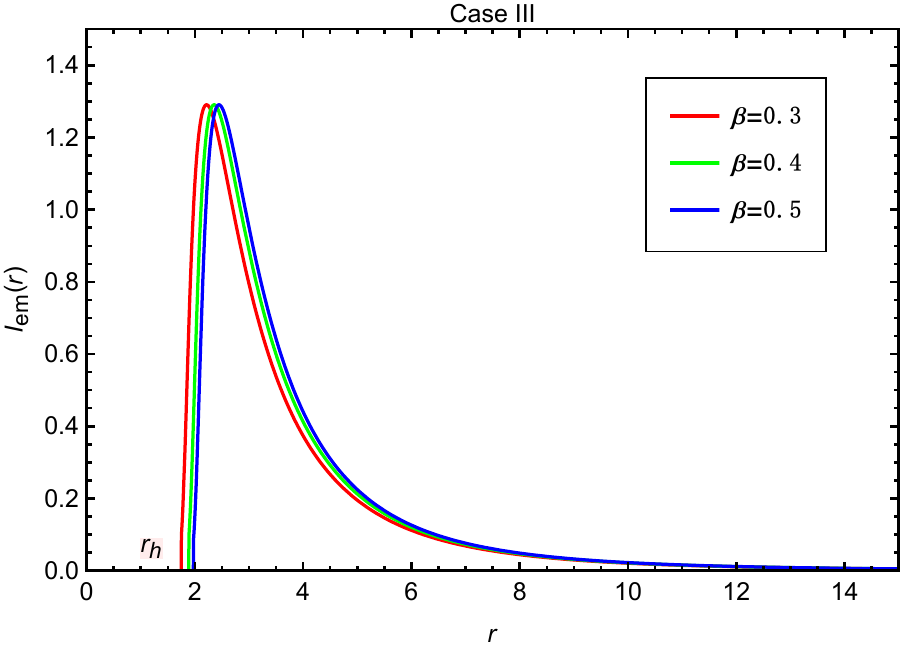}
\includegraphics[width=.34\textwidth]{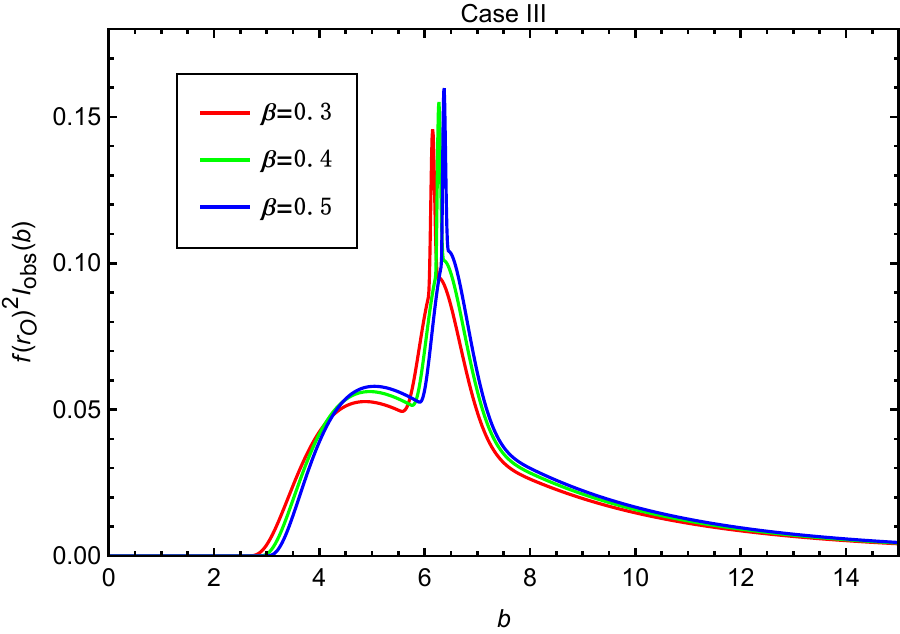}
\includegraphics[width=.28\textwidth]{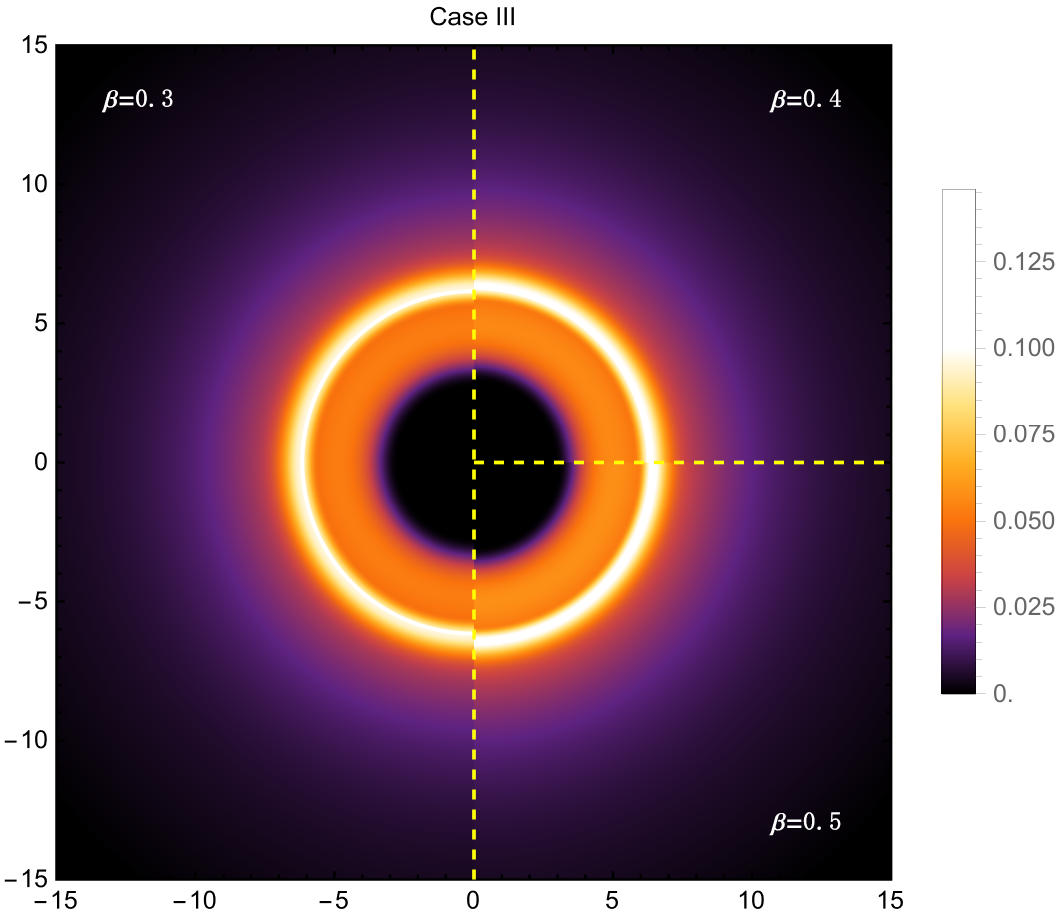}
% "\includegraphics" is very powerful; the graphicx package is already loaded
\caption{\label{Figvarybeita}Observational appearances of a geometrically and optically thin disk with different emissivity profiles near black hole with $M=1$, $Q=1.0$, $A=0.15$, $B=10^{-5}$ and $a=0.15$, viewed from a face-on orientation. The left column shows the profiles of various emissions $I_{\mathrm{em}}(r)$. The middle column exhibits the observed intensities $I_{\mathrm{obs}}(b)$ as a function of the impact parameter $b$. The red, green and blue curves correspond to $\beta=0.3$, $\beta=0.4$ and $\beta=0.5$, respectively. The right column shows the 2-dim density plots of the observed intensities $I_{\mathrm{obs}}(b)$.}
\end{figure*}
\begin{figure*}[htbp]
\centering % \begin{center}/\end{center} takes some additional vertical space
\includegraphics[width=.34\textwidth]{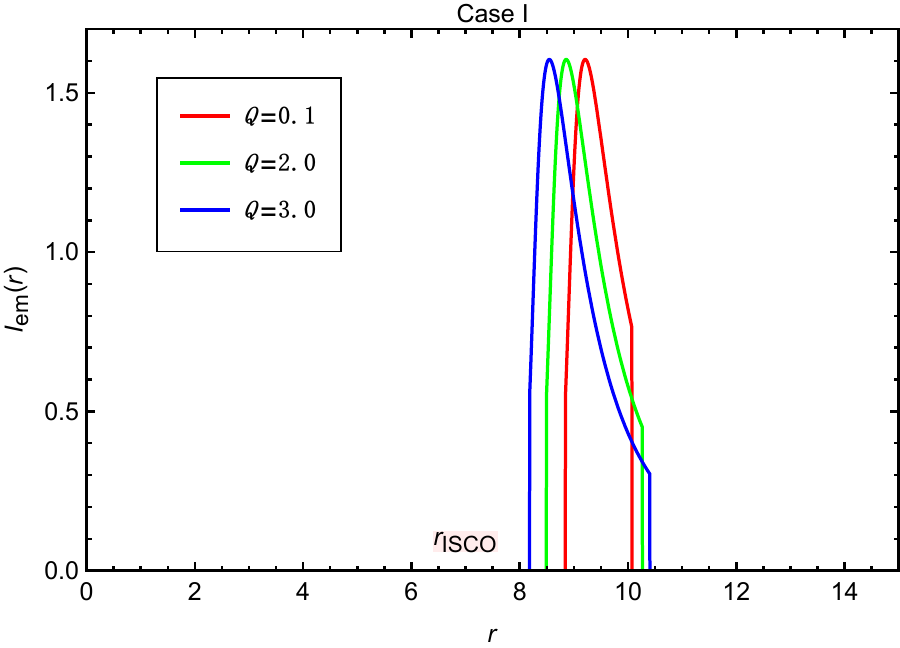}
\includegraphics[width=.34\textwidth]{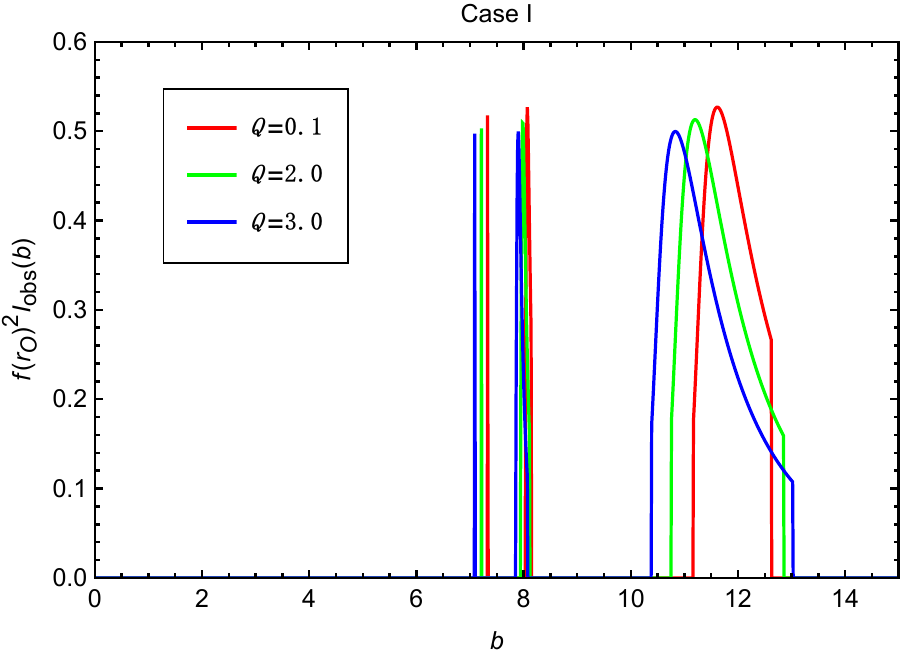}
\includegraphics[width=.28\textwidth]{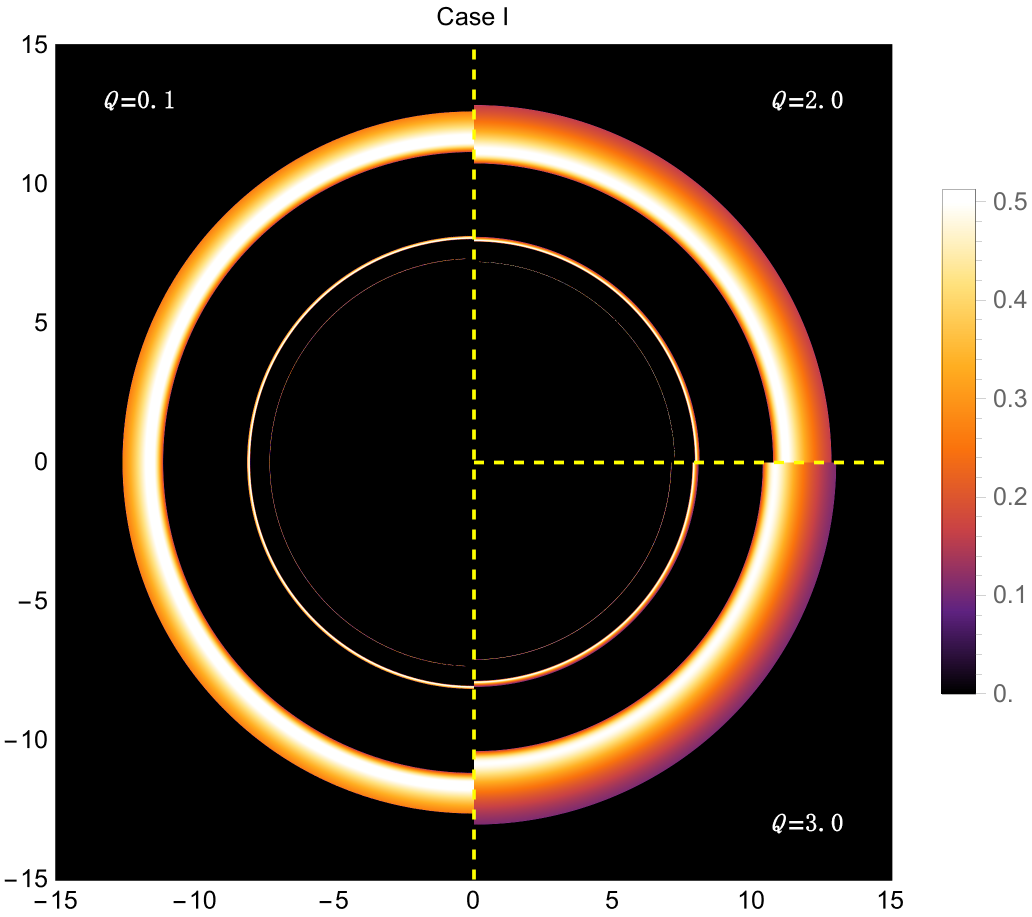}
\includegraphics[width=.34\textwidth]{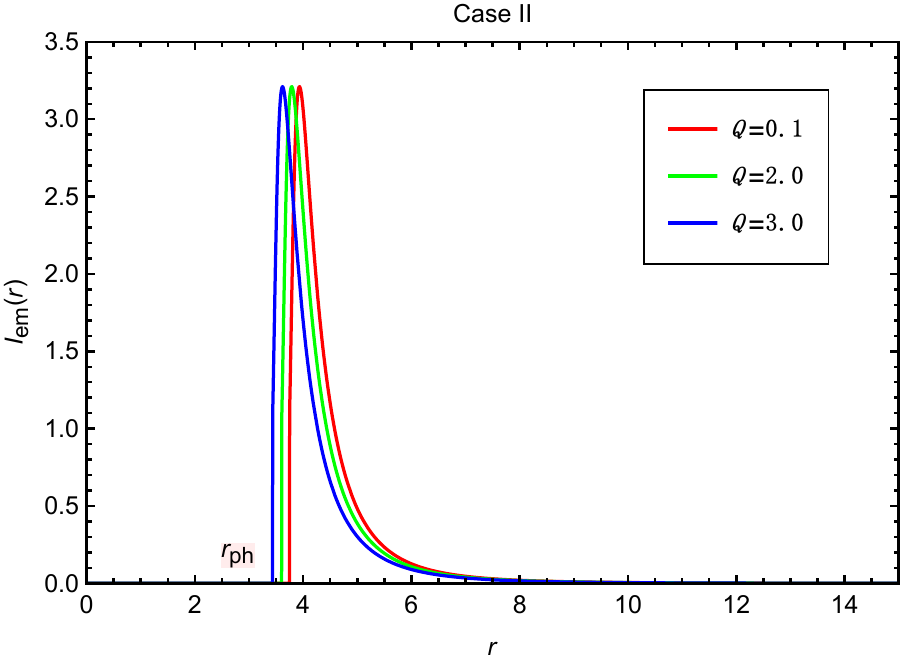}
\includegraphics[width=.34\textwidth]{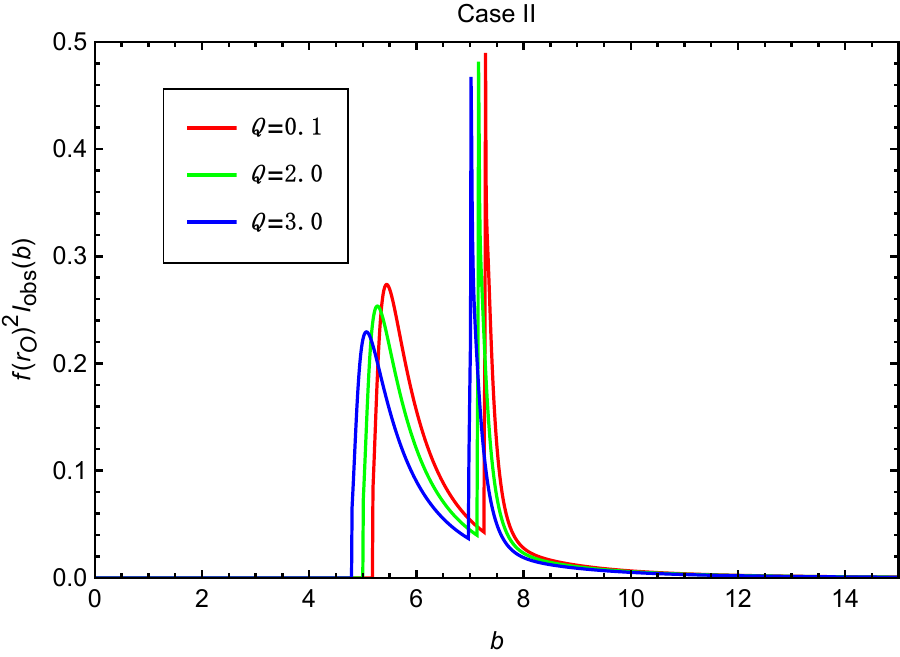}
\includegraphics[width=.28\textwidth]{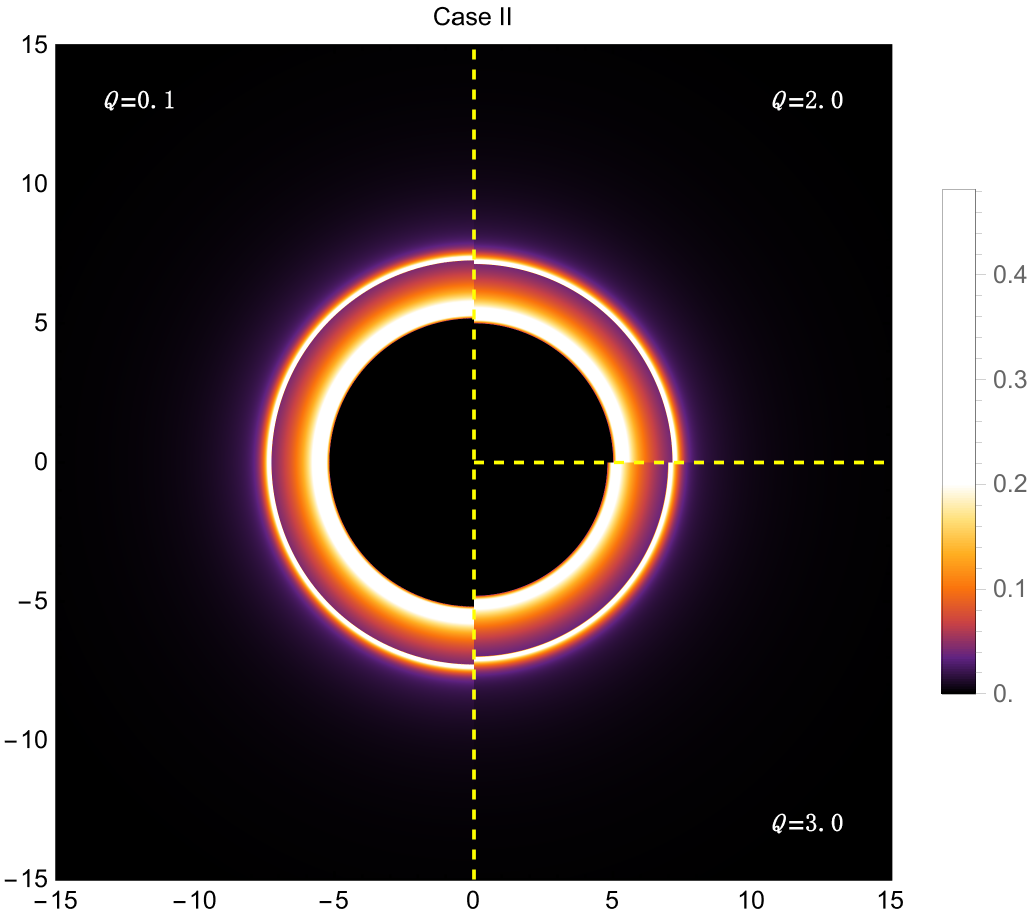}
\includegraphics[width=.34\textwidth]{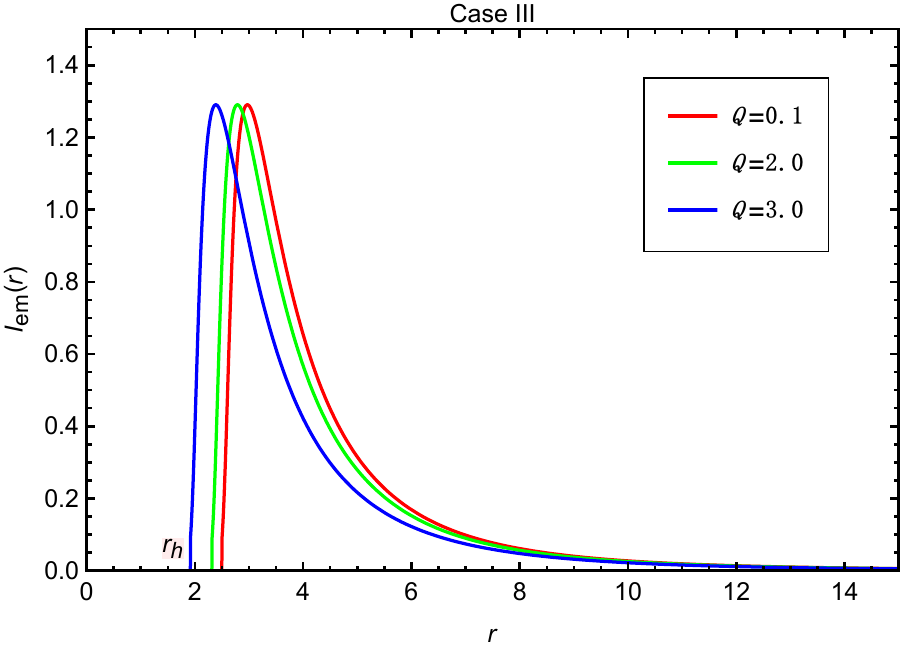}
\includegraphics[width=.34\textwidth]{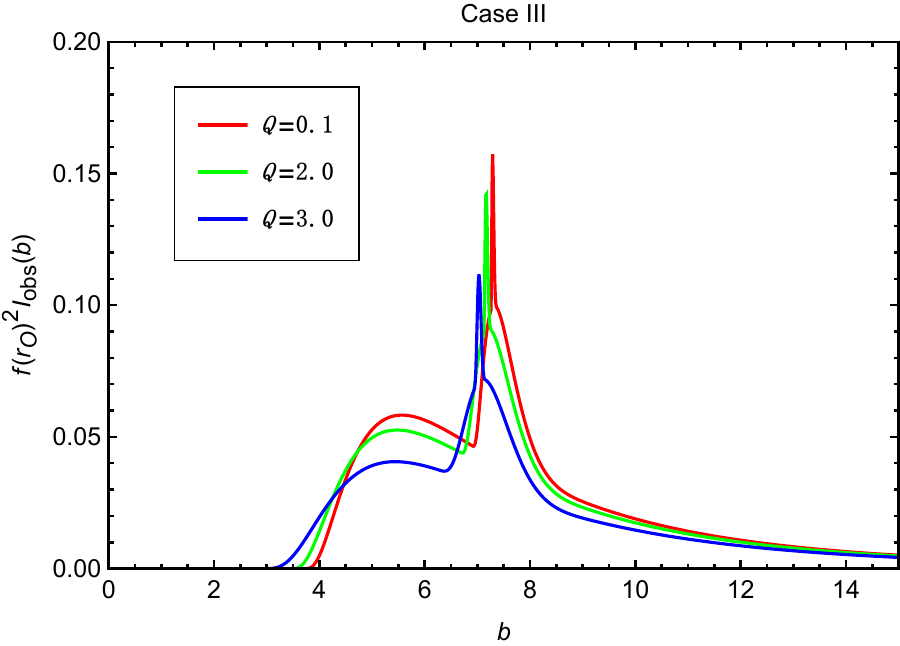}
\includegraphics[width=.28\textwidth]{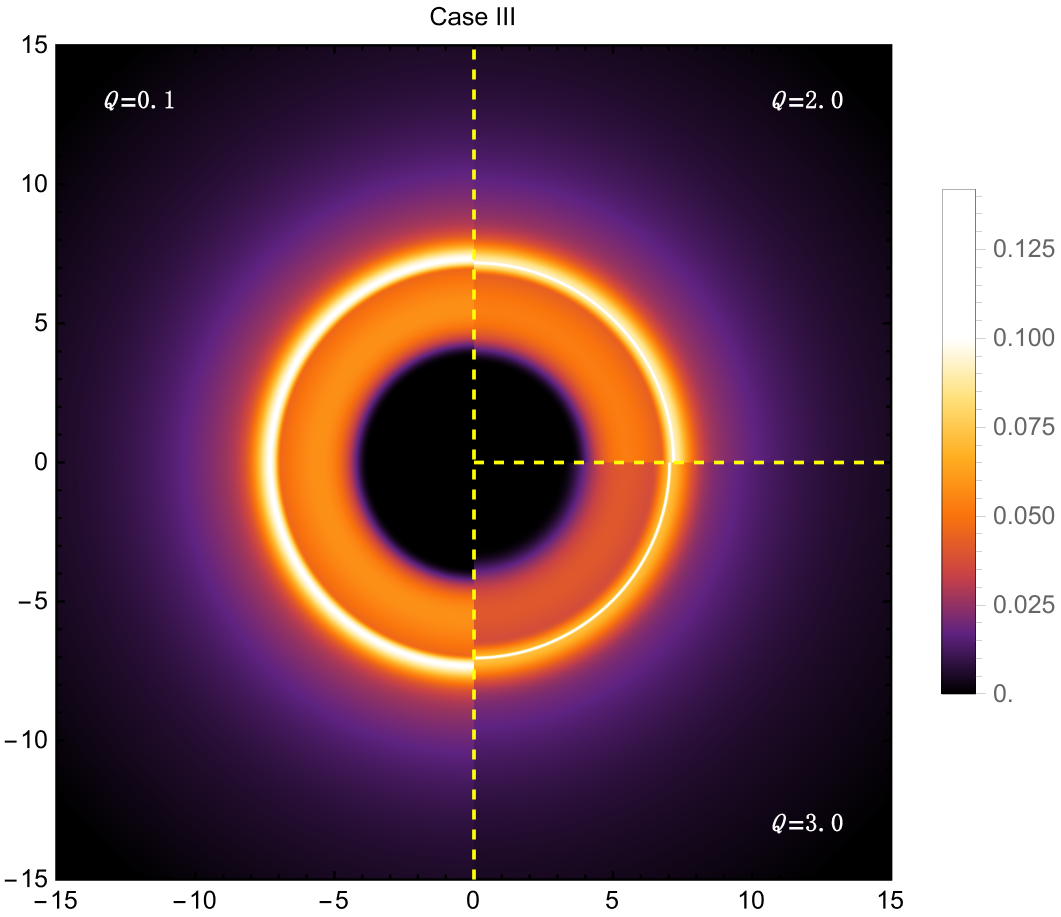}
% "\includegraphics" is very powerful; the graphicx package is already loaded
\caption{\label{FigvaryQ}Observational appearances of a geometrically and optically thin disk with different emissivity profiles near black hole with $M=1$, $A=0.4$, $B=10^{-4}$, $\beta=0.2$ and $a=0.2$, viewed from a face-on orientation. The left column shows the profiles of various emissions $I_{\mathrm{em}}(r)$. The middle column exhibits the observed intensities $I_{\mathrm{obs}}(b)$ as a function of the impact parameter $b$. The red, green and blue curves correspond to $Q=0.1$, $Q=2.0$ and $Q=3.0$, respectively. The right column shows the 2-dim density plots of the observed intensities $I_{\mathrm{obs}}(b)$.}
\end{figure*}
\begin{figure*}[htbp]
\centering % \begin{center}/\end{center} takes some additional vertical space
\includegraphics[width=.34\textwidth]{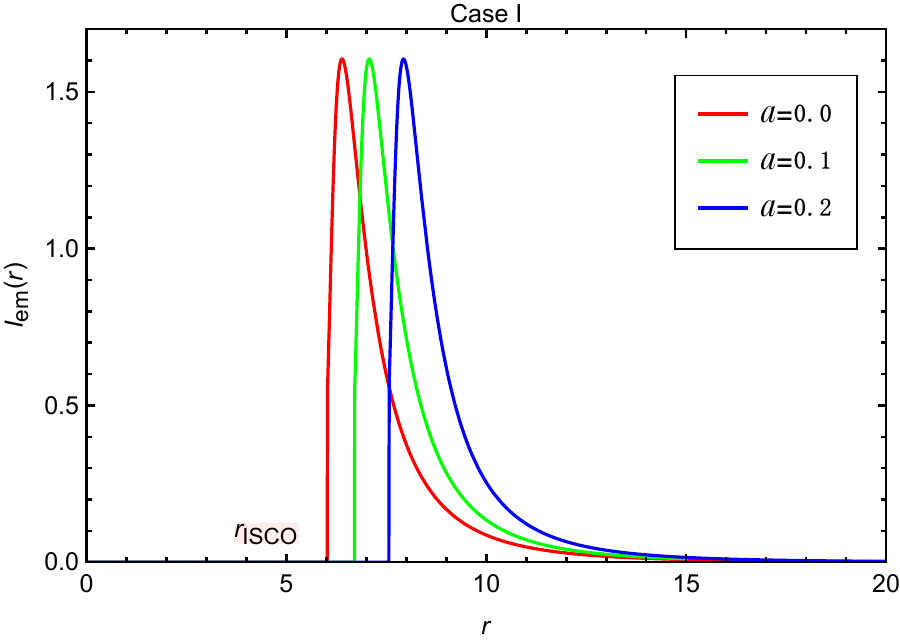}
\includegraphics[width=.34\textwidth]{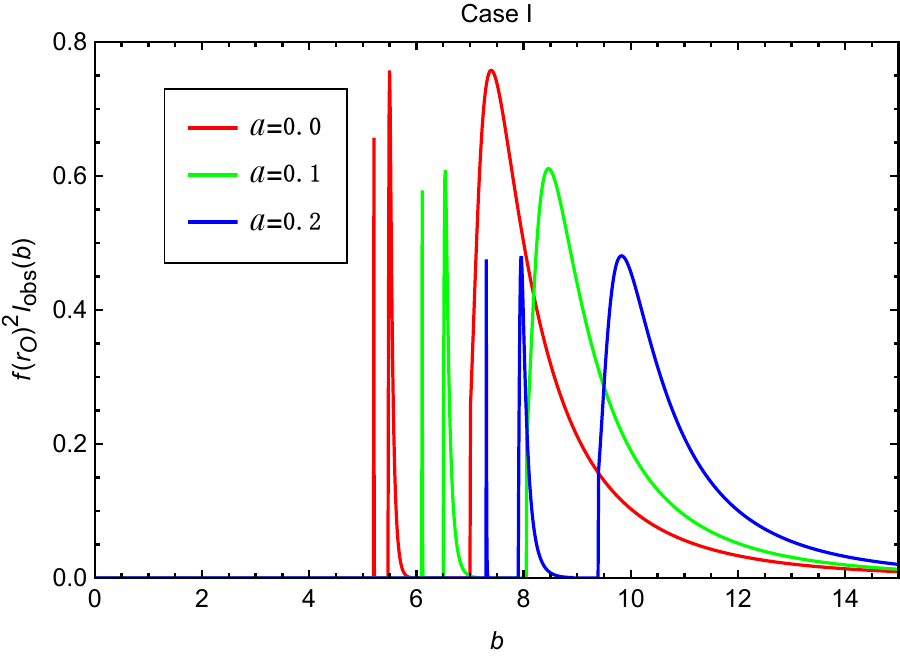}
\includegraphics[width=.28\textwidth]{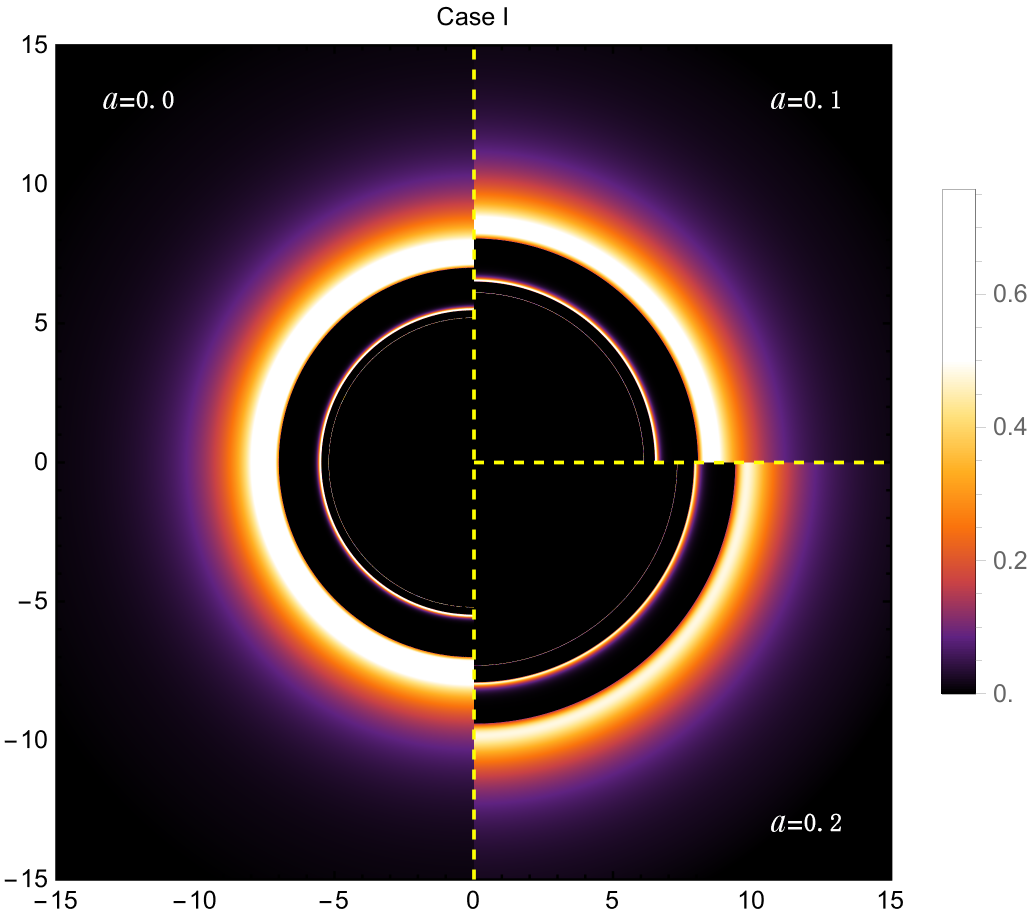}
\includegraphics[width=.34\textwidth]{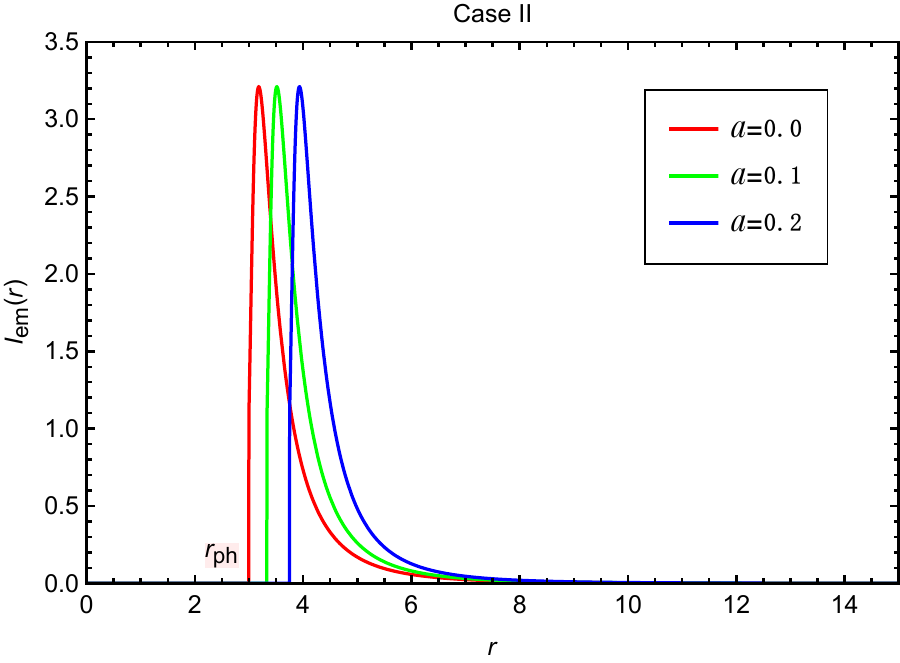}
\includegraphics[width=.34\textwidth]{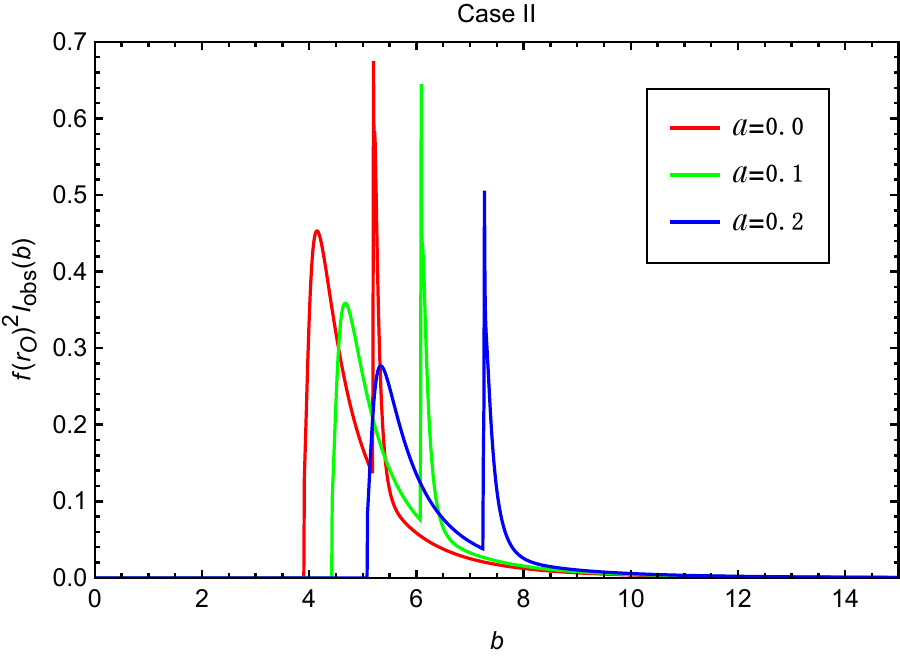}
\includegraphics[width=.28\textwidth]{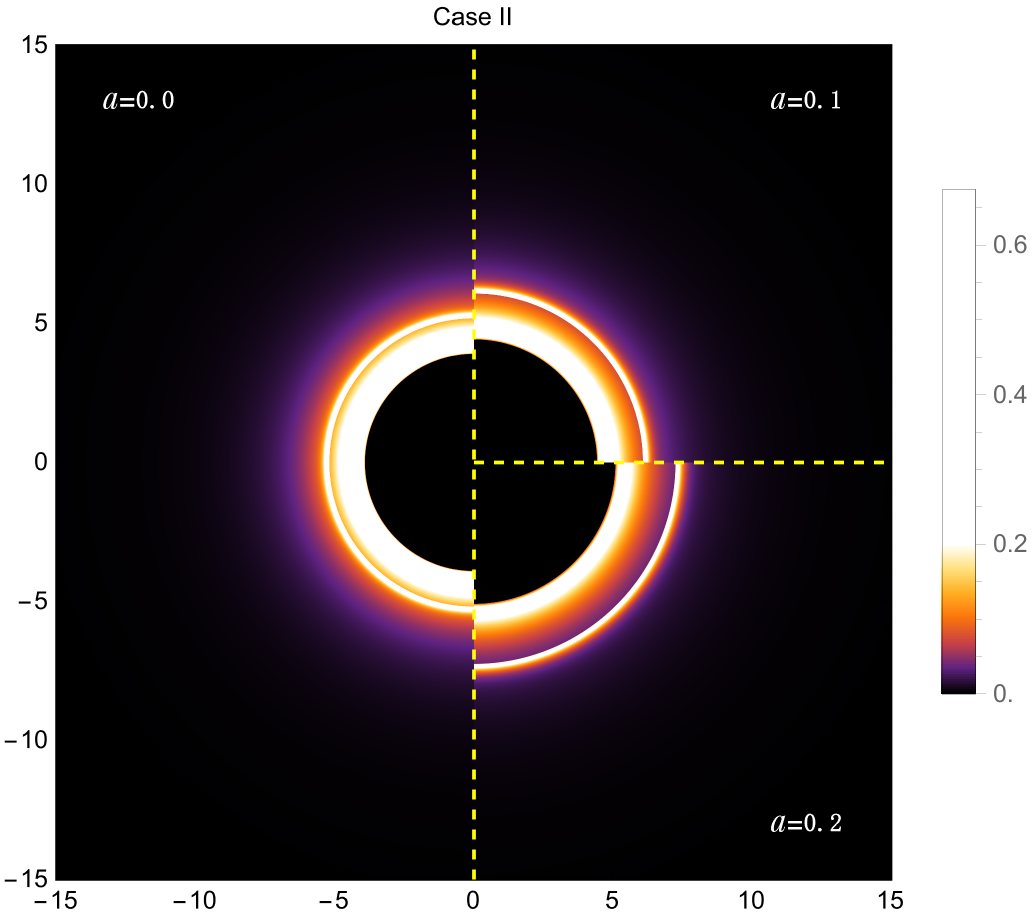}
\includegraphics[width=.34\textwidth]{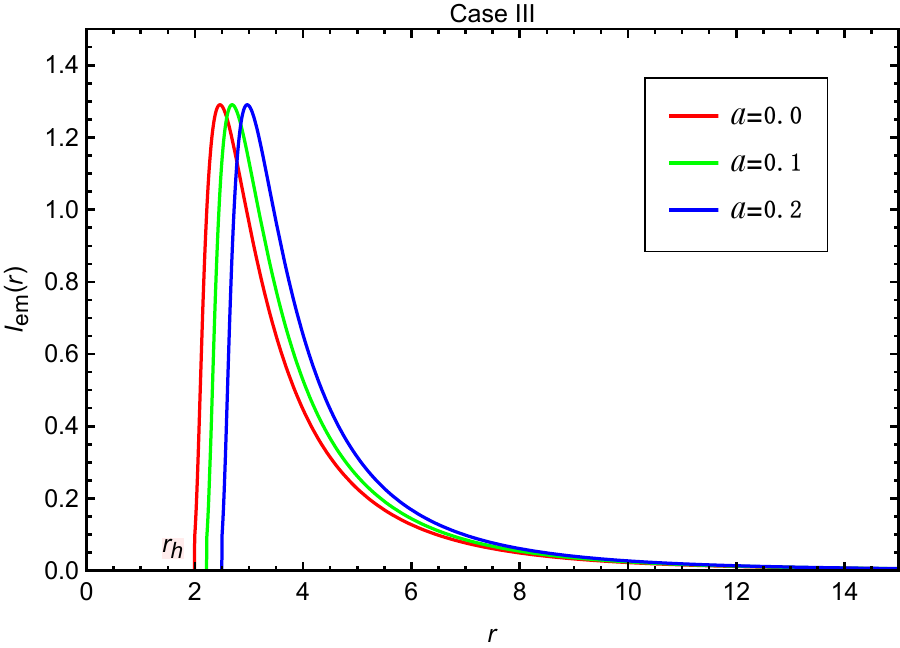}
\includegraphics[width=.34\textwidth]{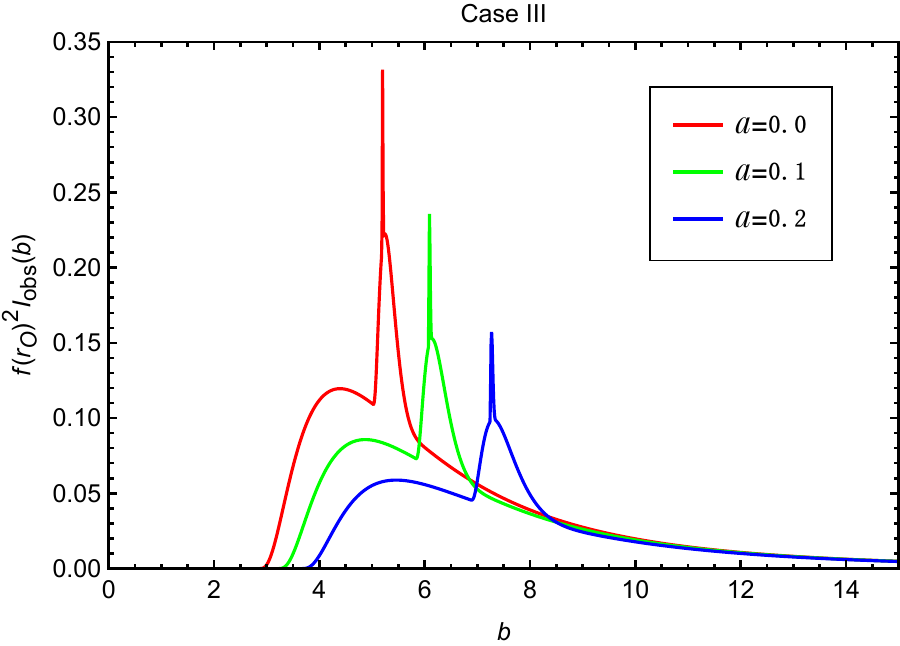}
\includegraphics[width=.28\textwidth]{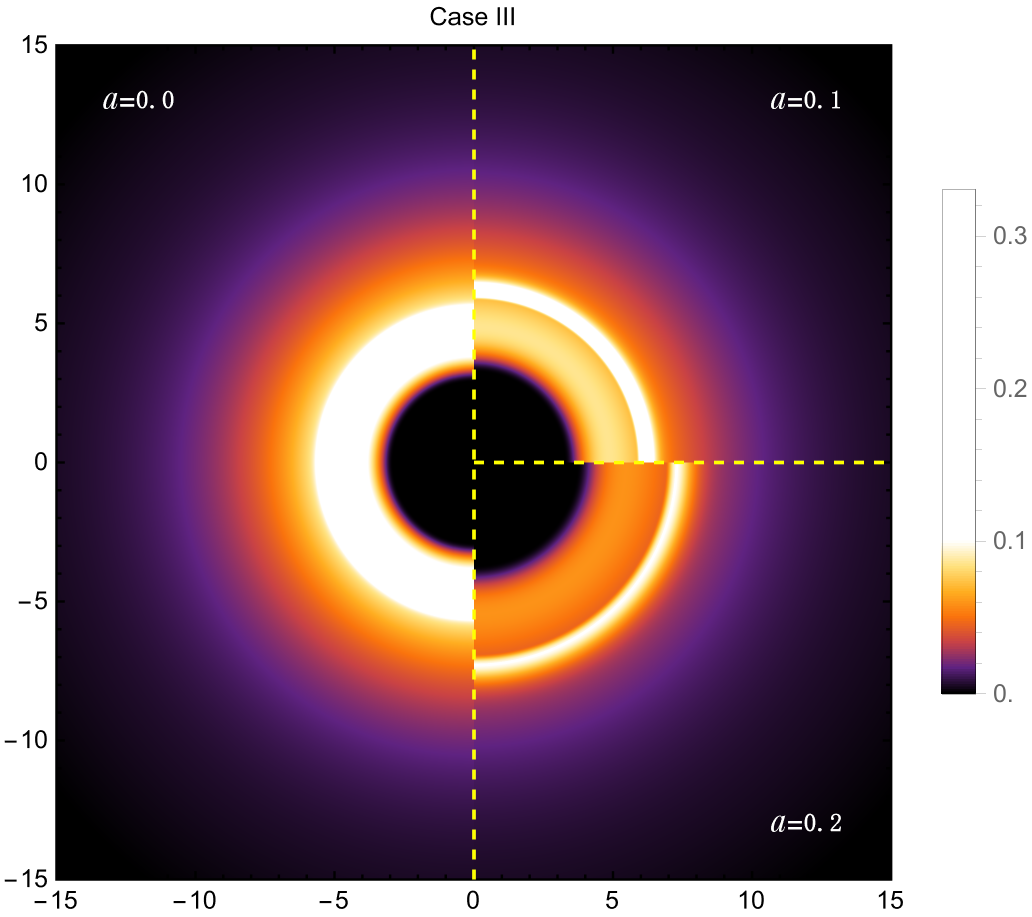}
% "\includegraphics" is very powerful; the graphicx package is already loaded
\caption{\label{Figvarya}Observational appearances of a geometrically and optically thin disk with different emissivity profiles near black hole with $M=1$, $Q=1.0$, $A=1.0$, $B=10^{-5}$ and $\beta=0.2$, viewed from a face-on orientation. The left column shows the profiles of various emissions $I_{\mathrm{em}}(r)$. The middle column exhibits the observed intensities $I_{\mathrm{obs}}(b)$ as a function of the impact parameter $b$. The red, green and blue curves correspond to $a=0.0$, $a=0.1$ and $a=0.2$, respectively. The right column shows the 2-dim density plots of the observed intensities $I_{\mathrm{obs}}(b)$.}
\end{figure*}
\subsection{Direct image, lensing ring and photon ring}
Upon acquiring the transfer functions, we can proceed to deduce the specific intensity based on Eq.~(\ref{lir}), given the emitted specific intensity. In this study, we parameterize the radiations intensity profile of the accretion disk as the recently introduced Gralla–Lupsasca–Marrone (GLM) disk model~~\cite{Gralla:2020srx}, which has been illustrated to be in a close agreement with
the observational predictions of general relativistic magnetohydrodynamics simulations of astrophysical accretion disks.
The radiations intensity profile of the GLM model is written
as~~\cite{Vincent:2022fwj}
\begin{equation}\label{GLMIem}
   I_{\text{em}}(r, \gamma, \alpha, \beta) =
\frac{
    \exp\left\{ -\frac{1}{2} \left[ \gamma + \operatorname{arcsinh}\left( \frac{r - \alpha}{\beta} \right) \right]^2 \right\}
}{
    \sqrt{(r - \alpha)^2 + \beta^2}
},
\end{equation}
where the shape of the radiation intensity profile \( I_{\rm em}(r) \) is characterized by three free parameters: \( \gamma \), \( \alpha \), and \( \beta \). Specifically, \( \gamma \) controls the steepness of the intensity increase from spatial infinity toward the peak; \( \alpha \) sets the radial location of the profile’s center; and \( \beta \) adjusts the overall width (dilation) of the profile. By tuning these parameters, we can construct suitable radiation intensity distributions tailored to different accretion scenarios. In this study, we focus on modeling the innermost emission regions of the accretion disk, corresponding to three representative radii: the innermost stable circular orbit (\( r_{\rm ISCO} \)), the photon sphere radius (\( r_{\rm ph} \)), and the outer event horizon (\( r_{\rm h} \)).
\begin{itemize}
  \item \textbf{Case I:} \\
  Parameters: \( \gamma = -2 \), \( \alpha = r_{\rm ISCO} \), \( \beta = M/4 \). \\
  We assume the ring-like accretion disk extends from the ISCO to the OSCO, with the emission profile modeled as a second-order power-law decay originating from \( r_{\rm ISCO} \), as described by Eq.~(\ref{RISCO}).

  \item \textbf{Case II:} \\
  Parameters: \( \gamma = -2 \), \( \alpha = r_{\rm ph} \), \( \beta = M/8 \). \\
  In this case, the emission region lies well outside the photon sphere, representing radiation from a more distant region of the disk.

  \item \textbf{Case III:} \\
  Parameters: \( \gamma = -3 \), \( \alpha = r_{\rm h} \), \( \beta = M/8 \). \\
  Here, the emission extends down to the vicinity of the event horizon, allowing us to explore the effects of near-horizon radiation.
\end{itemize}
In order to compare the observational results arising from different disk accretion models and various MCDF and cloud of strings parameters, we neglect the influence of the observer's position on the observed intensity, by presenting plots depicting $f(r_{\rm O})^2I_{\rm obs}(b)/I_0$.

As mentioned earlier, depending on the model parameters, black holes can be categorized into two types based on the trajectories of timelike particles: those that possess SCOs and those that do not. Considering the three GLM accretion disk parameter cases discussed above, all three cases can be applied to black holes with SCOs. However, for black holes lacking SCOs, only Case II and Case III are applicable. In the following analysis, we numerically study the observational appearances of the three accretion disk cases in spacetimes with SCOs, and present the results in Figs.~\ref{FigvaryAA}, \ref{FigvaryB}, \ref{Figvarybeita}, \ref{FigvaryQ}, and \ref{Figvarya}.

Figure~\ref{FigvaryAA} illustrates the emission profile \( I_{\rm em}(r) \) (left column), the observed intensity \( I_{\rm obs}(b) \) (middle column), and the two-dimensional observed image (right column) of the accretion disk for various values of the parameter \( A \).
We first analyze Case I shown in the first row. For a fixed value of \( A \), the observed intensity profile of the black hole consists of three distinct components arranged from left to right: the photon ring, the lensing ring, and the direct image. Both the lensing ring and the direct image exhibit well-defined inner and outer boundaries, which are clearly visible as concentric halos in the two-dimensional image. In contrast, the photon ring appears as a narrow spike in the intensity profile and is not readily discernible in the two-dimensional image without magnification. As the parameter \( A \) increases, the ISCO radius \( r_{\rm ISCO} \) first increases and then decreases, whereas the OSCO radius \( r_{\rm OSCO} \) exhibits an inverse trend, first decreasing and then increasing. This non-monotonic behavior is reflected in the left and right boundaries of the emission profile \( I_{\rm em}(r) \), and is also evident in Figures~\ref{FigEE} and~\ref{FigLL}. The structure and evolution of the observed intensities \( I_{\rm obs}(b) \) respond sensitively to these changes. The boundaries of the direct image component shift in accordance with the variations in \( r_{\rm ISCO} \) and \( r_{\rm OSCO} \), while the boundaries of the lensing ring and photon ring move outward monotonically as \( A \) increases, reflecting a more straightforward parameter dependence. It is also important to note that the direct image spans the widest range in \( b \) and contributes dominantly to the total observed intensity. The lensing ring makes only a minimal contribution to the overall brightness, and the photon ring's contribution is negligible. The two-dimensional observed image further visualizes this structural evolution. The left half, upper-right, and lower-right quadrants of the image correspond to \( A = 0.1 \), 0.15, and 1.0, respectively, with different ring-like features manifesting as distinct brightness distributions.

We then proceed to analyze Case II, presented in the second row of Fig.~\ref{FigvaryAA}. For a fixed value of \( A \), the observed intensity profile reveals a key distinction from the Case I model in terms of spatial structure: the direct image is consistently located inward of the lensing and photon ring intensities. From the center outward, the decline of the direct image intensity is followed by a narrow spike corresponding to the lensing ring. Notably, the photon ring manifests as an even sharper and narrower peak embedded within the span of the lensing ring, rendering the two effectively indistinguishable. It is evident that the lensing ring contributes only marginally to the total observed brightness, while the photon ring's contribution is negligible. As \( A \) increases, the photon sphere radius \( r_{\rm ph} \) increases, leading to a slight rightward shift of the \( I_{\rm em}(r) \) curve. Correspondingly, in the observed intensity profiles, both the peak of the direct image and the combined spike of the lensing and photon rings exhibit a slight rightward shift, accompanied by a modest increase in their amplitudes.

Finally, we analyze Case III with varying values of $A$ as shown in the third row of Fig.~\ref{FigvaryAA}. In this case, the lensing ring and photon ring are again superimposed on the direct image. The direct emission begins at the event horizon radius \( r_{\rm h} \), exhibiting an initial rise followed by a gradual decline. During this decline, a compound spike appears due to the overlapping contributions from the lensing ring and the photon ring. Among them, the lensing ring is relatively more prominent, yet the direct image remains the dominant contributor to the overall brightness. The photon ring continues to be entirely negligible due to its narrow width. As the parameter \( A \) increases, the entire observed image of the accretion disk shifts outward, reflecting the corresponding expansion of the emission structure in both the radial intensity profiles and the two-dimensional images.

We next examine the impact of parameter \( B \) on the optical appearance of the accretion disk models for the three cases, as shown in Fig.~\ref{FigvaryB}. For Case I, we observe that as \( B \) decreases, the ISCO radius shifts inward while the OSCO radius moves outward in the emission profile. This results in an overall leftward shift of the observed intensity curves. However, the right boundaries of both the direct image and the lensing ring move outward, a trend that is more intuitively reflected in the two-dimensional density plots. We also find that the gap between the direct image and the lensing ring narrows as \( B \) decreases. When \( B = 10^{-6} \), the lensing ring partially overlaps with the direct image. In contrast, for the Case II and Case III models, the effect of \( B \) is more straightforward: as \( B \) increases, the entire optical appearance of the accretion disk shifts outward in both models.

The influence of the parameter \( \beta \) on the optical appearance of the accretion disk is illustrated in Fig.~\ref{Figvarybeita}. It can be seen that, as \( \beta \) increases, the variation trends in the observed intensities for all three cases are similar to those observed with increasing \( B \). In fact, the changes in the lapse function \( f(r) \) resulting from variations in \( B \) and \( \beta \) exhibit the same qualitative behavior.

Fig.~\ref{FigvaryQ} presents the changes in the observed appearance of the accretion disk as the parameter $Q$ varies. For the Case I model, it is observed that as \( Q \) increases, the ISCO radius decreases while the OSCO radius increases in the emitted profile. In the observed intensity plots, the left edge of the direct image shifts leftward while the right edge moves rightward, resulting in a broader ring-like structure in the two-dimensional image. In contrast, the left and right edges of the lensing ring both shift to the left, as does the photon ring spike. For the Case II and Case III models, the behavior is more straightforward: as \( Q \) increases, the entire image shifts leftward, which corresponds to an inward shift in the two-dimensional brightness distribution.

Finally, the influence of the string cloud parameter \( a \) on the optical appearance of the accretion disk is examined, as depicted in Fig.~\ref{Figvarya}. In our earlier constraints on model parameters using EHT observational data, we have already found that string clouds significantly affect the black hole shadow radius. In fact, the string cloud parameter \( a \) appears to have the strongest effect on the optical appearance of the accretion disk among all five parameters analyzed. For the Case I model, as \( a \) increases, it leads to a larger ISCO radius and a smaller OSCO radius (in the emitted profile figure, the OSCOs lie beyond the visible range of the plot). Due to the relatively large OSCOs radii for timelike particles, the observed intensities of the direct image and the lensing ring drop to zero at low intensities, causing the outer boundaries of the corresponding rings in the two-dimensional density plots to appear less distinct. The ring structures corresponding to the direct image, lensing ring, and photon ring all shift outward with increasing \( a \), accompanied by a noticeable decrease in their peak intensities. The same trend is observed in Case II and Case III models, where larger values of \( a \) lead to outward shifts in the images and suppressed peak brightness.

\section{Conclusions and discussions}
\label{conclusion}
In this work, we have explored the geodesic structure, shadow, and optical appearance of a black hole immersed in both a cloud of strings and a cosmological dark fluid described by the modified Chaplygin-like equation of state (MCDF). The fluid is characterized by a generalized equation of state $p = A\rho - B/\rho^{\beta}$ and an additional parameter $Q$ that modifies the energy density. The influence of both the string cloud and MCDF parameters on the dynamics of particles and photons around the black hole has been analyzed in detail.

By studying the effective potential and epicyclic frequencies, we have demonstrated that the existence of innermost stable circular orbits (ISCOs) and outermost stable circular orbits (OSCOs) for timelike particles is significantly affected by both the MCDF and string cloud parameters. These parameters also affect the orbital conserved quantities and Keplerian frequency, modifying the motion of matter in the vicinity of the black hole. For null geodesics, we confirmed that the photon orbits are unstable, and the presence of a cosmological horizon further modifies the propagation of light rays in the spacetime.

Assuming that the influence of the MCDF at spatial infinity mimics that of a cosmological constant in a de Sitter background, we employed Event Horizon Telescope (EHT) observations of the shadow radii of Sgr A* and M87* to constrain the parameters of the MCDF and cloud of strings. This approach allows for a phenomenological connection between the properties of the black hole and the cosmological background.

We further investigated the optical images of the black hole, surrounded by various emissivity profiles of geometrically and optically thin accretion disks. By employing the method developed by Wald and collaborators, the light rays were classified into direct emission, lensing ring, and photon ring, based on their impact parameters. Our findings reveal that due to the existence of OSCO, outer boundaries may form in both the direct and lensing ring images. As in our previous study, the observed brightness is predominantly attributed to the direct emission component, while the lensing ring contributes marginally, and the photon ring contributes negligibly due to extreme demagnetization effects.

The parameters of the MCDF and string clouds significantly affect the key features of the black hole spacetime, including the radii of the event horizon, photon sphere, and ISCO/OSCO, which in turn influence the appearance of the shadow and the optical image structure. These findings highlight the rich phenomenology that arises when both stringy and dark fluid components are present.

Nevertheless, several limitations of our study must be acknowledged. In particular, we have not accounted for the dynamical evolution of the expanding universe and its influence on light propagation. While we approximated the asymptotic behavior of the MCDF spacetime by a de Sitter-like model, a full treatment would require incorporating the time-dependent cosmological background into the analysis of both black hole shadow formation and accretion imaging. Moreover, the relationship between the equation of state for dark energy in FLRW cosmology and its analog in black hole spacetimes remains unclear, due to differences in symmetry and field dynamics.

Future directions may include: (i) a systematic investigation of how expanding universe models impact the observable features of black holes, especially the shadow radius and ring structure; (ii) exploration of the connections between cosmological dark energy models and local black hole physics; and (iii) the study of rotating black holes or more general spacetime geometries immersed in multiple cosmic fluids. Such extensions would contribute toward a more comprehensive understanding of black hole phenomenology in realistic cosmological environments.

\section*{Funding information}
This work was supported by the China Scholarship Council (CSC) under Grant No. 202308140133, the National Natural Science Foundation of China under Grant No. 12305070, and the Basic Research Program of Shanxi Province under Grant Nos. 202303021222018 and 202303021221033. It was also supported by the National Research Foundation of Korea (NRF) under Grant No. RS-2020-NR049598 through the Center for Quantum Spacetime (CQUeST) at Sogang University.

\bibliography{apssamp}

\end{document}